\documentclass[aps,pra,10pt,notitlepage,nofootinbib,tightenlines,floatfix,
twocolumn,superscriptaddress]{revtex4-2}

\usepackage[english]{babel}
\usepackage{makecell}

\usepackage[margin=50pt]{geometry}

\usepackage[colorlinks=true, allcolors=blue]{hyperref}
\hypersetup{
    colorlinks=true,       
    linkcolor=red,          
    citecolor=magenta,        
    filecolor=magenta,      
    urlcolor=cyan,           
    runcolor=cyan
}
\usepackage{bm,bbm,amssymb,amsmath,amsfonts,amsthm,mathrsfs,MnSymbol,times}
\usepackage{natbib}
\usepackage{graphicx}
\usepackage{amsmath}
\usepackage{multirow}
\usepackage{color}
\usepackage{bbold}
\usepackage[utf8]{inputenc}
\usepackage{empheq}
\usepackage{soul}
\usepackage[ampersand]{easylist} 
\usepackage[normalem]{ulem}
\usepackage{braket}
\usepackage{enumerate}
\usepackage[shortlabels]{enumitem}
\usepackage{blkarray}

\newcommand{\eq}[1]{\begin{align}#1\end{align}}
\newcommand{\seq}[1]{\begin{align}\begin{split}#1\end{split}\end{align}}

\newcommand{\bel}{\begin{easylist}[itemize]}
\newcommand{\eel}{\end{easylist}}
\newcommand{\vsigma}{\vec{\sigma}}
\newcommand{\sx}{\sigma_x}
\newcommand{\sy}{\sigma_y}
\newcommand{\sz}{\sigma_z}

\newcommand{\vtheta}{\vec{\theta}}

\newcommand{\hn}{\hat{n}}

\newcommand{\mrm}{\mathrm}

\newcommand{\mC}{\mathcal{C}}

\newcommand{\tr}{\mathrm{Tr}}

\newcommand{\tH}{\tilde{H}}
\newcommand{\sX}{\sqrt{X}}

\renewcommand{\l}{\left}
\renewcommand{\r}{\right}
\DeclareMathOperator{\cosc}{cosc}

\DeclareMathOperator{\sinc}{sinc}

\definecolor{ao(english)}{rgb}{0.0, 0.5, 0.0}

\graphicspath{ {figures/} }

\begin{document}

\title{Sparse Non-Markovian Noise Modeling of Transmon-Based Multi-Qubit Operations}
\author{Yasuo Oda}
\affiliation{William H. Miller III Department of Physics \& Astronomy,
Johns Hopkins University, Baltimore, Maryland 21218, USA}
\author{Kevin Schultz}
\affiliation{Johns Hopkins University Applied Physics Laboratory, Laurel, Maryland 20723, USA}
\author{Leigh Norris}
\affiliation{Johns Hopkins University Applied Physics Laboratory, Laurel, Maryland 20723, USA}
\author{Omar Shehab}
\affiliation{IBM Quantum, IBM Thomas J Watson Research Center, Yorktown Heights, NY, USA}
\author{Gregory Quiroz}
\affiliation{William H. Miller III Department of Physics \& Astronomy,
Johns Hopkins University, Baltimore, Maryland 21218, USA}
\affiliation{Johns Hopkins University Applied Physics Laboratory, Laurel, Maryland 20723, USA}       
\begin{abstract}
    The influence of noise on quantum dynamics is one of the main factors preventing current quantum processors from performing accurate quantum computations. Sufficient noise characterization and modeling can provide key insights into the effect of noise on quantum algorithms and inform the design of targeted error protection protocols. However, constructing effective noise models that are sparse in model parameters, yet predictive can be challenging. In this work, we present an approach for effective noise modeling of multi-qubit operations on transmon-based devices. Through a comprehensive characterization of seven devices offered by the IBM Quantum Platform, we show that the model can capture and predict a wide range of single- and two-qubit behaviors, including non-Markovian effects resulting from spatio-temporally correlated noise sources. The model's predictive power is further highlighted through multi-qubit dynamical decoupling demonstrations and an implementation of the variational quantum eigensolver. As a training proxy for the hardware, we show that the model can predict expectation values within a relative error of 0.5\%; this is a 7$\times$ improvement over default hardware noise models. Through these demonstrations, we highlight key error sources in superconducting qubits and illustrate the utility of reduced noise models for predicting hardware dynamics.
\end{abstract}

\maketitle

\section{Introduction}

Superconducting qubits have emerged as one of the most promising quantum technologies for realizing practical quantum computing.
Many types of superconducting qubit devices have been developed, and the field is rapidly evolving~\cite{kjaergaard2020,krantz2019}. 
Among the assortment of superconducting qubits, the transmon is specifically designed to minimize sensitivity to charge noise~\cite{koch2007}, and has been at the core of recent breakthroughs~\cite{arute_quantum_2019,kim_evidence_2023,wang_towards_2022,barends2014superconducting,jurcevic2021demonstration,chen2021exponential}.
With the advent of enhanced control capabilities in cloud-based quantum computing such as IBM's Quantum Platform (IBMQP)~\cite{alexander_qiskit_2020}, transmon-based devices have become a mainstream architecture employed to test state-of-the-art quantum algorithms.

Despite advancements in scale and quality, noise in superconducting qubit architectures poses a crucial challenge. In order to manage noise in quantum applications, a wide range of techniques have been devised to suppress~\cite{
werschnik2007quantum,dalessandro2021introduction,viola1999dynamical,lidar2013quantum}, avoid~\cite{grassl1997codes,devitt2013quantum,lidar1998decoherence,kempe2001theory,lidar2014review}, correct~\cite{gottesman1997stabilizer,terhal2015quantum}, and mitigate~\cite{cai2023quantum,li2017efficient,temme2017error,endo2018practical,maciejewski2020mitigation,nachman2020unfolding,nation2021scalable,bravyi2021mitigating} errors. The benefits of many of these approaches have been showcased on superconducting hardware via demonstrations of dynamical decoupling (DD)~\cite{pokharel2018dd, pokharelDemonstrationAlgorithmicQuantum2022, Zhou2023, pokharel2024better, singkanipa2024demonstration}, decoherence-free subspaces~\cite{Mena-Lopez:2023aa, quiroz2024dynamically, han2024protecting}, quantum error correcting codes~\cite{harper2019ftgates, andersen2020repeated, chen2021qec, krinner2022realizing, miao2022overcoming, sivak2023real,ai2023suppressing}, and various quantum error mitigation protocols~\cite{Dumitrescu2018vqe,kandala2019error,kim2023scalable,kim2023evidence}.

Many error management (EM) protocols rely on knowledge of the underlying noise processes. Such information can be leveraged for the purposes of ensuring that specific assumptions are satisfied (e.g., Markovianity). It can also be crucial in assessing the inherent robustness and susceptibility of an EM protocol -- and more generally, any quantum algorithm -- to hardware-relevant noise sources. These can be key steps in properly devising targeted EM protocols and noise robust quantum algorithms.

The most common approaches to modeling noisy quantum dynamics rely on the quantum channel formalism, quantum master equations (MEs), or the stochastic Hamiltonian formalism. The first approach involves applying a composition of quantum error maps before or after an ideal gate operation. The channel representation aims to sacrifice modeling of intra-gate dynamics for simplicity and fewer model parameters~\cite{lidar2020lecture}. Furthermore, there is typically an assumption of Markovianity (i.e., memoryless environment) which in part results in neglecting potential inter-gate correlations. In contrast, MEs are based on physically motivated differential equations that arise from the theory of open quantum systems~\cite{campaioli2023tutorial,lidar2020lecture,breuerOQS}. They allow for more accurate descriptions of Markovian and non-Markovian system-environment interactions at the expense of added model and computational complexity. 

The stochastic Hamiltonian formalism relies on modeling system-environment interactions via a semi-classical Hamiltonian~\cite{crow2014shf,benedetti2015shf, rossi2017shf,halataei2017shf,Peng2022stochastic}. System operators couple to stochastic, time-dependent, variables as opposed to additional quantum degrees of freedom. This mean-field approach enables modeling of noisy quantum dynamics resulting from spatio-temporally correlated error processes by averaging time evolved quantum states over random realizations of the noise.
The advantage of the stochastic Hamiltonian approach lies in the trade-off between Hilbert space size and parametric complexity, which has proven particularly useful when non-Markovian noise is present~\cite{norris2016qns,Murphy2022}.

Each strategy has been investigated as a means for modeling noise in superconducting qubits. Models based on compositions of Markovian quantum channels have had variable success~\cite{dahlhauser2021modeling,Georgopoulos2021,payne2024lifetime,weber2024construction}, with sparse Pauli-Lindblad models~\cite{vandenberg2023probabilistic,jaloveckas2023efficient} providing the strongest agreement~\cite{kim_evidence_2023}. This comes at the cost of noise tailoring via Pauli twirling~\cite{knill2004twirling,dankert2009exact}, which can incur progressively larger sampling overheads as system size increases. 

ME-based techniques have achieved substantially more success in modeling superconducting qubit devices. Although the presence of Markovian environmental noise sources are prominently observed in superconducting qubits~\cite{krantz2019}, recent studies have highlighted substantial contributions from non-Markovian sources as well. For example, 
Post-Markovian~\cite{Zhang2022} and Redfield~\cite{tripathi2023modeling,tripathi2022} based MEs have shown to agree well with single- and two-qubit demonstrations. In addition, extensions of the Lindblad master equation (LME)~\cite{Manzano2020,lidar2020lecture} that include non-Markovian contributions via stochastic unraveling~\cite{DiBartolomeo2023} or the addition of quantum degrees of freedom~\cite{Shirizly2024dissipative} have also shown promise in the study of multi-qubit dynamics. Albeit successful, it is unclear if these approaches will remain viable as system sizes increase. Computational costs associated with obtaining solutions to multi-qubit dynamical equations and model training costs must be balanced to ensure scalability. 

The stochastic Hamiltonian formalism has also shown success in modeling non-Markovian dynamics in superconducting qubits. It has shown utility in capturing dephasing processes due to, for example, randomly fluctuating magnetic fields~\cite{krantz2019}. The formalism has been employed in studies of spatio-temporally correlated noise in single~\cite{bylander2011noise,Rower2023} and two-qubit~\cite{vonLupke2020}, as well as qudit~\cite{sung2021multi}, systems. Extensions to larger quantum systems poses challenges, specifically when attempting to perform complete characterization of the Hamiltonian. For this reason, the stochastic Hamiltonian formalism is most applicable for effective modeling where sparse characterization is sufficient.

In this study, we introduce a hybrid model that leverages aspects of the channel formalism, MEs and the stochastic Hamiltonian formalism to construct effective noise models for superconducting qubit hardware. Our method aims to balance the sparsity in model parameters, and thus the required number of characterization experiments, with predictive power. Through an extensive comparison against experimental demonstrations, we show that the model can predict a wide range of hardware dynamics and even be employed as a proxy to train variational quantum algorithms~\cite{cerezo2021variational}.

Specifically, the model is based on compositions of faulty single- and two-qubit gate operations. Each gate is constructed using the LME with extended degrees of freedoms to account for non-Markovian spatio-temporally correlated noise sources. Classical degrees of freedom are specifically included to capture temporal features of qubit dephasing and control noise during gate operations. Quantum degrees of freedom enable modeling of spatial correlations due to two-level system (TLS) interactions and quantum crosstalk~\cite{schwartzman2024modeling}. The model is parameterized by ten parameters per qubit and three parameters per qubit pair, that can be learned through a small number of characterization experiments.

With an eye towards balancing informational completeness with protocol efficiency, we focus on simplicity in the design of the characterization protocols. Thus, we do not depend upon tomographical methods that are traditionally resource intensive, such as gate-set tomography~\cite{greenbaum2015introduction, Nielsen_2021} or process tensor methods~\cite{white2022nonmarkovian,white_demonstration_2020} to provide a detailed analysis of the effective noise channels. Instead, we consider a suite of canonical noise amplification circuits to learn model parameters for both Markovian and non-Markovian noise sources. Key to our characterization procedure is quantum noise spectroscopy (QNS)~\cite{alvarez2011qns, norris2016qns, Paz-Silva2017, romach2015qns,frey2017application,chan2018qns,vonLupke2020, pazsilva2019qns, maloney2022qubit}. QNS draws on the filter function formalism (FFF)~\cite{cywinski2008ff, green_arbitrary_2013,Green2012,Paz-Silva2014,Kofman2001universal} and carefully tailored control sequences to probe the spectral properties of system-environment interactions via measurement of the system's evolution. 

It is through this suite of characterization experiments that we perform an extensive study of superconducting qubit devices offered by the IBMQP. Examining 39 qubits across seven devices, we comprehensively characterize Markovian and non-Markovian noise sources. Therein, we elucidate important details about correlated dephasing and control noise and the presence of TLSs on IBMQP devices.

The learned parameters are combined with the single and two-qubit models and shown to accurately predict a variety of quantum circuits. Randomized benchmarking (RB)~\cite{Knill2008} provides a benchmark to assess the model's ability to correctly predict single-qubit gate error rates captured by routine IBMQP calibrations. We go on to show that the model conveys strong agreement with state-dependent dynamics observed in multi-qubit DD sequences and a small-scale quantum algorithm. In particular, we train a variational quantum eigensolver (VQE) designed to find the ground state of molecular Hydrogen H$_2$ using our noise model as a surrogate to the hardware. It is shown that a relative error of 0.5\% is achieved between hardware and the noise model, a $7\times$ improvement over IBMQP's default hardware error models. 

Together, our model characterization and validation studies provide strong evidence for the viability of the proposed model. Moreover, they highlight key noise sources for the current generation of superconducting qubits. This work illustrates the potential utility of reduced noise models that draw on aspects of both MEs and channel representations to describe complex quantum dynamics in superconducting qubit systems.

The structure of this paper is as follows. 
In Sec.~\ref{sec:noise-model} we introduce the noise model, with emphasis in modeling via LME.
In Sec.~\ref{sec:characterization} we describe the characterization protocol and the relevant notation. More specifically, in Sec.~\ref{sec:characterization_experiments} we present a detailed description of the experiments used to characterize the device. Here, we also provide the analytical expressions obtained from solving the LME for each noise amplification circuit described in Sec.~\ref{sec:noise-model}. Section~\ref{sec:experiment_results} presents experimental results obtained from running the characterization experiments on the IBMQP, as well as model fits and validation through simulation. In Sec.~\ref{sec:applications} we apply our noise model on non-trivial applications in order to test its predictability. Section~\ref{sec:scaling} describes strategies for scaling the characterization and simulation of the noise model. We conclude in Sec.~\ref{sec:conclusion} with a summary of the protocol and results, as well as thoughts on limitations and future work.

\section{Noise Model}
\label{sec:noise-model}

We begin by defining the noise model utilized throughout this study. It includes both Markovian and non-Markovian noise sources that we effectively capture by the LME
\eq{
\dot{\rho}(t) = -i[H(t),\rho] + \sum_k \gamma_k \l( L_k\rho L_k^\dagger - \frac{1}{2}\{L_k^\dagger L_k, \rho\} \r).
\label{eq:LE_general}
}
The Hamiltonian $H(t)$ describes the unitary contributions, and will house specific contributions that generate non-Markovian noise. $H(t)$ acts on the Hilbert space $\mathcal{H}=\mathcal{H}_{\rm D}\otimes\mathcal{H}_{\rm Sp}\otimes \mathcal{H}_{\rm TLS}$, where the subspace $\mathcal{H}_{\rm D}$ defines the Hilbert space for the data qubits, i.e., the qubits that will be employed for a particular sequence of gate operations. Spectator qubits that impose unwanted external coupling on the data qubits are defined under $\mathcal{H}_{\rm Sp}$. Lastly, $\mathcal{H}_{\rm TLS}$ denotes the inclusion of fluctuating TLSs that couple to the data qubits.

More explicitly, the Hamiltonian is formally given by 
\begin{equation}
    H(t) = H_C(t) + H_N(t) + H_\mrm{XT} + H_\mrm{TLS},
    \label{eq:Htot}
\end{equation}
where $H_C(t)$ defines the control on the system and $H_N(t)$ designates local noise contributions; both act on $\mathcal{H}_{\rm D}$. $H_C(t)=H_{C,1}(t)+H_{C,2}(t)$ is composed of two terms defining single and two-qubit control, respectively. Parasitic quantum crosstalk interactions between data and spectator qubits are captured by $H_\mrm{XT}$ which acts on $\mathcal{H}_\mrm{D}\otimes \mathcal{H}_\mrm{Sp}$, with coupling between data qubits and TLSs being defined via $H_\mrm{TLS}$. The latter of which is defined within $\mathcal{H}_\mrm{D}\otimes \mathcal{H}_\mrm{TLS}$. Note that our model does not include local noise contributions for spectator qubits, nor coupling between spectators and TLSs. As we will show in Sec.~\ref{sec:experiment_results}, such additions to the model are not required to obtain strong agreement between model predictions and experimental results.

The latter part of our model includes jump operators $L_k$ and decay rates $\gamma_k$ that supply the dissipative effects. These terms are meant to capture Markovian effects of environmental noise sources that act upon the data qubits and are not associated with spectator or TLS coupling. As such, the model does not include dissipative contributions for spectator qubits. As in the case of $H(t)$, we do not observe a critical dependence on the inclusion of such terms and the predictability of our model.

We further distinguish between different types of noise based on their degree of correlation and modeling structure. We first define \textit{locally Markovian noise}, corresponding to those processes which are uncorrelated and local for each qubit. An accurate description of locally Markovian noise effects does not require specific considerations of the degrees of freedom that give rise to noise. 

Next, we define \textit{extended Markovian noise} to account for those processes which involve correlations, particularly spatial correlations, but whose effect on the system dynamics can be well modeled by a LME. This includes, for example, noise induced by TLSs or spectator qubits, which can be modeled via LME by enlarging the data qubit system to include the additional quantum degrees of freedom~\cite{Shirizly2024dissipative,schwartzman2024modeling}. Thus, extended-Markovian noise requires careful consideration of the degrees of freedom that cause it, but its dynamical modeling is analogous to the locally-Markovian case.

Lastly, we denote by \textit{stochastic noise} those processes which can be described by time-dependent Hamiltonians with well-defined statistical properties. Stochastic noise can introduce time-correlations into the dynamics~\cite{cywinski2008ff,Kofman2001universal,Paz-Silva2014}, and proves to be a powerful tool for studying and modeling non-Markovian noise. In the following sections, we provide details about our model that fit within this framework. First, we focus on single-qubit operations, defining both Hamiltonian and dissipative operators. We then move to two-qubit noise and operations, in which we specify a noise model for cross-resonance-based operations.

\subsection{Noiseless Control}

\subsubsection{Single-Qubit Control}
Arbitrary single-qubit operations can be parametrized using three Euler angles for which $x$- and $z$-rotations are sufficient~\cite{McKay2017_virtual}. 
Single-qubit microwave controlled $x$-rotations are represented by the $n$-qubit local control Hamiltonian 
\eq{
H_{C,1}(t) = \sum_{j=0}^{n-1} H_C^{(j)}(t),
\label{eq:control-H}
}
where $H_C^{(j)}(t)=\Omega_j(t) \sx^{(j)}/2$ is the single-qubit control Hamiltonian at site $j$. 
Here, $\sigma_\mu^{(j)}$, with $\mu=x,y,z$, is a Pauli operator acting on the $j$th qubit, and $\Omega_j(t)$ is the control amplitude. 
Note that the ideal total rotation at time $t$ is given by the angle $\Theta_j(t) = \int_0^t ds \Omega_j(s)$.
$z$-rotations are implemented virtually by performing instantaneous and noiseless phase shifts on the control~\cite{McKay2017_virtual}. 
The action of a virtual $Z$-gate is modeled via the unitary $U_{VZ}(\theta_0,...,\theta_{n-1})=\exp\l(-\frac{i}{2} \sum_{j=0}^{n-1} \theta_j \sz^{(j)} \r)$ for a set of rotation angles $\{\theta_j\}_{j=0}^{n-1}$.

\subsubsection{Two-Qubit Control}

In fixed-frequency transmon-based architectures, the cross-resonance (CR) gate is a microwave controlled operation that has shown promising potential as a two-qubit entangling gate~\cite{Paraoanu2006,Rigetti2010,Magesan2020,Malekakhlagh2020,Tripathi2019}. 
Leveraging native qubit-qubit couplings, the CR gate generates entanglement by driving the control qubit at the frequency of the target qubit, without the need for qubit or coupling tunability.

The simplicity of the CR gate implementation makes it an appealing candidate for a go-to entangling gate in these architectures. 
However, during the implementation of a CR gate, unwanted errors arise.
These errors are well known and documented~\cite{Sheldon2016,Wei2022}, and several of them can be corrected by the implementation of echo pulses~\cite{Sheldon2016}. 
These echo pulses aim to cancel both local errors, and crosstalk between control-target and spectator qubits. 
This gate is known as the echo cross-resonance gate (ECR) gate~\cite{Sundaresan2020}, and its main two-qubit operation consists of a $ZX_{\pi/2}$ rotation that can be modeled by the ideal CR Hamiltonian,
\seq{
H_{C,2} = \frac{\pi}{4\tau_{CR}} \sigma_z^{(c)} \sigma_x^{(t)}.
}
Here $\tau_{CR}$ is the duration of the ECR gate, typically much larger than the single-qubit gate duration $\delta t$.

\subsection{Locally Markovian Model}
\subsubsection{Local Dissipative Noise}
\label{sec:model-1Qdis}

Next, we describe the sources of dissipation that are included in the LME as jump operators $L_k$ with decay rates $\gamma_k$. These processes are locally-Markovian, and thus act only locally on each qubit. Here again the superscript $(j)$ will denote the qubit index number.

The first source of dissipative errors we consider is thermal relaxation in the form of generalized amplitude damping (GAD). GAD describes the effect of energy dissipation to an environment at a finite temperature $T_\mrm{env}^{(j)}$. Since $\{T_\mrm{env}^{(j)}\}$ represent effective bath temperatures, they are assumed to be different for each qubit. We define $0\leq q^{(j)}\leq 1$ as the excited state probability at thermal equilibrium, and $\gamma^{(j)}=1/T_1^{(j)}$ as the relaxation rate. The $T_1$ time is the characteristic decay time of the GAD process, where the relaxation probability $1-e^{-\tau/T_1^{(j)}}$ can be interpreted as the probability that a spontaneous emission occurs after a time $\tau$. Assuming that the equilibrium probabilities satisfy a Boltzmann distribution, we have $1-q^{(j)} = \l(1+e^{-\hbar \omega_{01}^{(j)}/k_B T_\mrm{env}^{(j)}}\r)^{-1}$, where $\omega_{01}^{(j)}$ is the energy of the first excited state, typically near $5~$GHz~\cite{kjaergaard2020}, and $k_B$ is the Boltzmann constant. In addition, GAD has a simple description in terms of two jump operators: $L_{\pm,j} = \sigma_\pm^{(j)}=\l(\sigma_x^{(j)}\pm i\sigma_y^{(j)}\r)/2$ with decay rates $\gamma_{+,j}=q^{(j)}\gamma^{(j)}$ and $\gamma_{-,k}=\l(1-q^{(j)}\r)\gamma^{(j)}$. 

Next, we consider exchange couplings with the environment leading to dephasing noise, which appears as \textit{phase damping} (PD). These are processes affecting the off-diagonal elements of the state density matrix. Analogously to the relaxation case, the characteristic decoherence time $T_\phi$ can be related to the dephasing rate via $\lambda^{(j)}=1/T_\phi^{(j)}$ and decay probability $1-e^{-\tau/T_\phi^{(j)}}$. In its Lindblad form, the action of PD is described with a single jump operator: $L_{z,j} = \sigma_z^{(j)}/\sqrt{2}$, with rate $\gamma_{z,j}=\lambda^{(j)}$.

Lastly, during the action of control operations, noise processes such as fluctuations in the control lines~\cite{wood_quantification_2018} can induce dissipation of information. To account for the presence of dissipation in control errors, we include a \textit{bit-flip} noise contribution with rate $\nu^{(j)}$ and probability $1-e^{-\nu^{(j)}\tau}$. The Lindblad form of control dissipation is $L_{x,j}=\sigma_x^{(j)}/\sqrt{2}$ with decay rate $\gamma_{x,j}=\nu^{(j)}$. Note that this form of error only acts during the implementation of $x$-rotations.

\subsubsection{State Preparation and Measurement (SPAM)}

We include SPAM errors in the error model as a bit-flip error in the quantum channel form acting before ideal state measurement operations. In general, the error probability can be different for the preparation and measurement of states. Previous work, e.g.,  Ref.~\cite{papic2023fast}, accounts for residual excited state population during state preparation by assuming that the initial state is the (unnormalized) thermal state $\exp(-H/k_{B}T_\mrm{env})$, rather than the ground state $\ket{0}\bra{0}$, where $H$ is the full system-bath Hamiltonian. However, in practice, we find that it is sufficient to group both processes into a single measurement channel, hence greatly simplifying the analytical derivations. This choice is standard practice in the literature, see for example~\cite{Georgopoulos2021}. Consequently, we choose to treat state preparation as an effectively ideal operation and assign all SPAM errors to measurement errors with rates $s_j$. 

The effect of SPAM errors arising from measuring qubit $j$
\eq{
\label{eq:E_Mj}
\mathcal{E}_M^{(j)}(\rho) = (1-s_j) \rho + s_j\, \sigma_x^{(j)} \rho \sigma_x^{(j)},
}
where $s_j$ is the probability of measurement errors on qubit $j$. Thus, the collective effect of data qubit measurement errors can be found by the application of maps $\mathcal{E}_M^{(j)}(\cdot)$ for $j=0,...,n-1$,
\seq{
\label{eq:E_M}
\mathcal{E}_M(\rho) &= \mathcal{E}_M^{(n-1)}\circ \cdots \circ \mathcal{E}_M^{(0)}(\rho).
}
To first order in the error parameters $s_j\ll1$, this map can be approximated as $\mathcal{E}_M(\rho)\approx \rho + \sum_{j=0}^{n-1}s_j\, \l(\sigma_x^{(j)} \rho \sigma_x^{(j)}-\rho\r)$.

\subsection{Extended Markovian Model}

In addition to local noise, we allow the data qubits to couple non-locally, both to other qubits and external degrees of freedom via $ZZ$ interactions. First, we consider qubit-qubit crosstalk noise resulting from always-on interactions that yield residual couplings~\cite{McKay2019,Li2020,Zhou2023}. We denote by $\mC$ the set of qubits that couple to the data qubits. 
In general, $\mC$ will depend on the connectivity of the device, and for fixed-frequency coupling transmons, the most  significant couplings will be induced by nearest-neighbor interactions.

Denoting by $\mC(j)\subseteq\mC$ the set of qubits connected to the $j$th data qubit, the crosstalk Hamiltonian is thus given by
\eq{
\label{eq:H_ZZ}
H_\mrm{XT} = \sum_{j=0}^{n-1} \sz^{(j)} \sum_{i\in \mathcal{C}(j)} J_{ij} \sz^{(i)},
}
for coupling strengths $J_{ij}$. To avoid double counting, we assume non-zero coupling only for $i<j$, while $J_{ij}=0$ for $i>j$. 

Analogously to crosstalk interactions, we can describe the TLS coupling via $ZZ$ interactions, where each data qubit is coupled to $n_\mrm{TLS}$ number of TLSs~\cite{Shirizly2024dissipative}. TLSs are defined on a computational basis $\{\ket{0}_\mrm{TLS},\ket{1}_\mrm{TLS}\}$ with Pauli operators $\sigma^{(j,k)}_{\mu,\mrm{TLS}}$, for $\mu=x,y,z$, qubit $j$ and TLS $k$. Then, the Hamiltonian that describes the coupling between data qubits and TLSs is 
\eq{
\label{eq:H_TLS}
H_\mrm{TLS} = \sum_{j=0}^{n-1} \sz^{(j)} \sum_{k=1}^{n_\mrm{TLS}} \xi_{jk} \sigma_{z,\mrm{TLS}}^{(j,k)},
}
where $\xi_{jk}$ are the coupling strengths between qubit $j$ and its $k$th TLS. At the beginning of each experiment, all TLSs are considered to be initialized in the $\ket{+}$ state. These are extended Markovian processes, and thus can be studied within the LME by including the additional qubits and TLSs as part of Eq.~(\ref{eq:Htot}). Note that both crosstalk and TLS interactions are included in single and two-qubit operations.

\subsubsection{Two-Qubit Gates and Noise}
\label{subsubsec:two-qubit-noise}

We allow two-qubit errors in the Hamiltonian, but interestingly, we find that single-qubit incoherent errors are sufficient to obtain excellent agreement with the decays observed in the experiments. In other words, no two-qubit dissipative terms are found to be necessary to explain the experiment results observed.  This remarkable fact significantly simplifies the modeling and characterization of the ECR gates, and is a promising feature looking towards multi-qubit modeling and characterization.

Following Refs.~\cite{wei2023characterizing,Sheldon2016}, and denoting the control and target qubits with $c$ and $t$ superscripts, we model the effective ECR Hamiltonian as
\begin{eqnarray}
H_\mrm{CR} &=& (1+\epsilon_{zx}) H_{C,2} \nonumber\\
&&+ \frac{\beta^{(t)}}{2} I^{(c)} \sigma_z^{(t)} +\frac{\zeta}{2} I^{(c)} \sigma_x^{(t)} + \frac{J}{2} \sigma_z^{(c)} \sigma_z^{(t)},
\label{eq:H_CR}
\end{eqnarray}
where $I^{(c)}$ denotes the identity operation on the control qubit. The first term corresponds to the main $ZX$ rotation of the ECR gate, allowing for an over-rotation fraction $\epsilon_{zx}$ due to implementation errors. This Hamiltonian also includes local noise rotations on the target qubit: a relevant detuning term $\beta^{(t)}$ and leftover single-qubit $x$-rotation $\zeta$. Lastly, qubit-qubit crosstalk is also present and accounted for in the last term with coupling strength $J$.

\subsection{Stochastic Noise Model}

\label{sec:noise-model-stoch}

We represent both local control and dephasing noise in the stochastic Hamiltonian formalism by the following time-dependent noise Hamiltonian, 
\eq{
\label{eq:stoch-noise-Hamiltonian}
H_{N,1}(t) = \sum_{j=0}^{n-1} \epsilon_j(t) H_{C}^{(j)}(t) + \frac{\beta_j(t)}{2} \sz^{(j)}.
}
Here, $\epsilon_j(t)$ and $\beta_j(t)$ are time-dependent random variables that represent the stochastic control and dephasing noise. These stochastic noise variables are assumed to be Gaussian and wide-sense stationary, and thus they are completely determined by their mean $\braket{f}=\overline{f}$ and two-point correlation function $C_f(t)=\braket{f(t)f(0)}$, for $f=\epsilon_j,\beta_j$. Here, $\braket{\cdot}$ denotes the average over noise realizations. From $C_f(t)$, we can define the noise power spectral density (PSD) $S_f(\omega) = \int_0^\tau C_f(t) e^{-i\omega t} dt$, which captures the information of the noise fluctuations in the frequency domain. Note that the limiting case of detuning corresponds to non-zero mean and $S(\omega)\equiv0$, while white noise is equivalent to $\braket{f}\equiv0$ and $S(\omega)\neq 0$.

In general, to evaluate the effect of stochastic noise on the system evolution in simulation, we first generate a number of different noise realizations, or trajectories. Then, the LME is solved for each realization, and lastly, the resulting density matrix is averaged over all generated trajectories. However, as we will discuss in Sec.~\ref{sec:scaling}, it is possible to circumvent noise averaging in certain scenarios.

\subsection{Model Solutions}

To find a solution for Eq.~(\ref{eq:LE_general}), we take a canonical approach~\cite{lidar2020lecture}. First, we expand the density matrix state on the $n$-qubit Pauli basis, $\rho(t) = \frac{1}{2^n} \l(I + \vec{v}(t)\cdot \vec{\mathcal{O}}\r)$, where $\vec{v}(t)$ is the generalized Bloch vector, and $\mathcal{O}_j\in\{I,\sigma_x,\sigma_y,\sigma_z\}^{\otimes n}$, for $j=0,...,2^{2n}-1$. In terms of the components of $\vec{v}(t)$, Eq.~(\ref{eq:LE_general}) becomes a system of coupled differential equations, 
\eq{
\label{eq:LE_v}
\dot{\vec{v}}(t) = \mathbf{G}(t) \cdot \vec{v}(t) + \vec{c}(t),
}
where the vector $\vec{c}(t)$ and the complex-valued coupling matrix $\mathbf{G}(t)$ depend on the circuit and noise parameters. In this form, Eq.~(\ref{eq:LE_v}) can be solved using standard coupled differential equations methods. When studying circuit-based evolutions in the Markovian regime, it is possible to separate $\mathbf{G}(t)$ into time-independent sections to be solved separately, and thus finding analytical solutions of Eq.~(\ref{eq:LE_v}) becomes feasible (see Appendix~\ref{app:LME_Mark_sol}). In this case, we can write the formal solution 
\eq{
\label{eq:LE_sol}
\vec{v}(t+\tau) = e^{\mathbf{G}\tau}\cdot \vec{v}(t) + \left(e^{\mathbf{G}\tau} - 1\right)\cdot \mathbf{G}^{-1}\cdot \vec{c},
} 
where $\tau$ denotes the duration over which $\mathbf{G}\equiv \mathbf{G}(t)$ remains constant. Note that, generally, under the condition of finite relaxation rates $\gamma_j>0$, $\mathbf{G}(t)$ is diagonalizable and invertible.

On the other hand, when time correlations are present through $\{\beta_j(t),\epsilon_j(t)\}_j$, finding general analytical solutions is challenging. Instead, the statistical properties of the stochastic noise processes can be used to generate realizations of noise trajectories. For each realization, assuming a discrete piecewise constant evolution of $\epsilon(t),\beta(t)$, Eq.~(\ref{eq:LE_v}) can be solved to find the final state $\rho(\tau;\{\beta_j,\epsilon_j\}_j)$. Lastly, the solution state is found by averaging over a sufficient number of noise realizations.

Gate-based experiments can be devised, where some sources of noise will contribute dominantly, while other types of noise will have limited effect. In the next section, we introduce a set of characterization experiments that aim to amplify and effectively isolate specific noise processes. In these cases, due to the relative simplicity of these circuits, we can solve  Eq.~(\ref{eq:LE_v}) analytically, and use the resulting expression to learn the model parameters.

\section{Characterization Protocol}
\label{sec:characterization}

\subsection{Overview}
\label{subsec:overview}

\begin{figure*}[t]
    \centering
    \includegraphics[width=\textwidth, trim={0cm 10cm 9cm 0cm},clip]{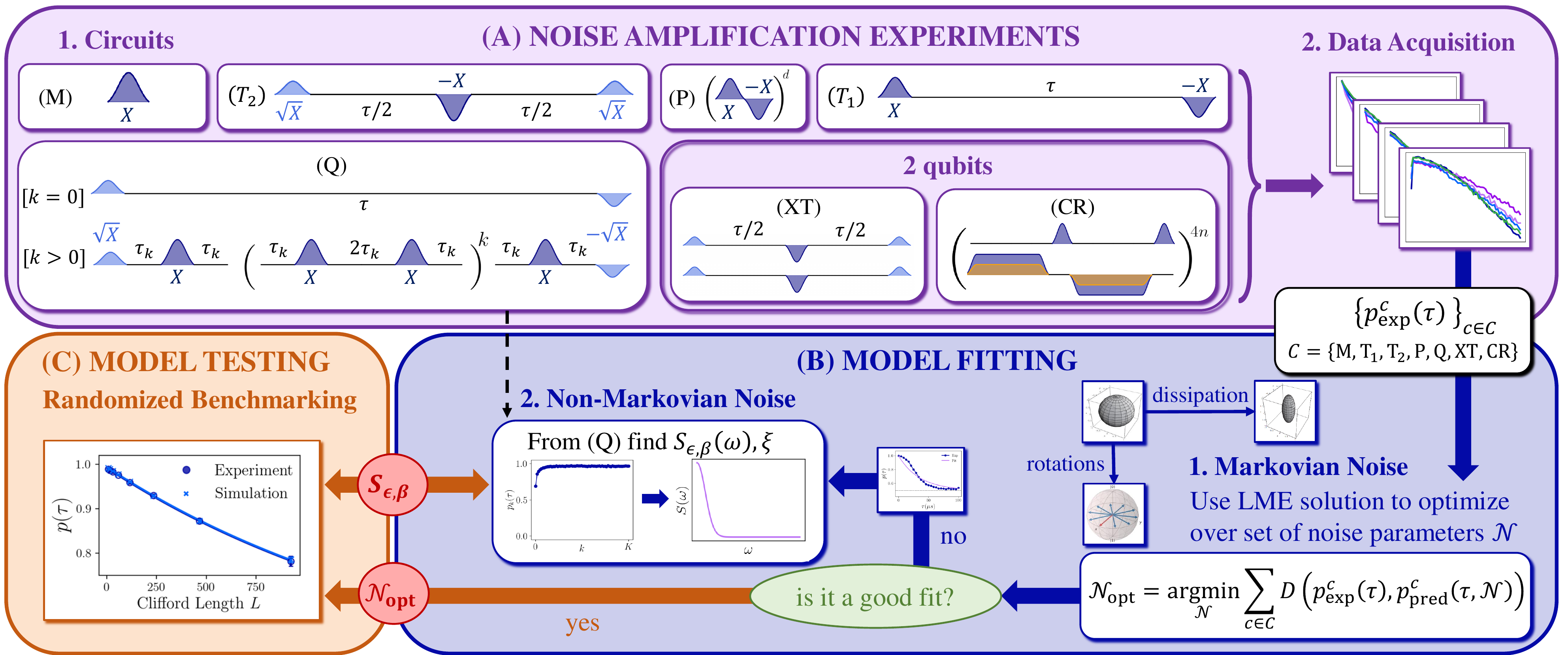}
    \caption{Noise characterization protocol used to learn the model parameters. (A) The state is assumed to be perfectly initialized in the ground state $\ket{0}\bra{0}$, and the characterization circuits $C=\{\mrm{M},T_1,T_2,\mrm{P,Q,XT,CR}\}$ are run on the device of interest. (B) The measurement data $\{p^c_{\text{exp}}(\tau)\}_{c\in C}$ is then used to fit the Markovian noise model parameters $\mathcal{N}=\{s,\gamma,q,\lambda,\beta,\epsilon,\nu\}$, based on analytical predictions obtained from solving the LME [Eq.~(\ref{eq:LE_sol})]. If deviations from the Markovian model are present in the data, the FTTPS (Q) protocol is used to extract dephasing and control PSDs, $S_\beta(\omega), S_\epsilon(\omega)$ respectively, as well as the TLS coupling $\xi$ obtained from the Ramsey experiment, denoted by Q$_{k=0}$. Two-qubit crosstalk $J$ can be obtained via analyzing the (XT) circuit results. (C) The model, either purely Markovian or extended with non-Markovian effects, is tested by predicting the RB decay rate.}
    \label{fig:main}
\end{figure*}

Figure~\ref{fig:main} presents a graphical overview of the characterization protocol. After selecting the qubit or pair of qubits of interest, the characterization experiments are run; this includes the set of single-qubit experiments $C_1=\{\mrm{M},T_1,T_2,\mrm{P},\mrm{Q}\}$, and two-qubit experiments $C_2=\{\mrm{XT},\mrm{CR}\}$, where each experiment may require the execution of multiple circuits. All noise amplification experiments are shown in Fig.~\ref{fig:main}(A), and described in detail in the next section. The association between model parameters and characterization experiments is summarized in Table~\ref{tbl:params}. In all cases, the qubits are assumed to be perfectly initialized in the ground state $\rho(0)=\ket{0}\bra{0}$. Each experiment results in estimates of the survival probabilities, denoted as $\{p_\mrm{exp}^c(\tau_c)\}_{c\in C}$, where $C= C_1 \cup C_2$ includes all characterization experiments performed. Note that the duration $\tau_c$ of each experiment will in general vary between experiments.

Next, the experimental data is compared to the analytical predictions, shown later in this section, obtained from solving the LME with the extended Markovian noise model for each of the experiments; see Fig.~\ref{fig:main}(B). The procedure used to find the analytical predictions and its formalism was outlined in the previous section, and is developed in detail in Appendix~\ref{app:LME_Mark_sol}.
The Markovian model predictions are denoted by $\{p_\mrm{pred}^c(\tau_c,\mathcal{N})\}_{c\in C}$, where $\mathcal{N}$ defines the set  of noise parameters. When performing these computations, stochastic dephasing and control noise are assumed to be static and time-independent.

The fitting procedure finds optimal error parameters $\mathcal{N}_\mrm{opt}$ that minimize the mean squared error (MSE) distance between vectors of length $N$, i.e., $D(\vec{x},\vec{y})=\frac{1}{N}\sqrt{\sum_{i=1}^{N}(x_i-y_i)^2}$, where $N$ denotes the total number of data points obtained from experiments. Note that, since this metric is used to compare distances between survival probabilities, the MSE distance is always bounded between 0 and 1. The quality of the fit can then be formalized as the distance between data and experiment being smaller than a certain value $\delta$, that is, $\frac{1}{|C|}\sum_{c\in C} D(p_\mrm{exp}^c(\tau_c),p_\mrm{pred}^c(\tau_c,\mathcal{N}_\mrm{opt}))<\delta$, where $|C|$ is the number of experiments. When a set of noise parameters satisfies this condition, we say it is $\delta$-optimal, and $0<\delta<1$ can be interpreted as a fractional error, with typical $\delta$ values below 1\%.

Model selection is determined by the success of the fitting. This process begins by assuming the validity of the extended Markovian model, as nearly all IBMQP qubits are subject to quantum crosstalk. Additional degrees of freedom accounting for TLSs are determined in part by a Ramsey experiment, or equivalently, the first sequence within the FTTPS protocol. The model is tested against RB experiments run (ideally) immediately after the characterization data is obtained in order to minimize drift in the error model parameters. The optimal parameters $\mathcal{N}_\mrm{opt}$ are used in simulation and evaluated against the experimental RB results. We employ a $\delta_\mrm{RB}$-optimal criterion analogous to the fitting procedure to validate the model. If the fit is satisfactory, we claim that the error present in the device is predominantly Markovian, with error parameters $\mathcal{N}_\mrm{opt}$; see Fig.~\ref{fig:main}(C).

Discrepancies between predicted and experimental RB data signify the presence of additional non-Markovian contributions. Such deviations are addressed by promoting the dephasing and control error processes from static to stochastic. QNS via the FTTPS protocol is then used to characterize the noise PSDs $S_\beta(\omega)$ and $S_\epsilon(\omega)$. We find that the addition of stochastic noise is sufficient for reconciling the noise model with RB experiments.

\begin{table*}[t]
\begin{tabular}{ccc}
\hline\hline
Parameter Name & Parameter Variable & Characterization Experiment \\
\hline
Relaxation Rate & $\gamma$ & $T_1$ Experiment [$T_1$]\\
Excited State Probability & $q$ & $T_1$ Experiment [$T_1$]\\
Dephasing Rate & $\lambda$ & Hahn Echo [$T_2$]\\
Detuning Rate & $\beta$ & Ramsey [$R$]\\
TLS Coupling Strength & $\xi$ & Ramsey [$R$]\\
Crosstalk Coupling Strength & $J$ & Crosstalk Experiment [$XT$]\\
Incoherent Control Error & $\nu$ & FPW Sequences [$P$]\\
Measurement Error Probability & $s$ & SPAM Experiment [$M$]\\
Coherent Control Error & $\epsilon$ & FTTPS [$Q$]\\
Dephasing Noise PSD & $S_{\beta}(\omega)$ & FTTPS [$Q$]\\
Control Noise PSD & $S_{\epsilon}(\omega)$ & FTTPS [$Q$] \\
CR Offset & $\zeta$ & CR Experiment [$CR$]\\
CR Coherent Control Error & $\epsilon_{zx}$ & CR Experiment [$CR$]\\
\hline\hline
\end{tabular}
\caption{Mapping between model parameters and characterization experiments.}
\label{tbl:params}
\end{table*}

\subsection{Characterization Experiments}
\label{sec:characterization_experiments}
In this section, we present the noise amplification experiments used to characterize the error model parameters. All qubits are initialized in the $\ket{0}$ state and measured in the computational basis. Since all experiments are performed on the IBMQP, we write all circuits in terms of native IBMQP single-qubit gates: $I, X,\sqrt{X}$ gates and arbitrary virtual $z$-rotations. Single-qubit gates are implemented over a duration $\delta t$ that depends on the device (e.g., $\delta t=0.035\,\mu$s for \textit{ibm\_algiers}). For each device, however, $\delta t$ is equivalent for all single-qubit gates.

Below, we describe which sources of noise are expected to contribute most significantly for each type of circuit. We provide solutions to the LME for these circuits when local and extended Markovian noise processes are dominant. Details on the specific calculations can be found in Appendix~\ref{app:subsec:exp_preds}. 

In the case of single-qubit experiments, the qubit index will be removed for clarity. For the measured qubit, the probability of finding the state in the ground state is computed in terms of the Bloch vector as $p(\tau) = \bra{0} \mathcal{E}_M(\rho(\tau)) \ket{0} = \frac{1}{2}\big(1+v_z(\tau)(1-2s)\big)$ where $\rho(\tau)$ is the density matrix at the end of the circuit of duration $\tau$, obtained from solving Eq.~(\ref{eq:LE_general}). Consequently, it suffices to find expressions for the pre-measurement Bloch vector component $v_z(\tau)$. For notational convenience, we will suppress the subscript $z$ from the scalar quantity $v_z(\tau)$, and denote the $z$ component of the Bloch vector for experiment $c\in C$ as $v^{c}(\tau)$. With the exception of the SPAM circuit, the noiseless control time propagators of the characterization circuits perform identity operations, i.e., $U_c(\tau)=I$. As a result, the expected ground state probabilities in the noiseless case are $p(\tau)=1$.

\subsubsection{\texorpdfstring{$T_1$}{T1} Experiments [\texorpdfstring{$T_1$}{T1}]}

In a $T_1$ experiment, the qubit is prepared in the excited state $\ket{1}$ by implementing an $X$ gate. It is then allowed to evolve freely for a period of time $\tau$. A measurement of the ground state population is performed after applying a second $X$ gate to complete the circuit. This experiment captures thermal relaxation, i.e., the decay of the excited state to thermal equilibrium. The Bloch vector decays exponentially with the relaxation decay rate $\gamma$, as obtained from solving the LME
\eq{
\label{eq:v_T1}
v^{T_1}(\tau) &\approx 1 - 2q\l(1-e^{-\gamma\tau}\r).
}
Note that for a zero-temperature bath, $q=1$, and $v^{T_1}(\tau\gg1/\gamma)\rightarrow -1$. 

\subsubsection{\texorpdfstring{$T_2$}{T2} Hahn-Echo (HE) Experiments [\texorpdfstring{$T_2$}{T2} ]}

The $T_2$ HE experiment is initialized by applying an $\sqrt{X}$ gate and preparing the qubit in the equator state $\ket{-i}=(\ket{0}-i\ket{1})/\sqrt{2}$. It is followed by a period of free evolution of duration $\tau/2$. Then an $X$ (echo) gate is applied, with the objective of reversing the effect of constant detuning. The echo gate is succeeded by a second idle period of duration $\tau/2$. Lastly, a measurement of the ground state population is performed after applying a $\sX^\dagger$ gate. 
This experiment aims to measure the decay rate of coherence of the Bloch equator states.
More specifically, Bloch vector decays as
\eq{
\label{eq:v_T2}
v^{T_2}(\tau) &\approx e^{-\tau \l(\frac{\gamma}{2}+\lambda\r)},
}
which depends on both relaxation and phase decay rates $\gamma$ and $\lambda$, respectively. 
To derive this expression, we have made the assumption that stochastic dephasing noise changes sufficiently slowly such that it can be decoupled with an echo pulse.

\subsubsection{Ramsey Experiments [R]}

The Ramsey experiment, also known as  $T_2^*$ experiment, consists of preparing the qubit with a $\sqrt{X}$ gate, letting it evolve freely for a time $\tau$, and applying $\sqrt{X}^\dagger$ to complete the circuit. During the idle period, the qubit is susceptible to dephasing and relaxation. However, it will also exhibit susceptibility to detuning, spectator qubit crosstalk, and TLSs. By solving the LME and tracing out the TLS degree of freedom, we find that the Bloch vector evolves as
\eq{
\label{eq:v_TLS}
v^{R}(\tau) = e^{-(\frac{\gamma}{2}+\lambda)\tau} \cos(\beta_\mrm{eff} \tau) \cos(\xi \tau).
}
where $\beta_\mrm{eff}=\beta+\sum_{i\in\mathcal{C}} J_i$ is the effective detuning. $J_i$ is the crosstalk coupling strength for the $i$th nearest-neighbors coupled with the data qubit. A single TLS is assumed to be present, with coupling strength $\xi$. Here, we have taken dephasing noise to be static over the duration of the experiment, i.e., $\beta(t)=\beta$. 
Note that when a TLS is present, the oscillations in these experiments will present two characteristic frequencies, rather than one.

\subsubsection{State Preparation and Measurement (SPAM) [M]}

Here, we characterize SPAM errors by preparing the qubit in a computational basis state and estimating the error probability. More specifically, we initialize the qubits in $\ket{1}$ and examine the probability of obtaining $\ket{0}$. To leading order in the error parameters, the prediction for the SPAM circuit yields $p^M(\delta t)=s$.

\subsubsection{Fixed Total-Time Pulse Sequences (FTTPS) [Q]}

The FTTPS are circuits used in QNS for estimating properties of correlated noise~\cite{Paz-Silva2014,Murphy2022}. FTTPS consist of $K$ distinct sequences (circuits) labeled by $0\leq k<K$. Each sequence contains $N=2$ $(K+1)$ gates for a total sequence duration $\tau=N\delta t$. The $k$th FTTPS sequence is composed of identity and $X$ gates, where the $\ell$th $X$ gate is located at $\lfloor\frac{(2 \ell +1) K}{2k}\rfloor$ for $0\leq\ell<2k$. Prior to each FTTPS sequence, the system is subject to a $\sqrt{X}$ gate to prepare the qubit in the $(x,y)$-plane of the single-qubit Bloch sphere; thus, allowing it to be strongly sensitive to dephasing noise. Upon the completion of the sequence, $\sX^\dagger$ is applied to return the qubit to the ground state prior to measurement. The dynamics generated by FTTPS leads to narrow qubit FFs centered around $\omega_k=2\pi k/\tau$ (see Appendix~\ref{app:subsec:QNS}) that are well-suited for QNS.

Intuitively, as shown in Fig.~\ref{fig:main}(A.Q), the $X$ pulses are spaced out with a time difference $\tau_k\approx \tau/2k$. This expression is derived in the instantaneous pulse limit, and the approximation symbol is meant to account for deviations due to finite pulse-width. 
Note that the $k=0$ FTTPS is a fixed-$\tau$ Ramsey experiment. For $0<k<K$, the LME for the FTTPS experiments predicts Bloch vectors
\eq{
\label{eq:v_Q}
v^{Q}_{k}(\tau) &\approx e^{-\tau \delta_k}
\cos(2 \pi k \epsilon),
}
where the FTTPS decay rate is $\delta_k= \frac{\gamma}{2}+\lambda +(\frac{\gamma}{2}-\lambda+2 \nu)\frac{k}{2K}$. This form is derived in Appendix~\ref{app:subsec:exp_preds}.
In finding these expressions, we have assumed that the over-rotation control noise is static, i.e., $\epsilon(t)=\epsilon$.

\subsubsection{Finite Pulse-Width (FPW) Sequences [P]}

The finite time duration $\delta t$ of the native IBMQP gates results in deviations from ideal, instantaneous pulses. In order to characterize FPW errors, we define the FPW circuits by $d$ repetitions of alternating pairs of $X$ and $-X$ pulses. These sequences are carefully chosen in order to cancel coherent control errors, while preserving FPW errors. In practice, each $-X$ pulse is implemented by bookending an $X$ gate with two $Z$ gates, i.e., $-X=ZXZ$. Consequently, FPW sequences are implemented as $(XZXZ)^{d}$. The total duration of an FPW circuit of $d$ repetitions is $\tau=2d\,\delta t$. The Bloch vector of the FPW experiment is obtained to be
\eq{
\label{eq:v_P}
v^{P}(\tau) &\approx e^{-\tau \l(\frac{3}{4}\gamma+\frac{\lambda}{2}+\nu\r)}\cos\l(\frac{4}{3\pi}\beta_\mrm{eff} \tau \r),
}
where $\beta_\mrm{eff}$ is the effective detuning computed from the Ramsey experiments. The factor of $4/3\pi$ arises from the specifics of a Gaussian pulse shape, as shown explicitly in Appendix~\ref{app:subsec:gaussian_FPW}. Note that, in addition to relaxation and phase damping, a dissipative process with decay rate $\nu$ contributes to decay due to the presence of control.

\subsubsection{Crosstalk Experiments [XT]}

The crosstalk noise amplification experiments involve simultaneously driving both qubits. As opposed to relying on the Joint Amplification of ZZ (JAZZ) protocol~\cite{Ku2020jazz,shirai2023jazz}, we employ joint $T_2$ HE circuits. This is, qubits are prepared in the plane of the single-qubit Bloch sphere with $\sqrt{X}$ gates, where they are most sensitive to $Z$ rotations. Then, the qubits are allowed to evolve freely for a time $\tau/2$ and subsequently subject to simultaneous $X$ gates on both qubits. Both qubits are then allowed to evolve freely for an additional time of $\tau/2$. Lastly, $\sqrt{X}^\dagger$ gates are applied before measurement.

To simplify the analysis of the experiment results, only one of the qubits is measured, denoted by qubit $M$, whereas the other qubit $S$ is traced over. Due to crosstalk coupling, the dynamics of qubit $M$ depend non-trivially on the characteristics of qubit $S$. By solving the two-qubit LME and tracing over qubit $S$, we find
\eq{
\label{eq:v_XT}
v^{XT}(\tau) \approx e^{-\alpha_M \tau} \left(\cos(\tau J_{MS})+\frac{\gamma_S}{2J_{MS}}\sin(\tau J_{MS}) \right),
}
where $\alpha_M = \frac{\gamma_M+\gamma_S}{2} + \lambda_M$. The decay rates $\gamma_i,\lambda_i$ are the relaxation and dephasing rates of qubit $i=M,S$, which can be learned by performing single-qubit characterization. Note that crosstalk between qubits $M,S$ and their neighboring spectators, as well as local detunings, are decoupled by the echo pulses, thus amplifying the effect of crosstalk between qubits $M$ and $S$. See Appendix~\ref{app:subsec:exp_preds} for details on this calculation.

\subsubsection{Cross-Resonance Experiments [CR]}

The experiments used to characterize the ECR gates consist of repeated applications of ECR gates on a chosen pair of qubits. Instead of measuring the survival probability, the expectation values $\braket{Y_t},\braket{Z_t}$ on the target qubit $t$ are measured. Here, the expectation value of a target qubit operator $O_t$ is defined by $\braket{O_t(\tau)} = \tr\l[ \rho(\tau) O_t \r]$, where $\rho(\tau)$ is the circuit's final two-qubit state. For initial control qubit states $q_c=0,1$, corresponding to the states $|0\rangle,|1\rangle$, the LME can be solved to leading order in the error parameters. The resulting solutions can be expressed in terms of the Bloch vector components, or equivalently, the target qubit expectation values
\eq{
\begin{split}
\langle Y_t (\tau)\rangle_{q_c=0,1} &\approx \pm e^{-\delta_\mrm{cr} \tau} \sin\left(\left(\omega_\mrm{cr}\mp\zeta\right)\tau\right), \\
\langle Z_t (\tau)\rangle_{q_c=0,1} &\approx e^{-\delta_\mrm{cr}\tau} \cos\left(\left(\omega_\mrm{cr}\mp\zeta\right)\tau\right),
\end{split}
}
where $\delta_\mrm{cr}=\frac{2\gamma}{5}+\frac{\lambda}{2}+\nu$ and $\omega_\mrm{cr}=\pi\left(1+\epsilon_{zx}\right)/2\tau_{CR}$. The experiment duration consists of multiples of the ECR gate duration, namely $\tau=n \tau_{CR}$, for $n=0,1,...$ repetitions of ECR gates. The $X_t$ expectation values are found to be proportional to $\braket{X_t}_{q_c=0,1}\propto(\beta_t\pm J)\tau_{CR}$ and are typically much smaller than $\braket{Y_t}_{q_c},\braket{Z_t}_{q_c}$. Here $J$ is the crosstalk coupling between the control and target qubits. From fitting these functions to experiment data, the over-rotation error $\epsilon_{zx}$ and the offset $\zeta$ can be obtained. The target qubit decay parameters $\gamma,\lambda,\nu$ are obtained through single-qubit characterization.

\section{Experiment Results}
\label{sec:experiment_results}

In this section, we showcase how the previous models and characterization protocols can be used to assess and predict a wide variety of experimental behavior on IBMQP devices. We begin by considering qubits that are dominated by Markovian noise and then extend the analysis to those that exhibit strong non-Markovian phenomena. Specifications for all devices studied can be found in Appendix~\ref{sec:device_cdf}.

\subsection{Locally Markovian Noise on the IBMQP}
\label{subsec:markov_exp_results}
\begin{figure}[t]
    \centering
    \includegraphics[width=0.48\textwidth]{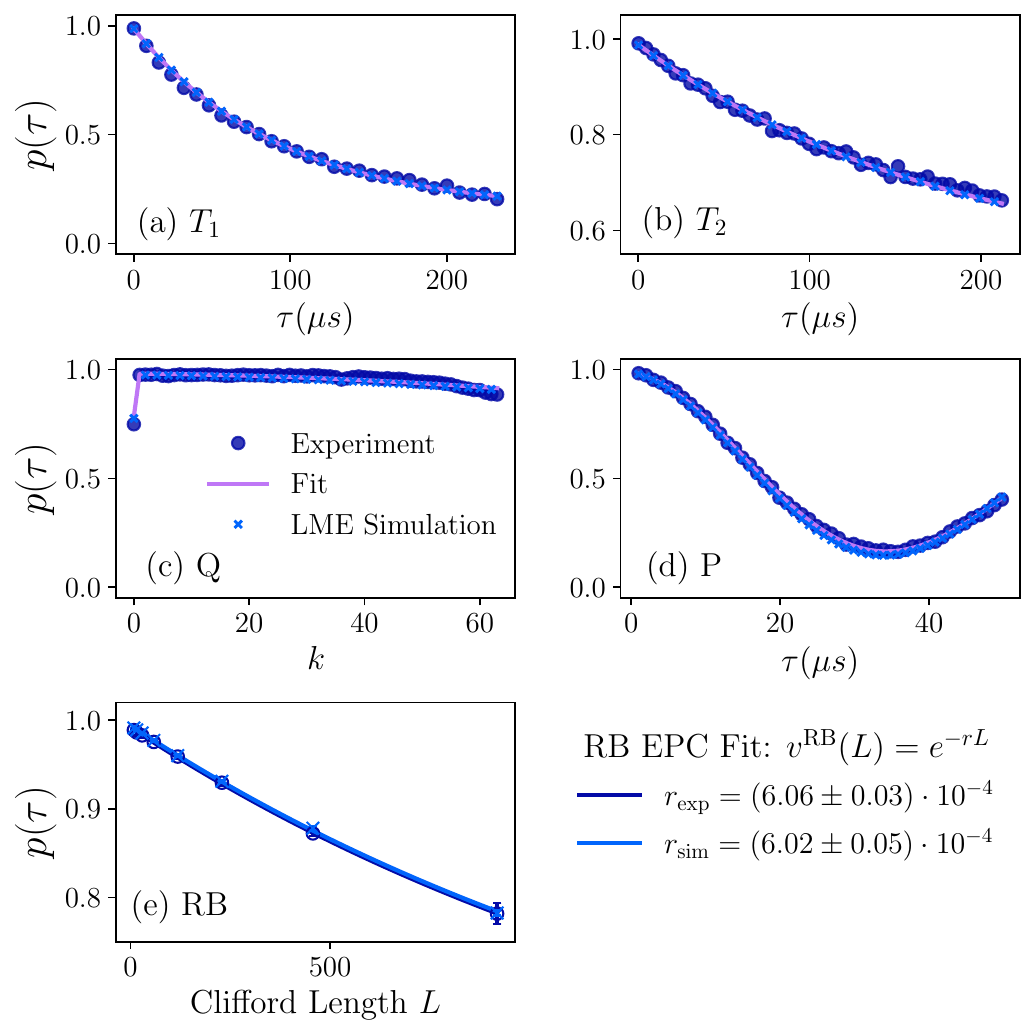}
    \caption{(a-d) Experimental implementation of the noise characterization protocol on qubit 8 of \textit{ibm\_algiers}. The results of the characterization experiments are shown (dark-blue dots). We perform a simultaneous fit of the experiments results to the Markovian model predictions given by Eqs.~(\ref{eq:v_T1})-(\ref{eq:v_P}) (purple), from where the noise parameters are obtained. Note that this set of experimental results agrees well with the Markovian model, including validation in LME simulation (light blue crosses). SPAM results, not shown for clarity, yield $s=1.2\%$. Other error parameters obtained from the fit are: $\gamma=0.0107(2)\,$MHz, $q=0.86(1)$, $\beta = 0.208(1)\,$MHz, $\lambda=0\,$MHz, $\epsilon=0.121(4)\%$, $\nu=0.005(1)\,$ MHz. (e) RB characterizations (with error bars) obtained from averaging over 10 gate realizations (dark-blue circles). The error parameters are used in combination with solutions to Eq.~(\ref{eq:LE_general}) to simulate the RB circuits (light-blue crosses). The experimental and simulated EPC, obtained from the standard exponential fit (solid lines) of the RB decay curve $p^{\mathrm{RB}}(L)= \l(1+(1-2s)e^{-rL}\r)/2$, agree within errorbars. In the RB fit, only the EPC $r$ is left as a free parameter. Hence, the RB error rate can be derived from the noise parameters without any a priori knowledge of RB results.}
    \label{fig:markov_exps}
\end{figure}

We begin by discussing qubits that exhibit dynamics consistent with the locally Markovian noise model. Figure~\ref{fig:markov_exps} presents results obtained from running the ($T_1$), ($T_2$), (Q), (P) characterization experiments on qubit 8 of \textit{ibm\_algiers}. These experiments are sufficient to characterize all locally Markovian processes, namely amplitude damping,  as well as Markovian dephasing and control noise. Panels (a)-(d) in Fig.~\ref{fig:markov_exps} show the results of the characterization experiments. 

A fitting procedure using Eqs.~(\ref{eq:v_T1})-(\ref{eq:v_P}) is  $(\gamma,q,\lambda,\beta,\epsilon,\nu,s)$ from the characterization experiments. To validate the model, Markovian simulations are carried out by numerically solving the LME via Eq.~(\ref{eq:LE_sol}) for each circuit element with the learned noise parameters. As can be seen in Fig.~\ref{fig:markov_exps}, the learned LME model conveys excellent agreement with experiment; thus successfully validating the noise model.

Next, we test the noise model's predictive capabilities by comparing it against experiments not used in the training process. In particular, we focus on RB experiments used to assess the single-qubit gate error rates. The inherent randomization of RB distinguishes it from the characterization experiments. Moreover, as shown in Appendix~\ref{app:sec:RB_dependence}, RB possesses a robustness to coherent errors while still remaining susceptible to all incoherent errors. This property makes RB a promising candidate to study the simultaneous action of the incoherent error processes. Deviations between RB predictions and experiments also serve as detection for time-correlated phenomena. The model can provide intuition on how each noise process contributes to RB decay rates. Interestingly, to first order in $\delta t$, we find that the RB decay rate depends linearly on the dissipative parameters $\gamma,\lambda,\nu$, and quadratically on the coherent error parameters $\epsilon,\beta$. 
See Appendix~\ref{app:sec:RB_dependence} for simulation details.

We run experiments and numerical simulations for a set of 10 RB circuits averaged and fit to an exponential decay model to extract the error-per-Clifford (EPC) RB decay rate. As shown in Fig.~\ref{fig:markov_exps}(e), the model is able to predict the RB decay rate with great accuracy. We again emphasize that the noise model does not have access to the RB experimental result, hence showcasing the predictive power of the Markovian error model. 

\subsection{Extended Markovian Noise on the IBMQP}

Although the locally Markovian model can be used to characterize any of the IBMQP devices, model violations are often found when reconciling simulation with the characterization experiments. To accurately describe the phenomena observed, we examine the characterization of extended Markovian processes introduced in Sec.~\ref{sec:noise-model}. Here, we analyze two distinct effects: multi-frequency oscillations induced by TLSs and crosstalk noise. These processes are non-Markovian effects when viewed locally, but are well modeled with a LME by extending the system Hilbert space to include TLS and spectator qubit degrees of freedom.  

\subsubsection{TLS in Ramsey Experiments}
\label{subsubsec:TLS_exp}

It is well established that Ramsey experiments performed on superconducting qubits often present clear deviations from the Markovian model~\cite{agarwal2023modelling,Thorbeck2023_TLS,burnett_decoherence_2019}. This is evidenced by oscillations that cannot be modeled as a single decaying and oscillating  function of the form $e^{-\alpha t}\cos(\beta t)$, but instead present multiple oscillation frequencies. A common approach to modeling this phenomenon consists of coupling the qubit to fluctuating or static TLSs~\cite{agarwal2023modelling,Thorbeck2023_TLS}. In this work, we treat the TLS as an effective qubit that couples to the main qubit via a static $ZZ$ interaction. Thus, TLS excitations translate as dephasing on the qubit. Note that alternative and more general approaches exist, such as considering additional couplings (e.g. $ZX$), varying the initial state of the TLS, or considering multiple TLSs; some of these approaches are discussed in Ref.~\cite{agarwal2023modelling}. Although the physical mechanism for TLS coupling is not entirely well understood -- whether it is induced by defects, quasiparticles, etc.~\cite{riste_millisecond_2013,deGraaf2020,deGraaf2021,Thorbeck2023} -- the TLS model is an attractive effective theory due to its mathematical simplicity and accurate description of experimental results. 

\begin{figure}[t]
    \centering\includegraphics[width=0.48\textwidth, trim={0cm 0.35cm 0cm 0cm},clip ]{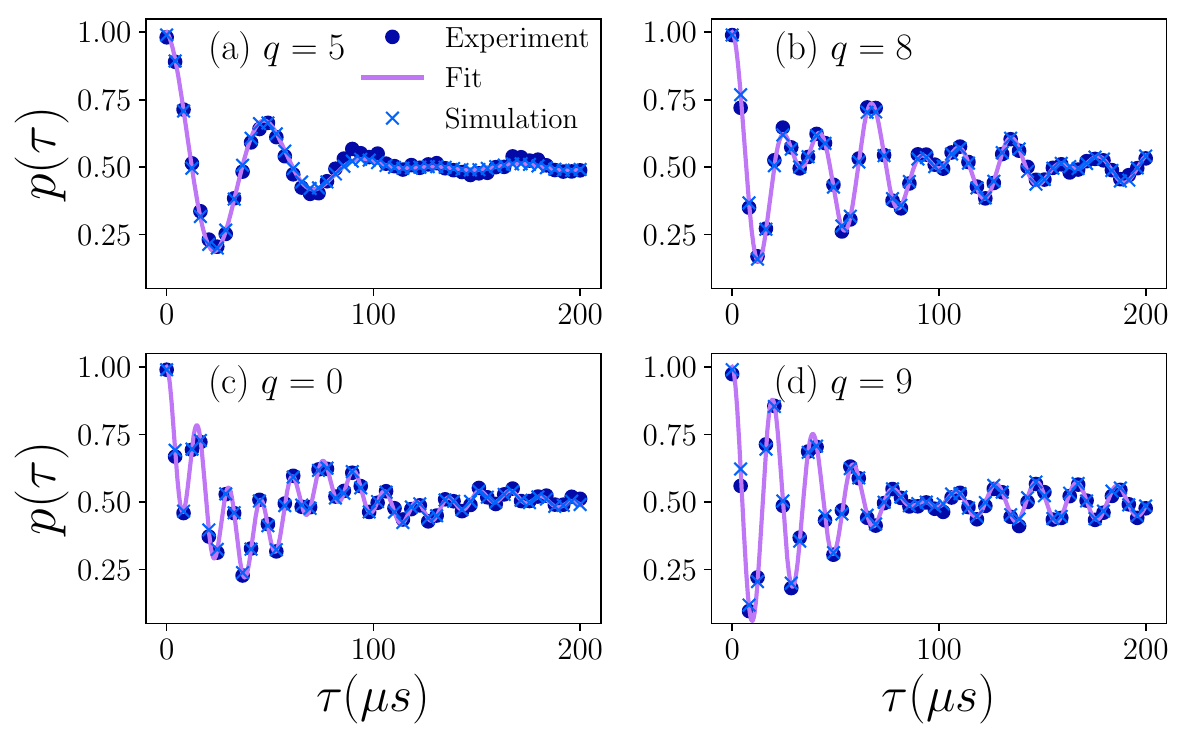}
    \caption{Ramsey experiment (R) results for qubits (a) 5, (b) 8, (c) 0, (d) 9 of \textit{ibm\_algiers}. Experiment results (dark-blue circles) are shown, along with the fits based on Eq.~(\ref{eq:v_TLS}) (solid purple lines). Markovian simulations (light-blue crosses) present excellent agreement with experiment. Detuning values $\beta$ are obtained from single-qubit characterization: (a) 0.12(9), (b) 0.04(8), (c) 0.16(9), (d) 0.017(0) MHz. TLS coupling strength $\xi$ values obtained from the fit of Eq.~(\ref{eq:v_TLS}): (a) 0.01(3), (b) 0.23(2), (c) 0.24(6), (d) 0.32(0) MHz. Note that for qubit 5 (see panel (a)), $\xi$ is approximately $\times$10 smaller than $\beta$, making it closer to a Markovian evolution. Intuitively, this is seen in the close to uniform oscillation frequency.}
    \label{fig:TLS}
\end{figure}

Figure~\ref{fig:TLS} presents experiments on four different qubits of \textit{ibm\_auckland}, where multiple oscillation frequencies are present. Rather than a single oscillating function $\cos(\beta t)$, these experiments can be well described by two oscillation frequencies inaccordance with Eq.~(\ref{eq:v_TLS}), from where both detuning $\beta$ and TLS coupling strength $\xi$ can be obtained. The qubits shown in Fig.~\ref{fig:TLS} were selected because they present different ratios of $\beta/\xi$, leading to seemingly qualitatively different phenomena. However, these experiments can be well described within the same TLS model, outlined in detail in Sec.~\ref{sec:noise-model}. The excellent agreement between experiment, theory, and simulation serves to validate the effectiveness of the simple yet strikingly predictive TLS model. Lastly, we note that the multi-frequency TLS behavior can be found in the vast majority of the qubits studied. However, cases exist where the presence of TLSs may be less obvious. For example, when a single oscillation frequency is observed in Ramsey experiments, this can be the result of a TLS rather than detuning, as shown in Appendix~\ref{app:sec:TLS_XT}.

\subsubsection{Characterization of Two-Qubit Crosstalk}
\label{subsubsec:XT_exp}

\begin{figure}[t]
    \centering\includegraphics[width=0.48\textwidth, trim={0cm 4.5cm 7cm 0cm},clip ]{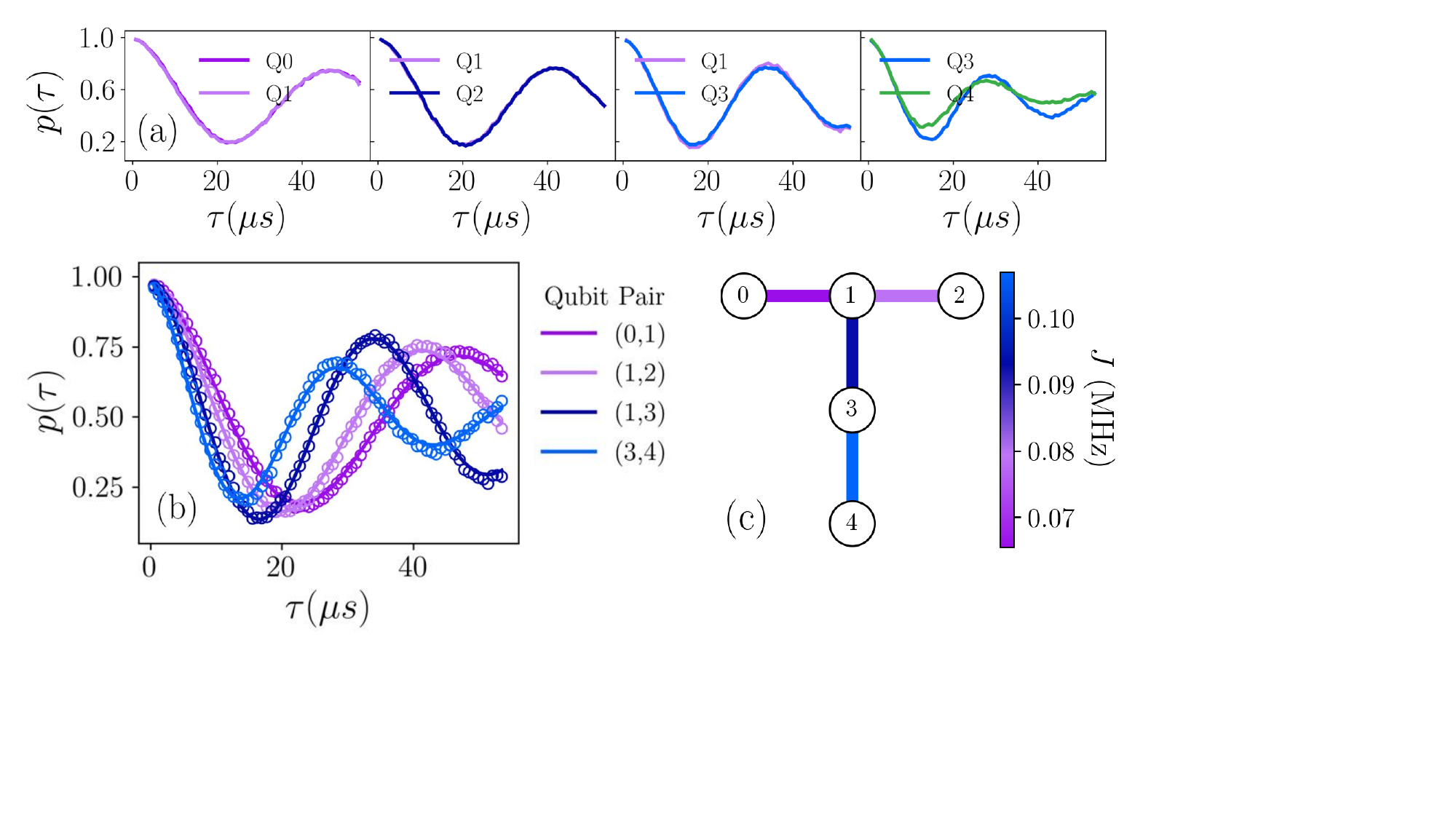}
    \caption{(a) Two-qubit crosstalk experiment (XT) results for all qubit pairs $(M,S)$ of \textit{ibmq\_lima}. (b) Experiment results of qubit $M$ (circles) and fit based on Eq.~(\ref{eq:v_XT}) (solid lines). (c) Diagram of crosstalk coupling strength values obtained from the fits shown in (b).}
    \label{fig:xt}
\end{figure}

Figure~\ref{fig:xt} shows the experimental results obtained from implementating the simultaneous HE protocol on \textit{ibmq\_lima}. Here, Eq.~(\ref{eq:v_XT}) is used to fit the crosstalk experiment results. Using the previously characterized single-qubit error parameters in the fit, the crosstalk coupling strength $J$ is found, showing excellent agreement between the model fit and experiment. Note [from Eq.~(\ref{eq:v_XT})] that the measured qubit's decay rate $\alpha_M$ contains a contribution from the spectator qubit's relaxation rate $\gamma_S$.

The noise model is able to accurately predict decay rates $\alpha_M$ with information obtained from single-qubit experiments. This result indicates that the presence of $ZZ$ crosstalk causes a incoherent mixed dissipation between the qubits, on top of additional coherent rotations. This observation provides additional evidence of the detrimental effect that crosstalk has on applications relying on simultaneous qubit manipulation and preservation. This is particularly interesting when considering error suppression techniques such as DD, which can only cancel the coherent rotations but not the incoherent contributions~\cite{tripathi2022,Zhou2023}. 

\subsection{Time-Dependent Correlated Noise on the IBMQP}
\label{subsec:corr-noise-assessment}

In addition to spatially correlated errors such as crosstalk, clear evidence of strong time-correlated noise on IBMQP processors is observed. This type of noise is ubiquitous in solid-state devices, and is known in particular to be widely present in transmon-based qubits~\cite{bylander2011noise,Paladino2014,yan2016flux,vonLupke2020,sung2021multi,Rower2023,burnett_evidence_2014}.  As described in Sec.~\ref{sec:noise-model-stoch}, properties of time-correlated stochastic noise are captured by the mean and PSD $S(\omega)$ of the noise. The FFF is the core framework used to investigate the effect of time-correlated errors on system dynamics. The FFF quantifies the noisy dynamics by computing an overlap integral in frequency space between the noise PSD and the filter functions (FFs) $F(\omega,\tau)$, which capture the sensitivity of the system. Furthermore, the FFF can be used to reconstruct spectral properties of the noise via QNS; see Appendix~\ref{app:subsec:FFF} for a detailed description of the FFF and QNS. In this section, we show that through selected characterization experiments, namely FTTPS and its variations, time-correlated non-Markovian dephasing and control noise can be detected and characterized on IBMQP systems.

\subsubsection{Correlated Dephasing Noise}
\label{subsubsec:corr_deph_exp}
We first focus on the characterization of correlated dephasing noise. We note an important distinction between two types of dephasing noise that are observed on hardware: white (uncorrelated) and colored (correlated). The former is Markovian, and is defined by a locally constant PSD, i.e., $S_\beta(\omega) = S^{(u)}_\beta$. This corresponds to dephasing noise contributing to the $1/T_2$ phase damping decay rate discussed in previous sections. Colored dephasing noise, on the other hand, is characterized by a changing PSD satisfying $S^{(c)}_\beta(\omega\rightarrow\infty)\rightarrow 0$, and commonly dominates at low frequencies. In practice, qubits generally experience a combination of both correlated and uncorrelated noise, which can be collectively notated by $S_\beta(\omega)=S^{(c)}_\beta(\omega) + S^{(u)}_\beta$. This distinction between uncorrelated and correlated will become useful in the present analysis. Since characterization of uncorrelated dephasing noise was discussed in the previous section in the context of HE experiments, it suffices to characterize the correlated contributions.

In the presence of low-frequency correlated noise, a maximum cutoff frequency $\omega_\mrm{max}$ can be generally defined such that $S_c(\omega<\omega_\mrm{max})\gg S_c(\omega>\omega_\mrm{max})$. The magnitude of this maximum frequency can be compared with the frequency resolution of a given circuit $\delta\omega=2\pi/\tau$, set by the circuit duration $\tau$. As will be shown below, this comparison gives rise to three distinct regimes. In turn, $\delta\omega$ is bounded by the qubit coherence time $1/T_2=1/2T_1+1/T_\phi$. This bound imposes a natural constraint on how finely the frequency features of the correlated dephasing PSD can be resolved. Namely, there exists a minimum frequency resolution $\delta\omega>\delta\omega_\mrm{min}=2\pi/T_2$.

First, when $\omega_\mrm{max}\gg\delta\omega_\mrm{min}$, it is possible to drive the qubit such that it becomes sensitive to a large number of frequencies in the range where correlated noise is strong. In this regime, the PSD can be characterized in detail, e.g., via standard QNS techniques~\cite{alvarez2011qns,norris2016qns} when $S(\omega)$ changes sufficiently slow compared to $\delta\omega$. Despite the ease in characterization, qubits in this regime were not observed in the course of our investigations and will not be further discussed in this work.

On the other hand, when $\omega_\mrm{max} \gtrsim \delta\omega_\mrm{min}$, the PSD can be learned partially by assuming a functional form for $S_c(\omega)$. Model parameters can be extracted by fitting to specific experiments designed to sufficiently probe low-frequency dynamics. This approach enables superresolution that is not afforded by standard QNS approaches. Lastly, if $\omega_\mrm{max} < \delta\omega_\mrm{min}$, qubit coherence is lost before enough information of long-time correlations can be collected, and the narrow spectral features of the PSD can not be resolved. This slowly varying noise is commonly denoted as quasistatic or DC noise, where the PSD $S^{(c)}_\beta(\omega)\propto \delta(\omega)$ is sharply concentrated around $\omega=0$. In addition, quasistatic noise can be thought of as fully correlated, the opposite limit to white noise.

\begin{figure}[t]
    \centering\includegraphics[width=\linewidth, trim={.4cm 0cm 15cm 0cm},clip]{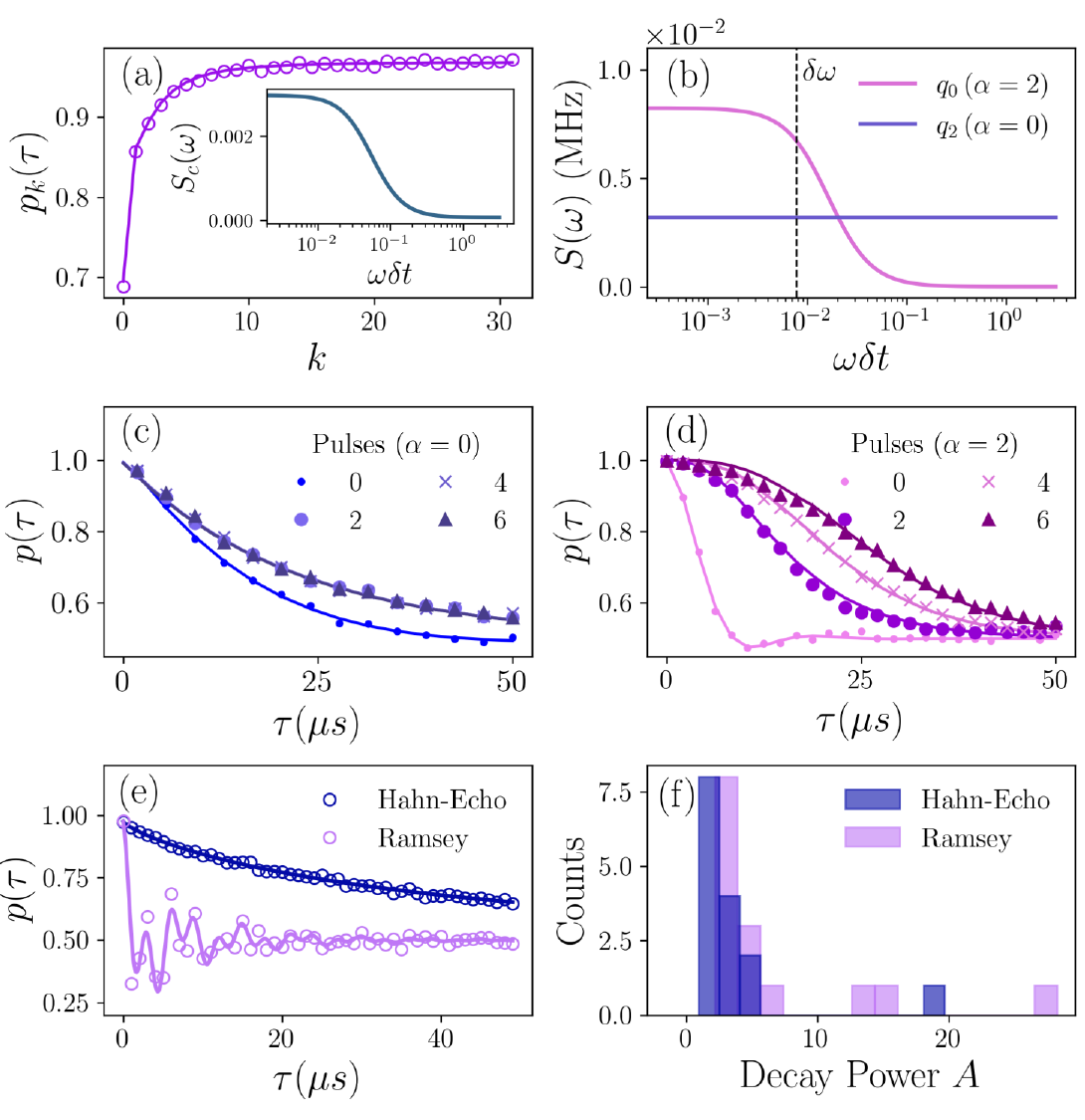}
    \caption{
    (a) FTTPS experiment results (circles) and predictions (solid line) obtained from the reconstructed PSD (inset) from qubit 4 of \textit{ibm\_hanoi}. 
    (b) Reconstructed PSDs from qubits 0 ($\alpha=2$) and 2 ($\alpha=0$) of \textit{ibmq\_belem}, showcasing two qualitatively different spectra.
    Experimental results (solid markers) of CPMG experiments with varying DD pulses for (c) qubit 0  and (d) qubit 2 of \textit{ibmq\_belem}.
    Solid lines represent the prediction for each experiment, obtained from using the corresponding PSDs shown in (b).
    (e) Ramsey and HE experimental results of qubit 7 of \textit{ibm\_algiers}, consistent with dephasing in the DC regime, showing a strong improvement in fidelity by applying a single decoupling pulse.
    (f) Histogram of decay powers $A$ on \textit{ibm\_algiers} for Ramsey and HE.
    }
    \label{fig:corr_deph}
\end{figure}

When the noise correlations satisfy $\omega_\mrm{max} > \delta\omega_\mrm{min}$, FTTPS experiments can be used to characterize correlated dephasing spectra. Figure~\ref{fig:corr_deph}(a) shows experimental results of FTTPS circuits executed on qubit 4 of \textit{ibm\_hanoi}, characterized by a $T_1\approx 70\mu s$ at the time of measurement and gate time $\delta t \approx 0.035\mu s$. For FTTPS, we set $N=128$, in order for the total evolution time $\tau\approx 9\,\mu s$ to be small compared to $T_1$. In turn, this provides a sufficiently fine frequency resolution $\delta \omega \approx 0.7\,$MHz. This combination of parameters allows us to work in the $\omega_\mrm{max} \gtrsim \delta\omega_\mrm{min}$ regime, where the noise PSD can be learned parametrically.

The inset of Fig.~\ref{fig:corr_deph}(a) presents the detected dephasing PSD obtained from QNS by fitting an auto-regressive moving average (ARMA) model, following the method introduced in Ref.~\cite{Murphy2022}. To validate the noise reconstruction protocol, the spectrum is used to obtain predicted values of FTTPS via numerical integration of the overlap integral. This is shown in the main panel as solid lines, displaying good agreement between the experimental and predicted FTTPS.

Through the ARMA fitting procedure, we identify an optimal number of model parameters based on the Akaike Information Criterion (AIC). The optimal model strongly overlaps with a Lorentzian-like spectrum 
\eq{
S_{L}(\omega;\alpha)=\frac{S_0}{1+(\omega/\omega_\mrm{max})^\alpha},
\label{eq:Sw-fun}
}
where $\alpha$ captures the color of the noise. This functional form for the correlated dephasing PSD proves to be applicable to a wide range of qubits observed on IBMQP. Furthermore, it is consistent with previous findings of $1/f^\alpha$ dephasing detected on superconducting qubits~\cite{bylander2011noise,yan2016flux, Rower2023,burnett_evidence_2014}.

By further inspecting FTTPS experimental results, we identify two examples that distinctly demonstrate the differences in ranges of correlation times. The resulting reconstructed PSDs of qubits 0 and 2 of \textit{ibmq\_belem} are shown in Fig.~\ref{fig:corr_deph}(b). The spectra show qubit 0 dominated by low-frequency noise, with non-zero contributions above the frequency resolution threshold $\delta\omega$ (see vertical dashed line), with noise color parameter $\alpha=2$. Qubit 2, on the other hand, presents Markovian dynamics with a constant PSD, consistent with $\alpha=0$. 

Next, we explore the efficacy of the characterized spectra and Eq.~(\ref{eq:Sw-fun}) through DD experiments. Experiments inspired by $T_2$ with varying duration and fixed number of DD pulses are selected, following the standard Carr–Purcell–Meiboom–Gill (CPMG) protocol~\cite{carr1954CPMG,meiboom1958CPMG}. CPMG and FTTPS probe different aspects of the same correlated noise dynamics. FTTPS vary the number of pulses for constant total time, thus changing the frequency location of maximum noise sensitivity $\omega_k$. The FF amplitudes are maintained and proportional to the circuit time squared, i.e., $F(\omega_k,\tau)\propto\tau^2$. On the other hand, CPMG$_d$ experiments with fixed number of pulses $d$ vary the inter-pulse delay times $\tau/d$, and consequently modify the FF amplitude without significantly changing the FF distribution in frequency. Analogous to the FTTPS, CPMG pushes the spectral weight of the FFs away from low frequencies, lowering the influence of low-frequency noise on the system evolution with increasing $d$.

The results of the CPMG experiments for qubits 0 and 2 are shown in Figs.~\ref{fig:corr_deph}(c) and (d), respectively. The experiments show that after applying $d=2$ pulses to qubit 2, no additional improvement is achieved, indicating that all low-frequency has been decoupled. For qubit 0 [see Fig.~\ref{fig:corr_deph}(d)], on the other hand, significant improvement is observed as more DD pulses are added. This behavior is consistent with the presence of low-frequency dephasing noise in the $\omega_\mrm{max} > \delta\omega$ regime. See Appendix~\ref{app:sec:time-correlated} for another example of a qubit with correlated dephasing noise in the $\omega_\mrm{\max}\gg\delta\omega$ regime.

We compare the CPMG experiment results with predictions obtained from the reconstructed spectra. The predicted curves are obtained from assuming that the dominant sources of errors in the DD experiments with $d>0$ pulses are amplitude damping and dephasing noise. Thus, the survival probability is described by
\begin{equation}
    p_{d>1}(\tau)\approx\frac{1}{2}\l(1+e^{-\chi_d(\tau)-\tau/2T_1}\r),
\end{equation}
with the decay parameter $\chi_d(\tau)=\int_0^{\pi/\delta t} S(\omega)F_d(\omega,\tau)d\omega$ [see Appendix~\ref{app:subsec:FFF}]. The $T_1$ time is obtained from separate $T_1$ experiments, following the steps outlined in Sec.~\ref{sec:characterization_experiments}.

The CPMG$_{d=0}$ corresponds to Ramsey $T_2^*$ experiments, and to find the predicted probabilities, we follow the Markovian noise derivation in Eq.~(\ref{eq:v_TLS}). We augment the decay to include correlated dephasing noise and including detuning as well as TLS coupling, namely 
\begin{equation}
    p_{d=0}(\tau)\approx\frac{1}{2}\l(1+e^{-\chi_R(\tau)-\tau/2T_1}\cos(\beta\tau)\cos(\xi\tau)\r),
    \label{eq:pd0-fun}
\end{equation}
where $\chi_R(\tau)$ is the Ramsey overlap integral. Note that FTTPS, $T_1$, and CPMG experiments are run in a single batch of circuits to minimize the effect of drift in the noise. In all cases, excellent agreement is found between the CPMG$_d$ experiment results and the predictions, further reinforcing the validity of the correlated dephasing model with the chosen PSD model.

Although qubits exhibiting behavior consistent with the $\omega_\mrm{max} \gtrsim \delta\omega_\mrm{min}$ regime are observed, they are not prominent. Based on our investigations, a majority of the IBMQP qubits suffering from correlated dephasing noise that can be well described within the DC regime, namely $\omega_\mrm{max} < \delta\omega_\mrm{min}$. The tradeoff between adequate frequency resolution and total circuit duration cannot be easily satisfied, thus limiting the utility of the FTTPS experiments. Noting the fact that in the quasistatic noise regime a single echo pulse suffices to decouple the noise, an alternative detection method can be devised through comparing decay rates of Ramsey and HE experiments.

In order to detect the presence of DC noise and quantify its strength, we fit the HE experiments to an ad-hoc exponentially decaying function 
\begin{equation}
    p_\mrm{HE}(\tau) = \frac{1}{2}\l(1 + \exp\l[-A(\tau/\tau_\mrm{max})^a-\tau/T_2\r]\r),
\end{equation}
where in the Ramsey case we also include the oscillating cosines as in Eq.~(\ref{eq:pd0-fun}). The fit parameters $a,A$, respectively, capture the color of the noise and the normalized decay power sensed by the qubit in each experiment setting; $\tau_\mrm{max}=50\mu s$ is the maximum experiment time, added for normalization.

An example of a qubit coupled to dephasing noise in the DC regime is shown in Fig.~\ref{fig:corr_deph}(e) via qubit 7 of \textit{ibm\_algiers}. Results are shown for both Ramsey and HE, where the improvement in decay rate after applying a single pulse is evident. This analysis can be further used to show that DC noise is pervasive in IBMQP systems. Figure~\ref{fig:corr_deph}(f) shows the decay power for 15 of the 27 qubits on the \textit{ibm\_algiers} device for both Ramsey and HE. Overall, the Ramsey experiments have larger decay powers, indicating the presence and ubiquity of correlated dephasing noise.

It is worth noting that even though it is challenging to characterize DC noise in detail via QNS, its simplicity proves advantageous in gate modeling. When dephasing noise correlations are predominantly DC, the contributions can be accurately separated into uncorrelated and correlated components. These two contributions can be modeled independently in a straightforward way, avoiding the need to compute FFs and/or average over a large number of noise realizations in simulation. This property will be exploited in Sec.~\ref{sec:scaling} to simplify the modeling task in a gate-based approach for computation that provides improved scalability.

\subsubsection{Correlated Control Noise}
\label{subsubsec:corr_ctrl_exp}

In addition to correlated dephasing, strong indications of correlated control errors were observed on IBMQP devices. Below, we discuss the characterization of this noise source and provide a few representative examples to convey the typical features found on these devices. Through this discussion, we give justification for the inclusion of control noise in the device noise model presented in Eq.~(\ref{eq:stoch-noise-Hamiltonian}). 

An approach to characterizing noise correlations in faulty control is through QNS. Previous work in this domain has focused on development of sophisticated control techniques based on functional expansions with the aim to thoroughly learn spectral features of control noise~\cite{frey2017application,maloney2022qubit}. Here, we take an alternative, minimalistic approach centered around detection rather than detailed characterization. In order to detect correlations, we rely on FTTPS, which can be used to extract information about control noise under reasonable assumptions, namely that the noise is concentrated around low frequencies.
As shown below, while originally designed to address dephasing noise, FTTPS can be tuned to either maximize or suppress sensitivity to low-frequency control noise. 

Crucially, FTTPS are maximally sensitive to low-frequency control noise, meaning that most of the weight of their FFs is concentrated around $\omega=0$. Intuitively, coherent and low-frequency errors accumulate after each pulse, aggregating to a non-zero total over- or under-rotation. Assuming control noise is dominant and ignoring for the moment other sources of noise, we can write the survival probability of the $k^\mathrm{th}$ FTTPS as 
\eq{
p_k(\tau) \approx \frac{1}{2}\l(1+e^{-\chi_k(\tau)}\cos(\bar{\epsilon} k)\r).
}
The coherent contribution appears as $\cos( \bar{\epsilon} k)$ oscillating periodically with the mean of the noise. Note that for weak noise $\bar{\epsilon}\ll1$, this term becomes a deviation from identity that increases quadratically with $k$. In addition, the effect of the stochastic contribution can be quantified from the explicit computation of the control FFs of FTTPS. The control FF can be approximated as $F_k(\omega,\tau)\approx 4 k^2 \delta(\omega)$, where the delta function is defined by $\int_0^{\delta \omega}\delta(\omega) d\omega = 1$ and $\delta(\omega>\delta\omega)=0$. Stochastic noise enters in the survival probability as an exponential decay contribution $e^{-\chi_k(\tau)}$, where in the low-frequency noise regime $\chi_k(\tau)=\int_0^\infty S_\epsilon(\omega) F_k(\omega,\tau) d\omega \propto \sigma^2 k^2$, with $\sigma$ the standard deviation of the noise. Thus, this analysis shows that FTTPS can be used to detect the presence of low-frequency correlated control errors, as well as distinguish between coherent and stochastic contributions. Further details on these FF expressions can be found in Appendix~\ref{app:subsec:corr_ctrl_noise}.
 
\begin{figure}[t]
\centering\includegraphics[width=\columnwidth, trim={0cm 4.1cm 0cm 0cm},clip]{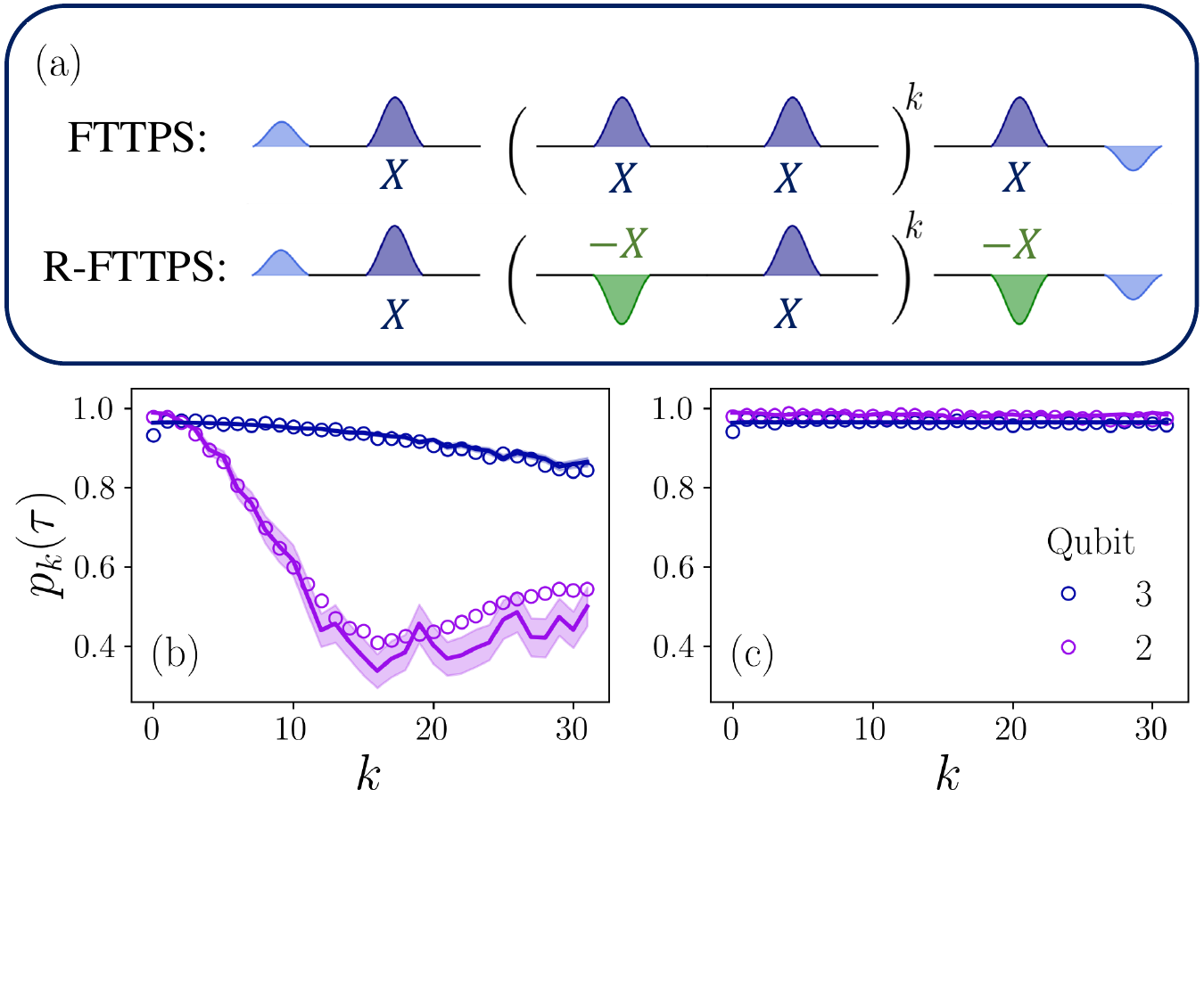}
    \caption{(a) FTTPS and R-FTTPS circuits, where the explicit distinction of alternating every even $X_\pi$ pulse has been explicitly highlighted. (b) FTTPS and (c) R-FTTPS examples of qubits presenting mostly coherent (qubit 3) and strong stochastic (qubit 2) control errors in \textit{ibmq\_lima}. Empty circles represent experimental results, and solid lines represent simulation of $1/f$ stochastic control noise with 100 Monte Carlo realizations. As shown, both the decay and oscillations can be approximately reproduced in simulation. 
    }
    \label{fig:ctrl_noise}
\end{figure}

In Fig.~\ref{fig:ctrl_noise}(b), we show examples of FTTPS experiments applied to two qubits influenced predominantly by control noise on \textit{ibmq\_lima}. Qubit 3 is mostly subject to coherent control errors, i.e., $\epsilon(t)\approx \bar{\epsilon}$ is constant. This characteristic is supported by the quadratic decay in the probability as a function of $k$. Qubit 2, in addition to oscillations, is subject to a decay, which is consistent with a correlated stochastic control noise description where $\chi_k(\tau)>0$. As shown in solid lines, the decay and oscillations can be approximately reproduced in simulation by $1/f$ stochastic control noise. Note that the loss of fidelity with increasing $k$ can be contrasted with the behavior previously observed due to low-frequency dephasing noise in Fig.~\ref{fig:corr_deph}(a), where the effect of the noise was more pronounced for small $k$. Apart from the qualitatively distinct phenomenon of resonances discussed in Appendix~\ref{sec:Sw_resonance}, high-frequency dephasing noise was not observed, and thus these large-$k$ features in FTTPS suggest that they are a signature of control noise.

To establish the connection between these FTTPS features and control noise with greater certainty, however, we need to evaluate the effect of control errors in isolation. This can be done, for example, by designing an alternative experiment where control noise effects are suppressed while the contributions of all other sources of errors are unchanged. Interestingly, this can be achieved with a simple modification through which the FTTPS are made insensitive to low-frequency control noise. To this end, we define the Robust-FTTPS (R-FTTPS) by alternating the sign of every other $X$ pulse in the standard FTTPS, i.e. $X\rightarrow-X$ [see Fig.~\ref{fig:ctrl_noise}(a)]. On the IBMQP, this is implemented by appending virtual $Z$ gates on each side of all even pulses.

In the absence of control errors, this sign change is undetectable, since the $Z$ gates commute with the native detuning and crosstalk errors, and $X$ and $-X$ pulses implement the same rotation on the Bloch sphere. However, when coherent or low-frequency control errors are present, even and odd pulses contribute approximately equal over-rotation angles but with opposite signs. The collective over-rotation error contribution thus cancels after each pair of pulses, supplying the R-FTTPS with robustness to coherent and low-frequency control noise. In the language of the FFF, the effect of phase alternation is to shift the peak sensitivity of the qubit away from low frequencies, i.e., $F_k^{R}(\omega=0,\tau)\approx 0$. Consequently, the R-FTTPS suppress low-frequency control errors on the survival probabilities, as seen by computing $\chi_k^R(\tau)=\int_0^\infty S_\epsilon(\omega) F_k^R(\omega,\tau)d\omega \approx 0$ for predominantly low-frequency $S_\epsilon(\omega)$. We further elaborate on R-FTTPS in Appendix~\ref{app:subsec:RFTTPS}.

The presence of correlated control errors can be established conclusively by comparing the experimental results of FTTPS and R-FTTPS. Figure~\ref{fig:ctrl_noise}(c) presents experimental results of R-FTTPS on qubits 2 and 3 of \textit{ibmq\_lima}. Notably, the features surmised to be associated with correlated control noise have been completely suppressed. This markedly distinct outcome from the standard FTTPS protocol agrees with expectations and further provides strong evidence for the presence of low-frequency correlated control errors on these qubits.

\subsubsection{A Word on the Prevalence of Single-Qubit Correlated Noise on IBMQP Devices}

The presence of correlated noise naturally gives rise to the following question: how prevalent are these noise sources on IBMQP devices? In order to provide insight into this question, we perform the characterization routines described in Sec.~\ref{sec:characterization_experiments} on seven different devices. We summarize our findings in Table~\ref{table:markov_table}.

We categorize qubits as exhibiting Markovian noise only, or including correlated noise processes as well. Our findings indicate that among the seven devices studied, approximately 64\% of the qubits exhibited pure Markovian noise. Deviations from this behavior were observed, where approximately 26\% and 10\%  experience correlated dephasing or control noise, respectively. Additional cases exist where both correlated noise processes are present. These qubits make up about 5\% of the total. Importantly, we observed that the presence of correlated dephasing noise was stable in time across different qubits. That is, qubits undergoing correlated dephasing noise are likely to remain under the influence of this type of noise. This observation becomes important when selecting qubits for different applications and error management protocols. For instance, those that experience a significant amount of correlated noise are potentially promising candidates for DD suppression protocols.

\begin{table}[t]
\centering
\begin{tabular}{c | c || c | c | c} 
\hline\hline
 \multicolumn{2}{c||}{Processor} &
\multicolumn{3}{c}{Type of Noise in Qubits} \\
 \hline
 Name & Type & \makecell{Pure\\ Markovian} & \makecell{Markovian + \\ Corr. Dephasing} & \makecell{Markovian + \\ Corr. Control} \\ 
 \hline
 belem & T & 0,1,2,3 & 4 &  \\[.5ex]  
 quito & T & 0,1,2,4 & 3 &  \\[.5ex] 
 lima & T & 3,4 &  & 0,1,2 \\[.5ex] 
 nairobi & I & 1,2,4,6 & 0,3,5 &  \\[.5ex] 
 jakarta & I & 0,1,3 & 2,4 & 4 \\[.5ex] 
 lagos & I & 1,2,3,5,6 & 0,4 & 4 \\[.5ex] 
 manila & --- & 0,1,3 & 2 & 4 \\[.5ex] 
 \hline\hline
\end{tabular}
\caption{Summary of single-qubit characterizations performed over seven IBMQP devices. Qubits found to be well described by the Markovian model produce a fitting error of at most $\delta=1\%$ for the overall MSE between data and fit. Those above this threshold are determined to be subject to correlated dephasing, control noise, or both. Approximately 64\% are found to exhibit purely Markovian behavior, while 26\% and 10\% experience correlated dephasing or control noise, respectively.}
\label{table:markov_table}
\end{table}

\subsection{Characterizing Entangling Two-Qubit ECR Gates}
\label{subsec:CR_gate}

\begin{figure}[t]
    \centering
    \includegraphics[width=\linewidth, trim={0cm 3cm 18cm 0cm},clip]{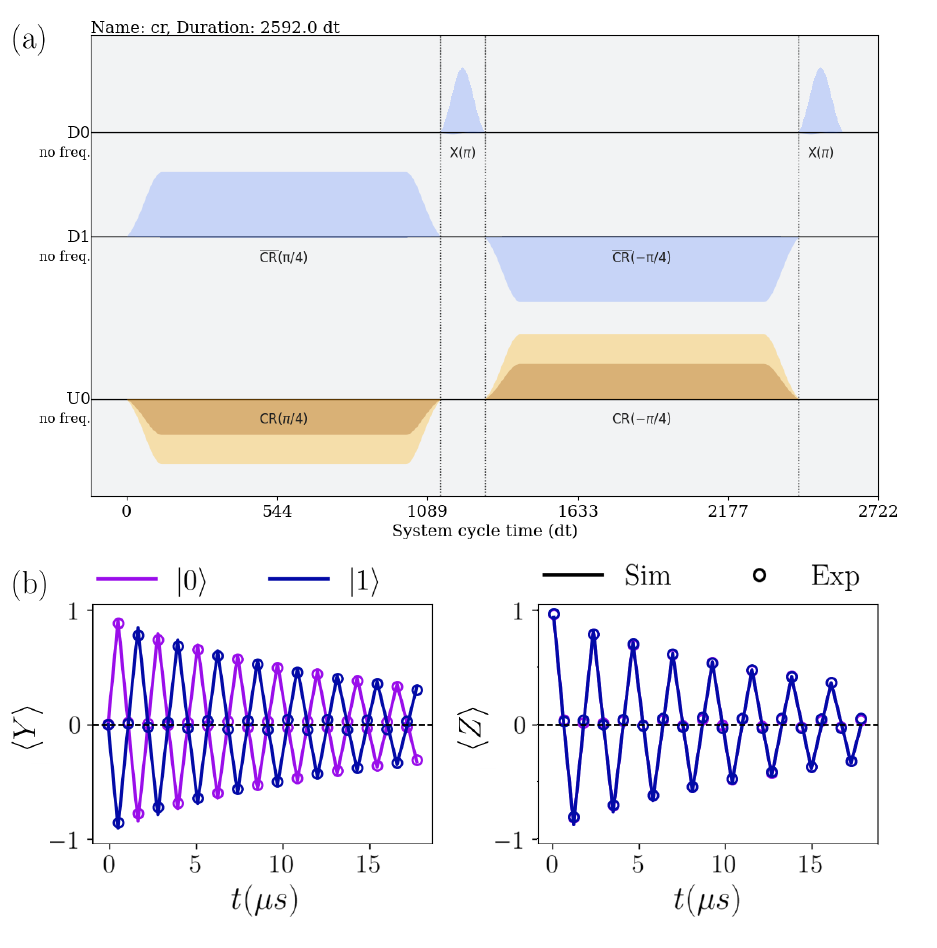}
    \caption{Comparison between experiment and the LME model for the ECR gate. (a) Pulse schedule between qubits 0 (control) and 1 (target) of \textit{ibm\_lagos} obtained from Qiskit Pulse. Here, $\tau_{CR}=0.576\mu s$ is much larger than the single-qubit gate duration $\delta t = 0.035\mu s$. (b) Experimental (empty circles) and simulated (solid lines) expectation values $\braket{Y}$ and $\braket{Z}$ for the target qubit obtained by $n$ repeated applications of the pulse shown in (a). The control qubit is prepared in $\ket{0}$ (purple) and $\ket{1}$ (blue). Time on the $x$ axis is computed as $n \tau_{CR}$.}
    \label{fig:CR_pulse_exp}
\end{figure}

Here, we turn our attention to entangling operations provided by the ECR gate. An example of the composite gate sequence is shown  Fig.~\ref{fig:CR_pulse_exp}(a). We employ the noise model of Eq.~(\ref{eq:LE_general}), leveraging single-qubit noise parameters in conjunction with additional fitting to the parameters of the effective ECR Hamiltonian [Eq.~(\ref{eq:H_CR})]. As we show below, this noise model is sufficient to account for errors observed during two-qubit operations on fixed coupler IBMQP devices. 

Experimental results for repeated ECR gates between qubits 0 and 1 on \textit{ibm\_lagos} can be seen in Fig.~\ref{fig:CR_pulse_exp}(b). Expectation value estimates for $\braket{Y}$ and $\braket{Z}$ are shown in the left and right panels, respectively, for control qubit states  $\ket{0}$ and $\ket{1}$. Through the characterization protocols, we estimate a gate over-rotation of $\epsilon_{zx}\approx 0.14$, while $\zeta \approx 0.01$~MHz. The duration of the CR gate is $\tau_{CR}=0.576\mu$s, which is approximately 16.5$\times$ larger than the single-qubit gate time.

Importantly, despite the complexity of the gate operation, the LME with single-qubit dephasing and amplitude damping proves to be sufficient for capturing the expectation value decay. As shown in Fig.~~\ref{fig:CR_pulse_exp}(b), we find strong agreement between the experiments and the model up to $n=16$ repetitions of the ECR gate. Although more experimental data may be needed in order to refine the model, we tested 7 pairs of qubits accross four different devices, and found that this minimal ECR model was successful in fitting the CR characterization experiments. In addition, for qubits $(c,t)=(0,1)$ of \textit{ibmq\_lagos}, the magnitude of $\braket{X}$ was non-negligible, and the model was not sufficient to fit $\braket{Y},\braket{Z}$. We attribute these deviations to a large crosstalk strength of $J\approx0.5\,$MHz, which is inconsistent with the approximations used in the present ECR gate characterization. For the other qubits, the maximum crosstalk strength found was $J\approx 0.09\,$MHz. Thus, we expect the model to fit the ECR experiments as long as crosstalk remains relatively small. 

\section{Applications}
\label{sec:applications}

Thus far, we have explored the efficacy of the single and two-qubit models, using the characterization experiments to convey model validation. Single-qubit RB experiments have been used to test the model and show agreement for predicted gate error rates when the qubit is subject to Markovian noise alone, or subject to additional non-Markovian noise sources. Here, we increase the complexity of model testing and examine two distinct multi-qubit scenarios. First, we examine multi-qubit DD and show that our model is capable of predicting state-dependent behavior in the ground state probability. Next, we consider an implementation of VQE in which the goal is to compute the dissociation curve of the H$_2$ molecule.

\subsection{Multi-Qubit Dynamical Decoupling}
\label{subsec:multi_qubit-DD}

In this section, we study the impact of simultaneous driving in the presence of parasitic crosstalk. We consider a \emph{main} qubit and investigate its dynamics in the presence of \emph{spectator} qubits that interact with the main qubit via $ZZ$ crosstalk. Following Refs.~\cite{tripathi2022,tripathi2023modeling}, we perform two types of simultaneous single-qubit experiments. 

\begin{figure}[t]
    \centering
    \includegraphics[width=0.9\linewidth, trim={0cm .58cm 0cm 0cm},clip]{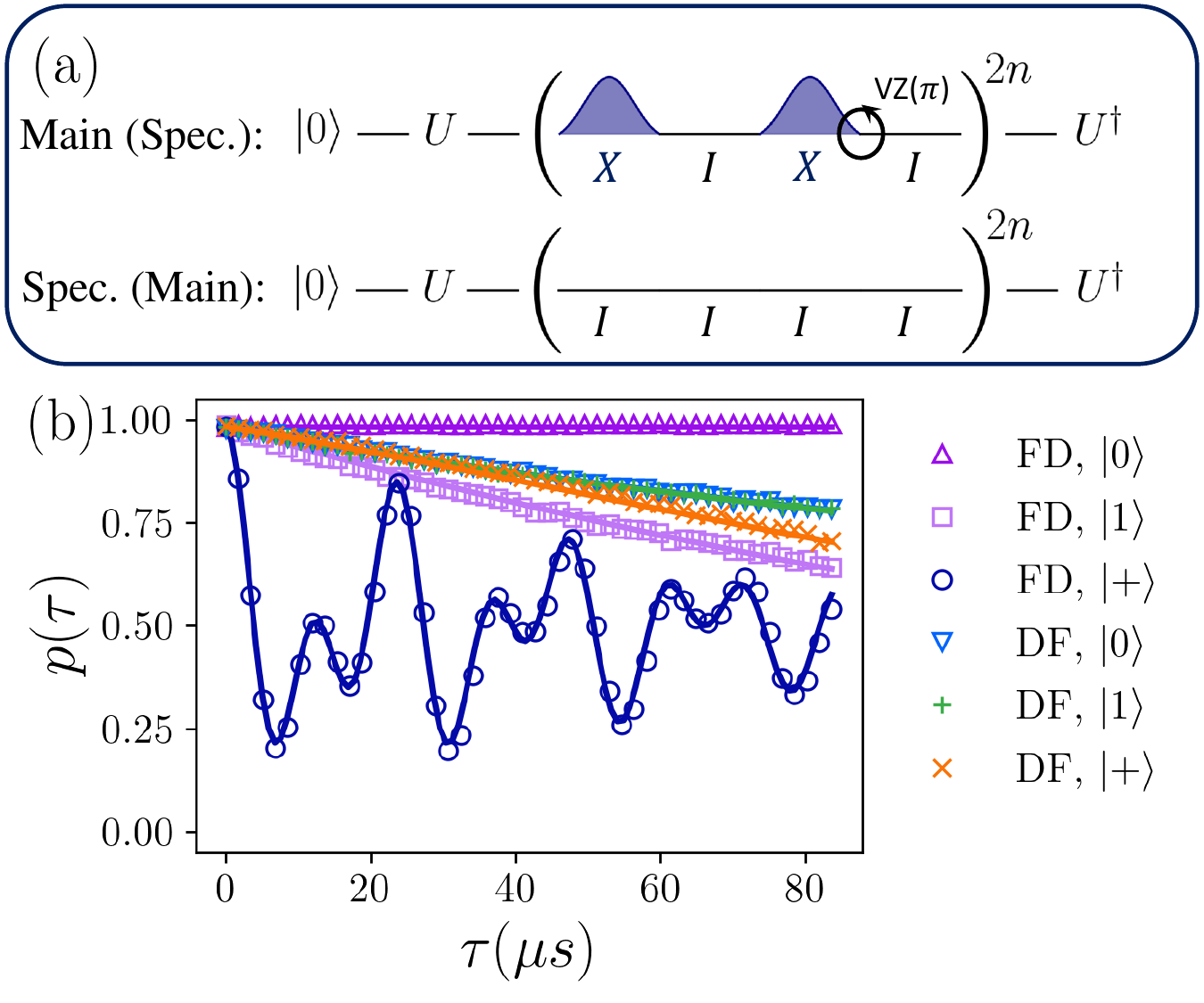}
    \caption{(a) Multi-qubit DD experiment schematics. The main (spectators) qubit is driven by XY4, implementing the $Y$ gate as $X$ followed by a $Z_\pi$ rotation, whereas the spectators (main) are left to evolve freely. The state preparation gates used were to prepare the states along $|\psi\rangle=\ket{0},\ket{1},\ket{+}$ we use $U=I,X,\sqrt{Y}$, respectively. (b) Experiment (circles) and LME simulation (solid lines) results performed on \textit{ibm\_cairo} using qubit 1 as the main qubit. Qubits 0,2 and 4 are taken to be spectators. Each curve is labeled as $MS,|\psi\rangle$, where $M(S)$ represents the evolution of the main (spectator) qubit(s), and $\ket{\psi}$ the initial states. F: Free Evolution; D: Dynamical decoupling XY4.}
    \label{fig:vinay}
\end{figure}

In Type 1 experiments, the main qubit is allowed to evolve freely. Meanwhile, the spectator qubits are subject to DD. In Type 2 experiments, the roles are reversed: the main qubit is subject to DD, whereas the spectators are allowed to evolve freely. Each experimental scenario utilizes the DD sequence ${\rm XY}_4=Yf_\tau Xf_\tau Yf_\tau Xf_\tau$, where $X$ and $Y$ are $\pi$-rotations along the direction of the $\sx$ and $\sy$ Pauli operators, respectively. $f_\tau$ denotes a period of free evolution, where the qubit is allowed to evolve according to its internal dynamics for time $\tau$. To ensure each pulse has equivalent time, $Y$ gates are compiled as $X$ followed by a virtual $Z_\pi$ rotation. All qubits, main and spectators, are prepared in either $|0\rangle, \ket{1}$ or $\ket{+}$ via simultaneous application of the single-qubit rotation operator $U$. Upon completion of $n$ repetitions of the DD sequence, the inverse state preparation unitary is applied to (ideally) return the qubits to the ground state prior to measurement. A schematic of the circuits is shown in Fig.~\ref{fig:vinay}(a).

In Fig.~\ref{fig:vinay}(b), we show an implementation of Type 1 and 2 experiments on \textit{ibm\_cairo}. The main qubit is designated by qubit 1, while spectators are qubits 0,2,4. Experimental data is collected up to a maximum time of $T=83.5\,\mu s = n \tau_c$, or equivalently, $n=1176$ repetitions of ${\rm XY}_4$. The inter-pulse delay is set to $\tau=0$, such that the total cycle time $\tau_c=4\delta t$ is determined solely by the $X$ pulse duration ($\delta t=0.035\,\mu$s). Measurement statistics are determined from 10,000 shots.

In comparing the experimental data with the LME model, we find excellent agreement. Simulations (solid lines) exhibit strong overlap with experimental data (open symbols). Most notably, the model captures state-dependent oscillatory behavior observed in Type 1 experiments. As described in Sec.~\ref{sec:characterization_experiments}, the presence of a TLS results in an additional oscillation frequency that accompanies ZZ crosstalk. Here, we observe this behavior for the main qubit, specifically for the equal superposition state. The extended LME model predicts these non-trivial dynamics over the range of time considered. 

Similarly, our model agrees well with Type 2 experiments. Intuitively, DD applied to the main qubit should result in a suppression of the TLS interaction and thus, an elimination of the oscillation frequency. Both experimental results and simulations corroborate this expectation. In addition, experimental results indicate a slight state-dependent performance in the main qubit. This behavior is also well described by the trained model, further conveying its efficacy.

\subsection{Variational Quantum Eigensolver for \texorpdfstring{H$_2$}{H2} Molecule}
\label{subsec:VQE}

In addition, we showcase the utility of our noise model in the prediction of hardware dynamics for a quantum algorithm. We focus on VQE~\cite{tilly_variational_2022}, a hybrid classical-quantum algorithm for simulating properties of a Hamiltonian. VQE relies on a classical optimizer to minimize an energy functional across a set of parameterized quantum circuits, where the energy is estimated via execution of the quantum circuits on hardware. When focused on finding ground states, the algorithm identifies a circuit that generates an approximation of the molecule's ground state and provides an estimate of its ground state energy.

\begin{figure}[t]
    \centering
    \includegraphics[width=\linewidth, trim={0cm 1cm 14cm 0cm},clip]{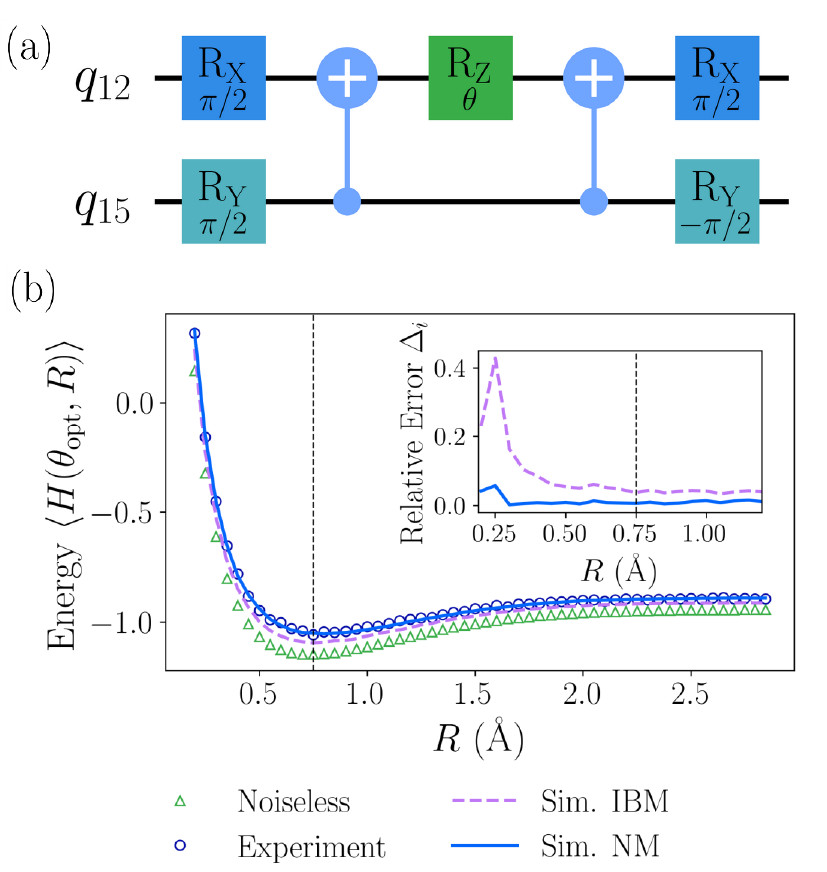}
    \caption{(a) Circuit implementing the VQE algorithm, from which the ground state energy $\langle H(\theta,R)\rangle$ is computed. (b) Experiment results (dark blue) after finding optimal $\theta_\mrm{opt}$ offline in an ideal simulator (purple), compared with IBM (pink) and our (light blue) simulations. The inset shows the relative error $\Delta(R)$ between the experimentally obtained energies and the two noise models. At the optimal atomic distance, our model presents a relative error of $0.5\%$, showing a $\times 7$ improvement compared to the $3.5\%$ relative error of the IBM model.}
    \label{fig:VQE}
\end{figure}

In this comparison, we study the dissociation curve of the H$_2$ molecule, using a previously implemented operator-to-circuit mapping~\cite{OMalley2016}. In the ideal case, this two-qubit implementation has shown to produce results compatible with the ideal ground state energy. The circuit is shown in Fig.~\ref{fig:VQE}(a). It consists of five single-qubit gates (one of which is parameterized) and two CNOT gates. Based on the mapping of Ref.~\cite{OMalley2016}, the average energy $\langle H(\theta,R) \rangle=\sum_i \braket{O_i}$ can be calculated via a linear sum of expectation values, where $O_i\in\{\sigma_x^{A}\sigma_x^{B},\sigma_y^{A}\sigma_y^{B}, \sigma_z^{A}\sigma_z^{B},\sigma_z^{A} I^B,I^A \sigma_z^{B}\}$. The optimization is carried out by varying the angle $\theta$ to identify the minimum of $\langle H(\theta,R) \rangle$ for each bond length $R$.

We implement the above algorithm on \textit{ibm\_algiers}, where $A$ and $B$ are given by qubits 12 and 15, respectively. Both qubits are first subject to the single and two-qubit characterization protocols described in Sec.~\ref{sec:characterization_experiments}. 
We find that qubit 15 possesses a strong contribution of correlated dephasing, while qubit 12 is predominantly characterized by Markovian processes. Upon learning the noise model parameters for each qubit, we leverage the model to perform offline training of the VQE via an exhaustive search. Experimental demonstrations are then performed to estimate the average energy using the optimal $\theta$ parameters obtained for each $R$ using 10,000 shots. The dissociation curve obtained from the quantum hardware is compared against the IBM device noise model and the learned LME model. The results of this comparison are shown in Fig.~\ref{fig:VQE}(b).

The IBM device noise model draws on characterization data imported from the backend properties. This data is updated periodically, typically once over the course of 24 hrs. In the simulations shown in Fig.~\ref{fig:VQE}, the characterizations were performed approximately 9 hours prior to the experiments. The device noise model consists of single and two-qubit depolarizing errors followed by thermal relaxation errors. Readout errors for all measurements are also included. Assuming pure Markovian dynamics, the default device noise model does not fully capture prominent noise sources. As a result, we find that our noise model affords stronger agreement with the hardware.

We quantify the difference in performance between the models via the relative error in energy
\begin{equation}
    \Delta(R)= \l| \frac{E_{\rm sim}(R) - E_{\rm exp}(R)}{E_{\rm exp}(R)} \r| \times 100\%.
\end{equation}
Shown in the inset of Fig.~\ref{fig:VQE}(b) is the relative error as a function of $R$ for both noise models. The vertical dotted line denotes the optimal atomic distance $R_{\rm opt}=0.75\,\mathring{A}$. Here, the default IBM device noise model yields $\Delta(R_{\rm opt})\approx 3.6\%$. In contrast, our noise model obtains a relative error of $\Delta(R_{\rm opt})\approx 0.5\%$, a $7\times$ improvement over the default model. This improved agreement speaks to the efficacy of our model and its potential applicability to larger quantum systems implementing more complex quantum circuits.

Lastly, we test the reliance of these results on the correlated dephasing noise contribution of our noise model. To this end, we substitute the correlated dephasing term in the Hamiltonian with a modified phase damping channel. The corresponding $T_\phi$ is obtained from fitting the HE experiment with Eq.~(\ref{eq:v_T2}), following the steps outlined in Sec.~\ref{subsec:markov_exp_results}, from where we obtain $T_\phi\approx 29\mu s$. This ``Markovianized'' version of our noise model resembles closely the IBM one, had it been characterized shortly before the VQE experiment was run, while including more realistic sources of noise. Using the ``Markovianized'' model, we calculate a relative error $\Delta(R_\mrm{opt})\approx 3.8\%$, consistent with the IBM model. Thus, the non-Markovian model outperforms the ``Markovianized'' model in the VQE application. 

This result highlights the importance of including noise correlations for accurate predictions of complex quantum applications influenced by non-Markovian noise. The non-Markovian model rests on more physically motivated principles. Enforcing a Markovian fit often leads to model violations, where parameters such as $T_\phi$ are obtained from a HE characterization experiment presenting clear deviatinos from exponential decays (e.g., see Fig.~\ref{fig:corr_deph}(d)). Standard IBM calibrations appear to rely on assumptions of exponential decay for \emph{all} qubits. This approach can yield unreliable predictions of model parameters when non-Markovian noise is present, and ultimately, improper modeling of quantum circuit dynamics. Consequently, when non-Markovian noise is present, the explicit modeling of correlations is paramount.

\section{Model Scalability}
\label{sec:scaling}

Here, we comment on the scalability of the characterization protocol and the model. In the former, we address the number of characterization experiments required as the system size increases. Model scalability is tackled by reducing the LME models to composite channel approximations that effectively capture stochastic and extended degrees of freedom when applicable.

\subsection{Scaling of Characterization Protocol}
\label{subsec:char-scaling}

The scaling of the characterization protocol is tightly coupled to the required noise model. If we assume that all model parameters are needed then the model is determined by 10 single-qubit parameters and 3 parameters associated with qubit-qubit interactions. Thus, for a processor consisting of $Q$ qubits, $10Q$ parameters are required to describe single-qubit dynamics. As such, an equivalent number of characterization circuits are required at a minimum. Although, more may be desired to obtain more accurate estimates. 

Parallelization can assist in reducing the number of circuits. This is common practice amount existing IBMQP devices, where simultaneous characterization experiments are performed on next-nearest neighbor qubits to avoid unwanted crosstalk interactions. Note that parallelization in QNS is also possible using crosstalk robust characterization protocols such as those introduced in Ref.~\cite{Zhou2023}. As a consequence, the inherent overhead in the number of qubits with regards to characterization time can be substantially reduced.

In the case of the two-qubit parameters, the number of required characterization experiments is highly dependent upon the device topology. For the heavy-hex topology, each unit cell contains 12 qubits with an average qubit degree of 2.4~\cite{chamberland2020qec}. The ring-like topology results in approximately 12 interactions per unit cell. For $L$ unit cells, there are approximately $9L$ distinct interactions that must be characterized. The CR and crosstalk characterization protocols are sufficient to perform this characterization, where a lower bound of $9L$ circuits must be executed. Parallelization can again be utilized to further reduce the overhead, where non-nearest neighbor interactions are simultaneously characterized.

\subsection{Scaling the Model}
\label{subsec:channel_rep}

The LME models described above exhibit strong agreement with hardware in both single and two-qubit demonstrations. Despite this success, one cannot overlook the analytical and numerical challenges associated with solving the LME for multi-qubit systems beyond a few qubits. In order to address this challenge, we draw on the quantum channel formalism. Each noise source is reduced to a quantum map such that the composition of said maps approximates the single or two-qubit error channel. The model accounts for both Markovian and non-Markovian behavior.

The effect of noise on the system is described via the operator sum representation (OSR), where a dynamical map is defined by the action on a density matrix $\mathcal{E}(\rho)= \sum_{\alpha} E_\alpha \rho E_\alpha^\dagger$, for some Kraus operators $\{E_\alpha\}_\alpha$.
The total map $\mathcal{E}(\cdot)$ propagates the state in time from time $t$ to time $t+\tau$, that is, $\rho(t+\tau)=\mathcal{E}(\rho(t))$. 
This map includes both noise and control operations and can be written as a composition of noise and control maps $\mathcal{E}= \mathcal{E}_{N}\circ\mathcal{E}_C$, where the subscripts $C,N$ denote the control and noise processes, respectively. The noise map can itself can be separated into a number of independently acting noise channels 
\begin{equation}
    \mathcal{E}_{N} = \mathcal{E}_{N_r}\circ \cdots \circ \mathcal{E}_{N_1}
\end{equation}
where 
\begin{equation}
    \mathcal{E}_{N_k}\in\{ \mathcal{E}_\mrm{GAD}, \mathcal{E}_D, \mathcal{E}_{CN}, \mathcal{E}_{ZZ}, \mathcal{E}_M\}.
    \label{eq:channel_list}
\end{equation}
Note that $r$ will depend on the number of qubits, and also on the operation implemented $\mathcal{E}_C$; namely, whether it is identity, $z$- or $x$-rotations, or ECR gates. Throughout this section, we assume that evolution times $\tau$ are short, typically in the range of a native gate duration $\delta t$, and noise processes are weak, meaning that error rates are much smaller than $1/\tau$.
Longer evolution times can be achieved in a straightforward manner by channel composition. 

Among the error maps listed in Eq.~(\ref{eq:channel_list}) is the GAD channel. Ignoring the qubit label and dropping the initial time $t$ dependencies for notational simplicity, the GAD quantum map consists of $\mathcal{E}_\mrm{GAD} (\rho) =  \sum_{\alpha=0}^3 E_\alpha \rho E_\alpha^\dagger$, where the four Kraus operators are given by
\eq{
\begin{split}
E_0&=\sqrt{q}\,\l( \ket{0}\bra{0} + \sqrt{1-p_\gamma}\ket{1}\bra{1}\r),\\
E_1 &= \sqrt{p_\gamma q}\ket{0}\bra{1},\\
E_2 &= \sqrt{1-q}\,\l( \sqrt{1-p_\gamma}\ket{0}\bra{0} + \ket{1}\bra{1}\r), \\
E_3 &= \sqrt{p_\gamma (1-q)}\ket{1}\bra{0}.
\end{split}
}
The jump probability is $p_\gamma=1-e^{-\gamma \tau}$, where $\gamma,q$ are defined as in Sec.~\ref{sec:noise-model}.

In addition, we can define a model-reduced dephasing channel $\mathcal{E}_D$ that accounts for time-varying noise and Markovian processes relevant to the stochastic LME. Specifically, for a stochastic process $\beta(t)$ as defined in Eq.~(\ref{eq:stoch-noise-Hamiltonian}) with mean $\overline{\beta}$ and noise PSD $S_\beta(\omega)$, we construct the composite channel $\mathcal{E}_D(\rho)=U_{\beta}(\tau)\mathcal{E}_{\rm \beta}(\rho)U^\dagger_{\beta}(\tau)$. The unitary $U_{\beta}(\tau)=e^{-i \overline{\beta}\tau \sz/2}$ captures drift due to the static components of the dephasing. In contrast, the channel
\begin{equation}
    \mathcal{E}_{\beta}(\rho) =\left[1-p_\beta(\tau)\right] \rho + p_\beta(\tau) \sz \rho \sz,
    \label{eq:dephasing-channel}
\end{equation}
characterizes the decay component of the noise. The error probability $p_\beta(\tau)=1/2(1+e^{-\chi_\beta(\tau)})$, where
\begin{equation}
    \chi_\beta(\tau) = \frac{1}{\pi}\int^{\infty}_0 d\omega S_\beta(\omega) F_\beta(\omega, \tau)
    \label{eq:chi-beta-overlap}
\end{equation}
is the dephasing overlap integral determined via time-dependent perturbation theory~\cite{Kofman2001universal,cywinski2008ff,biercuk2011dynamical,green_arbitrary_2013,Paz-Silva2014}. Based on the assessment of temporally-correlated noise on IBMQP devices [Sec.~\ref{subsec:corr-noise-assessment}], the noise PSD is well-approximated by a DC component with a white noise floor. Hence, one can approximate the decay component of the noise using $\chi_\beta(\tau)=\Gamma_\beta F_\beta(0, \tau)/\pi + S^{(u)}_{\beta} \tau$, where $S_\beta(\omega)=\Gamma_\beta \delta(\omega) + S^{(u)}_{\beta}$. $\Gamma_\beta$ denotes the variance for the DC component. Note that the channel presented in Eq.~(\ref{eq:dephasing-channel}) is a noise-averaged map evaluated by performing a statistical average over the classical noise process. We elaborate on the specifications for the FF and a derivation of the above channel in Appendices~\ref{app:sec:time-correlated} and \ref{app:sec:stochastic_channel}.

Similar to the dephasing case, we can define a control noise channel with static and dissipative contributions. Namely, $\mathcal{E}_{CN}(\rho)=U_{\epsilon}(t)\mathcal{E}_{\epsilon}(\rho) U^\dagger_{\epsilon}(t)$, where
\begin{equation}
    \mathcal{E}_{\epsilon}(\rho) =\left[1-p_\epsilon(\tau)\right] \rho + p_\epsilon(\tau) \sx \rho \sx
\end{equation}
defines dissipation with error probability $p_\epsilon(\tau)=1/2(1+e^{-\chi_\epsilon(\tau)})$; the decay parameter $\chi_\epsilon(\tau)$ is defined similar to Eq.~(\ref{eq:chi-beta-overlap}). The coherent error evolution operator $U_{\epsilon}(t)=e^{-i\pi\bar{\epsilon}/2\sx}$ accounts for amplitude error and therefore, over-rotations. Through our characterization experiments, we find that the control noise is predominately described by white noise with a DC component. Thus, equivalent to the dephasing channel, the error probability decay rate can be simplified as $\chi_\epsilon(\tau)=\Gamma_\epsilon F_\epsilon(0,\tau)/\pi + S^{(0)}_{\epsilon}\tau$. Additional information on the control noise channel can be found in Appendix~\ref{app:sec:stochastic_channel}.

In the case of TLS and crosstalk noise, the multi-qubit state $\rho$ is evolved under Hamiltonians $H_\mrm{XT}$ and $H_\mrm{TLS}$ defined in Eqs.~(\ref{eq:H_ZZ}) and (\ref{eq:H_TLS}), respectively. These Hamiltonians are time-independent, and consequently the system evolves via 
\begin{equation}
    \mathcal{E}_\mrm{ZZ}(\rho)=U_{ZZ}(\tau) \rho U^\dagger_{ZZ}(\tau),
\end{equation}
with $U_{ZZ}(\tau) = e^{-i(H_\mrm{XT}+H_\mrm{TLS})\tau}$. Since the Hamiltonians are proportional to $\sigma_z$ operators, they are diagonal in the $n$-qubit+TLSs computational basis, and thus the action of $U_{ZZ}$ on a density matrix can be found efficiently.

Lastly, we comment on the channel implementation of the ECR gate. This operation is performed via the control operation $\mathcal{E}_{CR}(\rho)=U_{CR}(\tau_{CR})\rho U_{CR}^\dagger(\tau_{CR})$, where $U_{CR}(t)=e^{-i H_{CR}t}$ with $H_{ZX}$ defined in Eq.~(\ref{eq:H_CR}). Noise contributions, beyond those inherent in the CR Hamiltonian enter via single-qubit GAD and dephasing error channels in accordance with the results of our characterization experiments. Additional sources of ZZ noise can be incorporated via $\mathcal{E}_{ZZ}$.  

In the above discussion, we present one variant of the error channels based on our LME model and characterization experiments. However, additional model reductions are also possible. For example, effective error maps can be derived by moving into the rotating frame with respect to the control operations and utilizing time-dependent perturbation theory to extract error channels. It is not known \emph{a priori} which approach is more relevant, as the efficacy of the model may be dependent upon the application~\cite{schwartzman2024modeling}. A thorough examination of various reductions of our LME models is beyond the scope of this study and left for future work.

\section{Summary and Conclusions}
\label{sec:conclusion}

In this work, we address the challenge of balancing model sparsity with predictive power. To this end, we propose a model for single and two-qubit gate operations on IBMQP fixed-frequency superconducting transmons. The model is a modified LME that includes both Markovian and non-Markovian contributions. The latter includes extending the model to include coupling between the system and classical or quantum environments. Using only a few additional degrees of freedom, the model is capable of capturing spatio-temporally correlated noise processes observed on hardware. The model is specified by at most 10 parameters per qubit and three parameters per qubit pair that can be learned via seven characterization experiments, each potentially comprised of multiple circuits depending on the desired accuracy.

Through demonstrations of the characterization protocol on the IBMQP, we develop effective noise models for 39 qubits across seven devices within the \textit{Falcon} and \textit{Eagle} processor generations. While we find that a significant subset of qubits are described by Markovian noise alone (64\%), approximately 26\% and 10\% are subject to correlated dephasing or control noise, respectively. Quantum crosstalk and coupling to TLSs are also commonly observed. 

The efficacy of the learned models is explored through characterization, error suppression, and quantum computing experiments. Single-qubit models are tested against RB demonstrations and shown to accurately predict gate error rates. Multi-qubit models are examined through simultaneous single-qubit DD and a VQE designed to find the ground state of molecular Hydrogen. The former exhibits strong agreement with state dependent fidelity decay observed in previous studies. The latter highlights the predictive power of our model particularly for qubits that display non-Markovian noise. We show that our model is capable of achieving relative errors that are $7\times$ better than Qiskit's default error model. 

Lastly, we present a composite channel approximation to the LME model that seeks to address model scalability. In addition to constructing error maps for Markovian processes, we develop effective error channels for non-Markovian processes, most notably temporally-correlated dephasing and control noise. Key to their design is the use of the FFF commonly used in control theory.

Although the noise model introduced here is targeted towards fixed-frequency transmons, the processes discussed are general and pervasive to all quantum devices. Thus, we expect this work to provide valuable insight in the development of different superconducting-based qubits as well as other architectures.  In addition, accurate modeling is essential not only for quantum computing but also for all quantum applications. For instance, the scalability of this noise characterization model allows for efficient optimization of multi-qubit sensor arrays in real-time noise monitoring, enhancing the performance of quantum magnetometers and QND-based photon detectors in dynamic environments.

Further examinations are required to determine if the model presented here continues to yield predictive power as system size increases. This is true for both the LME approach, which suggests that at most two-qubit error operators are required to predict hardware behavior, and the reduced model based on quantum channels. Nevertheless, this work emphasizes the viability of effective non-Markovian noise models for describing complex multi-qubit dynamics on superconducting qubits in a variety of applications. 

\acknowledgments
This work was supported in part by the U.S. Department of Energy (DOE), Office
of Science, Office of Advanced Scientific Computing Research (ASCR), the Accelerated Research in Quantum Computing program under Award Number DE-SC0020316. This research used
resources of the Oak Ridge Leadership Computing Facility,
which is a DOE Office of Science User Facility supported
under Contract DE-AC05-00OR22725.

\appendix

\section{Solutions to the LME in the Markovian Regime}
\label{app:LME_Mark_sol}

\subsection{Single-Qubits} 
\label{app:subsec:1Q}

In the Markovian limit, the dynamical evolution under the above mentioned processes can be studied directly using the LME. For simplicity, we suppress the qubit index $j$. We assume constant $x$-control with duration $\delta t$ that executes a rotation $\theta$, i.e., $\Omega(t)=\theta/\delta t$. Thus, the control Hamiltonian [Eq.~(\ref{eq:control-H})] during the implementation of a gate takes the time-independent form $H_C=\frac{\theta}{2\delta t}\sigma_x$. Furthermore, we assume the noise Hamiltonian [Eq.~\ref{eq:stoch-noise-Hamiltonian}] consists of detuning and coherent control errors, i.e., $\beta(t)=\beta$ and $\epsilon(t)=\epsilon$. Since SPAM errors act only at the end of the circuit, they are represented as a quantum map after the circuit's control sequence. Defining the noisy control rotation frequency $\omega = (1+\epsilon)\theta/\delta t$, the LME becomes
\eq{
\label{eq:LE_1Q}
\begin{split}
\dot{\rho}(t) &= \mathcal{L}(\rho, \mathcal{N}_1) \\
&= -i \frac{\omega}{2}[\sigma_x,\rho] - i \frac{\beta}{2} [\sigma_z,\rho] \\
&+ \frac{\lambda}{2} \l( \sigma_z \rho \sigma_z - \rho\r) + \frac{\nu_\theta}{2} \l( \sigma_x \rho \sigma_x - \rho\r)  \\
&+ \sum_{\pm} \gamma^\pm \l( \sigma^\pm\rho \sigma^\mp - \frac{1}{2}\{\sigma^\mp \sigma^\pm, \rho\} \r),
\end{split}
}
where $\mathcal{L}$ is the Lindbladian superoperator, and $\mathcal{N}_1$ the set of all single-qubit error parameters. Note that the bit-flip noise with rate $\nu$ only acts when a gate is implemented, i.e., $\nu_\theta=0$ if $\theta=0$, and $\nu_\theta=\nu$ otherwise.

In order to solve Eq.~(\ref{eq:LE_1Q}), we follow the steps of Ref.~\cite{lidar2020lecture}. Moving to the Bloch vector equation in Eq.~(\ref{eq:LE_v}), we find that $\vec{c} = (0,0,\gamma(2q-1))$, and the coupling matrix is
\eq{
\mathbf{G} = \begin{pmatrix}
-\l(\frac{\gamma}{2} + \lambda\r) & -\beta & 0 \\
\beta & -\l(\frac{\gamma}{2} + \lambda+\nu\r) & -\omega \\
0 & \omega & -(\gamma+\nu) \\
\end{pmatrix}.
}
In this form, Eq.~(\ref{eq:LE_v}) can be solved using standard coupled differential equations methods. This equation can be solved numerically for the identity gate and native $X$ rotations, and subsequently stored for simulation. Moreover, in the weak noise regime, Eq.~(\ref{eq:LE_sol}) can be written analytically (see next section).

An analytical solution of the LME can be a powerful way to provide insight into the effects of noise on the evolution of the system. However, computing the exponential $U(\tau)=e^{\mathbf{G}\tau}$ for a general $\mathbf{G}$ can be challenging. Furthermore, $\mathbf{G}$ may not be invertible. In this section, we provide analytical solutions to Eq.~(\ref{eq:LE_sol}) in the two cases of interest: identity gates and $X$ rotations. 

\subsection{Identity Gates}
\label{app:subsec:I}

In the case of identity operations, the LME can be solved exactly. Following the steps outlined in the previous section, we set $\theta=\nu=0$ since no control noise is present in the absence of qubit drive. Defining for notational convenience $\alpha=\gamma/2+\lambda$, we write the LME solution in Bloch vector form
\eq{
\label{eq:bloch_sol_I}
\vec{v}(t_0+\tau) &= \begin{pmatrix}
e^{-\alpha \tau} \cos(\beta\tau) &  -e^{-\alpha \tau} \sin(\beta\tau) & 0 \\
e^{-\alpha \tau} \sin(\beta\tau) & e^{-\alpha \tau} \cos(\beta\tau) & 0 \\
0 & 0 & e^{-\gamma \tau}
\end{pmatrix} \cdot \vec{v}(t_0) \nonumber
\\
&+ \frac{1-e^{-\gamma\tau}}{\gamma} \vec{c},
}
which holds in general for a free evolution of duration $\tau$. In the practical case of weak noise and short gate time regime $\gamma\tau\ll1$, it is possible to approximate the inhomogeneous term with $\frac{1-e^{-\gamma\tau}}{\gamma}\approx \tau$.

\subsection{Perturbative Solution for \texorpdfstring{$X$}{} control}
\label{app:subsec:1QX}

In this section, we provide a solution to the LME in the presence of $X$ control in the weak noise regime. We follow the perturbative approach of the FFF~\cite{green_arbitrary_2013}. For a non-zero rotation, such as those corresponding to $X,\sqrt{X}$ gates, the generator $\mathbf{G}$ can be written as a perturbation from the noiseless generator $\mathbf{G_0}$. Writing this explicitly as $\mathbf{G} = \mathbf{G_0} + \mathbf{g}$, we define
\eq{
\mathbf{G_0} = 
\begin{pmatrix}
0 &   & 0 \\
0 & 0 & -\omega \\
0 & \omega & 0 \\
\end{pmatrix}, \qquad
\mathbf{g} = 
\begin{pmatrix}
-\alpha & -\beta & 0 \\
\beta & -\mu & 0 \\
0 & 0 & -\eta \\
\end{pmatrix},
}
with $\mu=\alpha+\nu,\eta=\gamma+\nu$. In the following we will assume $||\mathbf{g}||\ll||\mathbf{G_0}||$, which holds as long as the noise is sufficiently weak, where $||\cdot||$ is the Schatten 1-norm. Distinguishing between $\mathbf{g}$ and $\mathbf{G_0}$ is further motivated by the fact that it is easy to compute the effect of an ideal rotation on the Bloch vector, i.e.,
\eq{
U_0(\tau)=e^{\mathbf{G_0}\tau} = 
\begin{pmatrix}
1 & 0 & 0 \\
0 & \cos(\omega \tau) & -\sin(\omega \tau) \\
0 & \sin(\omega \tau) & \cos(\omega \tau) \\
\end{pmatrix},
}
which enables the perturbative approach.

First, we take the problem to the toggling frame, i.e., perform a change of basis with respect to $\mathbf{G_0}$, and we write $U(\tau)=U_0(\tau) \tilde{U}(\tau)$. Here, $\tilde{U}(\tau)$ can be thought of as analogous to the noise propagator in the toggling frame, and is the solution to the differential equation $\frac{d}{dt}\tilde{U}(t) = \mathbf{\tilde{g}}(t)\tilde{U}(t)$, where $\mathbf{\tilde{g}}(t)=U_0^\dagger(t) \mathbf{g} U_0(t)$. Note that both $U(\tau)$, $U_0(\tau)$ can be written as solutions to the same differential equation, with generators $\mathbf{G}$ and $\mathbf{G_0}$, respectively. Using the first order Magnus expansion~\cite{Moan_1999_magnus}, we can write $\Tilde{U}(\tau) \approx e^{\Phi(\tau)}$, where
\eq{
\Phi(\tau) &=\int_0^\tau U_0^\dagger(t) \mathbf{g} U_0(t) dt
}
is first order in the noise parameters. 

Next, the inverse of $\mathbf{G}$ needs to be computed. In order to ensure that $\mathbf{G}$ is invertible we require that $\alpha\neq0$ and $\omega\neq0$. The latter is satisfied for both $\pi$ and $\pi/2$ rotations of interest, as long as $|\epsilon|\ll1$, whereas the former will hold as long as $T_1,T_2$ are finite. Then, keeping the first two orders in the noise parameters,
\eq{
\mathbf{G}^{-1} = \frac{-1}{\alpha \omega^2}
\begin{pmatrix}
\omega^2 & -\beta\eta & \beta\omega \\
\beta\eta & \alpha\eta & -\alpha\omega \\
\beta\omega &\alpha\omega & \alpha\mu+\beta^2 \\
\end{pmatrix}.
}
Lastly, a first order approximation can be taken where $\tilde{U}(\tau) \approx I + \Phi(\tau)$, which yields $e^{\mathbf{G}\tau} \approx e^{\mathbf{G_0}\tau}(I + \Phi(\tau))$. With $e^{\mathbf{G}\tau}$ and $\mathbf{G}^{-1}$ calculated explicitly, we can find an analytical expression for Eq.~(\ref{eq:LE_sol}). This is, by approximating $e^{\mathbf{G}\tau}\approx U_0(\tau) \l(I+\Phi(\tau)\r)$, we obtain
\eq{
\vec{v}(t+\tau) &= e^{\mathbf{G}\tau}\cdot \vec{v}(t) + \left(e^{\mathbf{G}\tau} - I\right)\cdot \mathbf{G}^{-1}\cdot \vec{c} \nonumber\\
&\approx \mathbf{L}(\tau) \cdot
\vec{v}(t) + \vec{u}(\tau)
}
where, defining $\cosc(x)=\frac{1-\cos(x)}{x}$ for notational convenience, we have
\eq{
\vec{u}(\tau) &= \gamma\tau(2q-1)\begin{pmatrix}
0 \\
\cosc(\omega\tau) \\
-\sinc(\omega \tau)
\end{pmatrix},
}
and
\begin{widetext}
\eq{
\label{eq:LME_X_L}
\mathbf{L}(\tau) &= \begin{pmatrix}
e^{-\alpha \tau} & 
-\beta\tau \sinc(\omega \tau) & 
\beta \tau \cosc(\omega \tau) \\
\beta\tau \sinc(\omega \tau) & 
e^{-\frac{(\mu+\eta)\tau}{2}} \cos(\omega \tau) -\frac{(\mu-\eta)\tau}{2}\sinc(\omega\tau)& 
e^{-\frac{(\mu+\eta)\tau}{2}} \sin(\omega \tau) \\
\beta \tau \cosc(\omega \tau) & 
e^{-\frac{(\mu+\eta)\tau}{2}} \sin(\omega \tau) & 
e^{-\frac{(\mu+\eta)\tau}{2}} \cos(\omega \tau) + \frac{(\mu-\eta)\tau}{2}\sinc(\omega\tau)
\end{pmatrix}.
}
\end{widetext}

This analytical expression is useful to perform single-qubit simulations, and examine the effect of noise when the qubit is being driven.

\subsection{Prediction of Characterization Circuits}
\label{app:subsec:exp_preds}

The results from the above sections can be used to compute predictions for the characterization experiments. Here, we show how it can be used to compute the results shown in Eqs.~(\ref{eq:v_T1}$-$\ref{eq:v_P}). Denoting by $\vec{v}(\tau)$ the Bloch vector state at the end of a given circuit, the effect of SPAM errors on the Bloch vector is to take 
\seq{
\vec{v}(\tau) =
\begin{pmatrix}
    v_x(\tau)\\
    v_y(\tau)\\
    v_z(\tau)
\end{pmatrix}
\xrightarrow[]{\mathcal{E}_M(\cdot)}
\begin{pmatrix}
v_x(\tau)\\
v_y(\tau)(1-2s)\\
v_z(\tau)(1-2s))
\end{pmatrix}.
}
For the corresponding density matrix $\rho(\tau)=(I +\vec{v}(\tau)\cdot \vec{\sigma})/2$, the survival probability of the $\ket{0}$ state is computed in terms of the Bloch vector as 
\seq{
\bra{0}\rho(\tau)\ket{0} &= \frac{1}{2} \bra{0} I +\vec{v}(\tau)\cdot \vec{\sigma}\ket{0} \\
&= \frac{1}{2} \l( 1 + \sum_{j=x,y,z} v_j(\tau) \bra{0} \sigma_j\ket{0} \r)\\
&= \frac{1}{2} \l( 1 + v_z(\tau) \r).
}
Consequently, the Bloch vector component of interest is $v_z(\tau)$. Combining these two results, we can see that the survival probability becomes
\eq{
\bra{0}\rho(\tau)\ket{0} = \frac{1}{2} \l( 1 + (1-2s) v_z(\tau) \r).
}
Throughout this section, we focus on calculating the $z$ component of the Bloch vector at the end of the circuit for the characterization experiments, of duration $\tau$. Moreover, we remove the subscript and denote $v_z(\tau)$ by $v(\tau)$, for convenience of notation.

\subsubsection{(M) SPAM}

The SPAM experiment consists of a single $X$ gate of duration $\delta t$. Assuming weak noise, particularly $s, \gamma \delta t\ll 1$, it is straightforward to see that before the measurement:
\begin{itemize}
    \item Coherent errors contribute quadratically, i.e., $v(\delta t)=-1+O(\epsilon^2,(\beta \delta t)^2)$;
    \item Dissipative effects contribute linearly, i.e., $v(\delta t)=-1+\delta t \, O(\gamma,\lambda,\nu)$.
\end{itemize}
Then we apply the measurement error map $\mathcal{E}_M$ with rate $s$. To first order in the noise parameters,
\eq{
v(\delta t) = s + \delta t\,O(\gamma,\lambda,\nu).
}
Finally, we make the experimentally verifiable assumption that the dissipative effects are weak compared to the duration of a single qubit gate, i.e., $\delta t \ll \frac{1}{\gamma},\frac{1}{\lambda},\frac{1}{\nu}$. Thus, we obtain $v(\delta t) \approx s$.

\subsubsection{(\texorpdfstring{$T_1$}{}) Thermal Relaxation}

The $T_1$ experiment result is easy to compute following a similar analysis. The experiment starts with a single $X$ gate, which, as was discussed in the SPAM experiment case, only contributes error terms from dissipative effects, i.e., $v(\delta t)=-1+\delta t \, O(\gamma,\lambda,\nu)$. Since there are no more gates applied until the end of the experiment, and the state will be primarily along the $\ket{1}$ direction, neither $\lambda$ nor $\nu$ will accumulate significantly. Consequently, we focus on the effect of GAD. Since the experiment consists of successive applications of $I$ gates, we can use Eq.~(\ref{eq:bloch_sol_I}), with $t_0=\delta t$ and $\vec{v}(\delta t)\approx (0,0,-1)$, where the approximation holds up to terms proportional to $\gamma\delta t,\lambda\delta t,\nu\delta t\ll1$. Hence, the $z$ component of the Bloch vector at time $\tau\gg\delta t$ becomes
\eq{
v(\tau) &= -e^{-\gamma \tau} + (1-e^{-\gamma \tau}) (2q-1) \nonumber\\
&= -1 + 2q(1-e^{-\gamma\tau}),
}
up to a term $\delta t \, O(\gamma,\lambda,\nu)$. Note that a result of the above equation is that, after applying the final $X$ measurement gate, the asymptotic probability of finding the state in the $\ket{0}$ state is $p(\tau\rightarrow\infty) = 1-q + \delta t \, O(\gamma,\lambda,\nu)$.

\subsubsection{(\texorpdfstring{$T_2$}{}) Hahn-Echo}

The $T_2$ experiment state can be computed in a similar fashion to that of $T_1$. Ignoring constant order effects that do not scale with $\tau$, we prepare the state on the plane with a $\sqrt{X}$ gate, thus taking $\vec{v}(0)=(0,0,1)$ to $\vec{v}(\delta t)\approx (0,-1,0)$. Evolving with Eq.~(\ref{eq:bloch_sol_I}) for a time $\tau/2$ yields $\vec{v}((\tau/2)^-)\approx e^{-\alpha \tau/2}(\sin(\beta\tau/2),-\cos(\beta\tau/2),\psi_1)$, where we defined $\psi_1 = e^{\alpha\tau/2}(1-e^{-\gamma\tau/2}) (2q-1)$. Next, an $X$ gate is applied with the effect of changing signs of the $y,z$ components, i.e., $\vec{v}((\tau/2)^+)\approx e^{-\alpha \tau/2}(\sin(\beta\tau/2),\cos(\beta\tau/2),-\psi_1)$. The superscripts $()^\pm$ indicate whether the vector is evaluated before or after the echo pulse. Following the echo, the qubit is left to evolve freely for a time $\tau/2$. Once again drawing on Eq.~(\ref{eq:bloch_sol_I}), we find $\vec{v}(\tau)\approx (0,e^{-\alpha \tau},\psi_2)$, where the specifics of $\psi_2$ are irrelevant for the final result. Once the final $\sqrt{X}^\dagger$ is applied, the $y$ and $z$ components of the Bloch vector are exchanged, yielding 
\eq{
v(\tau) = e^{-\alpha \tau},
}
up to terms proportional to $\delta t\,O(\gamma,\lambda,\nu)$.

\subsubsection{(R) Ramsey Experiments}

The multi-frequency oscillations observed in Ramsey experiments can be readily modeled by coupling the system qubits to a TLS initialized in the $\ket{+}$ state. The single-qubit system is enlarged to a qubit+TLS system interacting via a $ZZ$ coupling, which in the absence of control can be represented by the Hamiltonian 
\begin{equation}
    H = \frac{1}{2}(\beta Z_Q I_\mrm{TLS} + \xi Z_Q Z_\mrm{TLS}).
\end{equation}
Since this Hamiltonian is diagonal, and momentarily ignoring dissipative processes, the evolution is dominated by a diagonal unitary operator 
\begin{eqnarray}
    U(t) &=& e^{-iH t}\nonumber \\
        &=&\mrm{diag}(e^{-i\frac{t}{2}(\beta+\xi)},e^{-i\frac{t}{2}(\beta-\xi)},e^{i\frac{t}{2}(\beta+\xi)},e^{i\frac{t}{2}(\beta-\xi)}).\quad\quad\quad
\end{eqnarray}
Upon executing the Ramsey protocol, the joint state is initialized in $\ket{\Psi(0)}=\ket{0}_Q\otimes \ket{+}_\mrm{TLS}$, followed by a $\sqrt{X}$ gate on the qubit. Applying the operator $U(\tau)$ and tracing over the TLS degrees of freedom, the reduced density matrix is given by
\eq{
\rho(\tau) = \frac{1}{2} \begin{pmatrix}
    1 & i e^{-i\beta \tau} \cos(\xi \tau) \\
    -i e^{i\beta \tau} \cos(\xi \tau) & 1
\end{pmatrix}
}
Upon applying the final gate $\sqrt{X}^\dagger$, we obtain a Bloch vector $z$ component of $v(\tau)= \cos(\beta\tau)\cos(\xi\tau)$, which shows the clear multi-frequency dependence. Lastly, since the evolution is mostly free except state preparation and measurement operations, the exponential decay rate is given by $\alpha$ as in the $T_2$ experiment. This is confirmed numerically, as well as via solving the LME analytically. As before, all other errors, including gate errors, will contribute factors of order $\delta t$, and are consequently omitted. 

\subsubsection{(Q) FTTPS}

FTTPS are a set of probe sequences used in QNS to access frequency dependent decay rates of a single-qubit system~\cite{Murphy2022}. Following the protocol described in Sec.~\ref{app:subsec:QNS}, they can be used to learn the spectral properties of correlated dephasing noise. However, to obtain accurate characterization of correlated noise, it is essential to have a thorough understanding of the effects Markovian noise processes have on these experiments.

The $k=0$ sequence is a Ramsey experiment, and it thus follows the discussion in the previous section. For $k>0$, it is easy to first show the results of each type of noise independently. Each Markovian dissipative error contributes an exponential decay of the form $e^{-\delta_k \tau}$ to the Bloch vector. The decays obtained from each source of dissipative noise acting individually are listed below:
\begin{itemize}
    \item Amplitude Damping: $\delta_k^{AD}=\frac{\gamma}{2}\l(1+\frac{k}{2 K}\r)$,
    \item Phase Damping: $\delta_k^{PD}=\lambda\l(1-\frac{k}{2 K}\r)$,
    \item Control Noise: $\delta_k^{X}=\nu \frac{k}{K}$.
\end{itemize}
Note that $\delta_k$ is independent of $q$ in the AD case. 

\begin{figure}[t]
    \centering\includegraphics[width=\columnwidth, trim={0cm .5cm 0cm 0cm},clip]{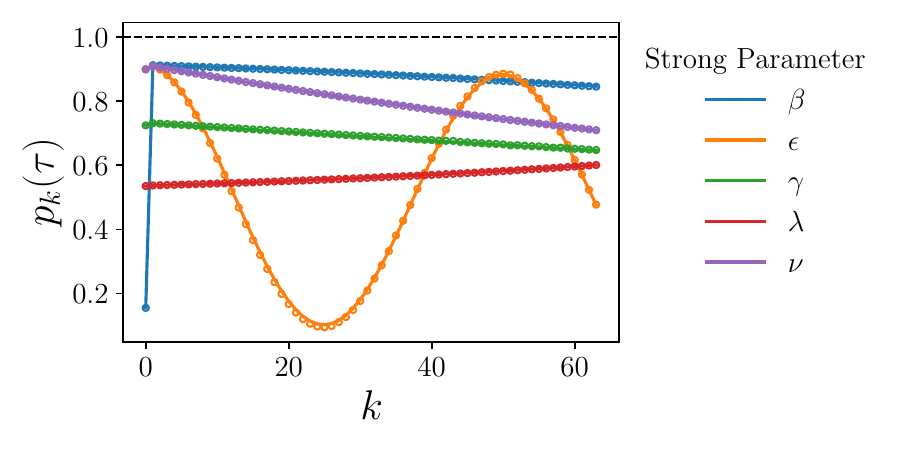}
    \caption{Comparison between FTTPS simulations (dots) and analytical prediction (solid lines) obtained from Eq.~(\ref{eq:v_Q_k}). Each curve corresponds to a different set of noise parameres. The Markovian noise parameters $q=1$ and $\beta,\epsilon,\lambda,\gamma,\nu$ were obtained from Table~\ref{table:noise_params} by selecting one process to be strong while maintaining the others as weak. The FTTPS were chosen with $K=64$. In all cases, we find excellent agreement between theory and simulation. Dashed lines represent the ideal probabilities, obtained in the noiseless scenario.}
    \label{fig:FTTPS_markovian}
\end{figure}

Additionally, it is straightforward to see that the effect of an under/over-rotation angle contributes a cosine to the Bloch vector. The Hamiltonian of a single gate is $H_C=\frac{\pi}{2\delta t}(1+\epsilon) \sigma_x$, and the associated control propagator is found to be 
\begin{equation}
    U_C = \exp(-i H_C \delta t) = X e^{-i \frac{\pi}{2}\epsilon\sigma_x}.
    \label{eq:ctrl-prop}
\end{equation} 
Since the $k^\mrm{th}$ FTTPS has $2k$ number of $X$ gates, and we assume that no other noise is acting, the total propagator can be computed as $U_k=(U_c)^{2k}=X^{2k} e^{-i \pi k \epsilon\sigma_x} = e^{-i \pi k \epsilon\sigma_x}$. Thus, the survival probability of the $\ket{0}$ state is $p_k(\tau) = |\langle 0| U_k |0\rangle|^2 = (1+\cos(2\pi\epsilon k))/2$. Combining these results, we can see that the analytical prediction of the Bloch vector becomes,
\eq{
\label{eq:v_Q_k}
v^{Q}_{k}(\tau) &= e^{-\tau \delta_k}
\begin{rcases}
\begin{dcases*}
\cos(2 \beta \tau) &, if $k=0$\\
\cos(2 \pi k \epsilon) &, if $k>0$
\end{dcases*}
\end{rcases},
}
where FTTPS decay rate is $\delta_k= \frac{\gamma}{2}+\lambda +(\frac{\gamma}{2}-\lambda+2 \nu)\frac{k}{2K}$. However, the simultaneous action of all noise processes may not combine trivially. To test the validity of Eq.~(\ref{eq:v_Q_k}), we compare it to numerical simulation. In Fig.~\ref{fig:FTTPS_markovian}, we present the results of performing LME simulations with various combinations of Markovian noise parameters $(\beta,\epsilon,\lambda,\gamma,q=1,\nu)$. The specific values of the noise parameters are shown in Table~\ref{table:noise_params}. In these simulations, all noise parameters except one were set by default to the ``weak'' values shown on the left column of Table~\ref{table:noise_params}. Then, for each curve of Fig.~\ref{fig:FTTPS_markovian}, one parameter at a time was set to the values corresponding to ``strong'' noise, shown on the right column of Table~\ref{table:noise_params}. In all cases, we find excellent agreement between Eq.~(\ref{eq:v_Q_k}) and the numerical results. This provides strong support for the use of Eq.~(\ref{eq:v_Q_k}) to compute predictions when Markovian noise is dominant. 

\begin{table}[t]
\centering
\begin{tabular}{c || c | c} 
\hline\hline
 Parameter & Weak & Strong \\ 
 \hline
 $\beta$ (MHz) & 0.06 & 0.6\\[.5ex]  
 $\epsilon$ (\%) & 0.1 & 2 \\[.5ex] 
 $\gamma$ (MHz) & 0.03 & 0.3 \\[.5ex] 
 $\lambda$ (MHz) & 0.03 & 0.6 \\[.5ex] 
 $\nu$ (MHz) & 0.03 & 0.15 \\[.5ex] 
 \hline\hline
\end{tabular}
\caption{Noise parameters used in FTTPS and RB simulation.}
\label{table:noise_params}
\end{table}

\subsubsection{(P) FPW}
Here, we examine the lowest order noise contribution to the FPW experiments under the square pulse approximation. The ideal FPW experiment is built with $d$ repetitions of the unitary building block $G=XZXZ$, i.e., the circuit unitary is $U_{FPW}(\tau=2d\delta t)=G^d$. The $x$ control Hamiltonian and propagator follow from Eq.~(\ref{eq:ctrl-prop}), while the noise Hamiltonian $H_N = \frac{\beta}{2}\sigma_z$ accounts for detuning. The effect of the noise on the ideal $X$ operator can be understood intuitively in the Hamiltonian formulation. We transform $H_N$ to the interaction picture with respect to $H_C$, known as the toggling frame:
\begin{eqnarray}
\tilde{H}_N(t) &=& U_C^\dagger(t) H_N U_C(t) \nonumber\\
&=& \frac{\beta}{2} \l(\sigma_z \cos\l(\theta \tfrac{t}{\delta t}\r) + \sigma_y \sin\l(\theta \tfrac{t}{\delta t}\r)\r).
\end{eqnarray}
As a result, the rotated frame noise propagator is
\begin{eqnarray}
\tilde{U}(\tau) &=& \mathcal{T}_+ e^{-i\int_0^\tau dt \tilde{H}_N(t) } \nonumber\\
&\stackrel{(1)}{\approx}& e^{-i\frac{\beta}{2}\int_0^\tau dt \l(\sigma_z \cos\l(\theta \frac{t}{\delta t}\r) + \sigma_y \sin\l(\theta \frac{t}{\delta t}\r)\r) },
\end{eqnarray}
where $\mathcal{T}_{+}$ is the time ordering operator and in (1) we have approximated the time-ordered dynamics by the first order Magnus expansion. More specifically, for a single $X$ gate, where $\tau=\delta t$, we obtain
\seq{
\tilde{U}(\delta t) &\approx e^{-i\frac{\beta \delta t}{2 \theta}\l(\sigma_z \sin\theta + \sigma_y \l(1 - \cos\theta\r)\r) }\\
&\approx e^{-i\frac{\beta \delta t}{\pi} \sigma_y},
}
where we used the weak noise conditions $\epsilon\ll1$ and $\beta\delta t\ll 1$ in approximating $\sin\theta\approx-\pi\epsilon$, $1-\cos\theta\approx 2-\frac{(\pi\epsilon)^2}{2}$ and $\frac{\beta}{\theta}\approx\frac{\beta}{\pi}$ to first order.

In the lab frame, the noisy $X$ gate can be written in terms of the control and toggling frame propagators as
\begin{eqnarray}
X' &=& U_C(\delta t) \tilde{U}(\delta t) \nonumber\\
&\approx& X e^{-i\frac{\pi \epsilon}{2}\sigma_x} e^{-i\frac{\beta \delta t}{\pi} \sigma_y}.
\end{eqnarray}
Consequently, the noisy implementation of the building block $G$ becomes
\seq{
G' &= X' Z X' Z \\
&\approx X e^{-i\frac{\pi \epsilon}{2}\sigma_x} e^{-i\frac{\beta \delta t}{\pi} \sigma_y} Z X e^{-i\frac{\pi \epsilon}{2}\sigma_x} e^{-i\frac{\beta \delta t}{\pi} \sigma_y} Z \\
&= e^{-i\frac{\pi \epsilon}{2}\sigma_x} X e^{-i\frac{\beta \delta t}{\pi} \sigma_y} X Z e^{-i\frac{\pi \epsilon}{2}\sigma_x} Z Z e^{-i\frac{\beta \delta t}{\pi} \sigma_y} Z \\
&= e^{-i\frac{\pi \epsilon}{2}\sigma_x} e^{i\frac{\beta \delta t}{\pi} \sigma_y} e^{i\frac{\pi \epsilon}{2}\sigma_x} e^{i\frac{\beta \delta t}{\pi} \sigma_y}\\
&\approx e^{i\frac{\beta \delta t}{\pi} \sigma_y} e^{i\frac{\beta \delta t}{\pi} \sigma_y} \\
&\approx e^{2 i\frac{\beta \delta t}{\pi} \sigma_y},
}
where we used that $X$ and $Z$ anti-commute, as well as $e^{-i\frac{\pi \epsilon}{2}\sigma_x}$ and $e^{-i\frac{\beta \delta t}{\pi} \sigma_y}$ commute to first order; operator equalities hold up to a global phase. Additionally, we used $\sigma_j e^{-i \phi \sigma_k} \sigma_j =  e^{-i \phi \sigma_j\sigma_k\sigma_j} =  e^{i \phi \sigma_k}$ for $k\neq j$. This result indicates that the interleaved $Z$ gates cancel the $X$ over-rotation errors given by $\epsilon$, but maintain the first-order detuning errors, thus causing FPW errors.

Finally, the full intended operation consists of $d$ repetitions of the $G'$ operator, yielding
\eq{
(G')^d \approx e^{i 2 d\frac{\beta \delta t}{\pi} \sigma_y}.
}
The survival probability for $\tau=2d\delta t$ can then be computed as 
\seq{
p(\tau) &= \l|\bra{0} (G')^d \ket{0} \r|^2 \\
&\approx \cos\l(\frac{2\beta}{\pi} \tau \r)^2,
}
from where the functional form of Eq.~(\ref{eq:v_P}) becomes clear. However, the numerical factor of the oscillation frequency of $2\beta/\pi$ depends on the specific shape of the pulse. Interestingly, the difference between the square pulse and the Gaussian pulse is a factor of $2/3$, as discussed in Sec.~\ref{sec:characterization_experiments}. Lastly, the decay rate can be found by successive implementation of Eq.~(\ref{eq:LME_X_L}), and is given by $(\mu+\eta)/2 = 3/4\gamma + \lambda/2 +\nu$.

\subsubsection{(XT) Crosstalk}

The (XT) circuits can be computed similarly to the Ramsey experiments earlier in this section. In fact, the calculation is analogous, by renaming qubit $Q$ to $A$ and promoting the TLS to qubit $B$. The latter involves considering a detuning $\beta_B$ for qubit $B$, as well as dissipative error contributions. Thus, the crosstalk Hamiltonian becomes,
\eq{
H_\mrm{XT} = \frac{\beta_A}{2} \sigma_z^A + \frac{\beta_B}{2} \sigma_z^B + \frac{J}{2} \sigma_z^A \sigma_z^B.
}
Like in the TLS case, $H_\mrm{XT}$ is diagonal, and so is the time propagator it induces. More specifically, $U_\mrm{XT}(t) = e^{-iH_\mrm{XT} t}$, where the diagonal entries correspond to
$e^{-i\frac{t}{2} (\beta_A+\beta_B+J)}$,
$e^{-i\frac{t}{2} (\beta_A-\beta_B-J)}$,
$e^{-i\frac{t}{2} (-\beta_A+\beta_B-J)}$,
$e^{-i\frac{t}{2} (-\beta_A-\beta_B+J)}$.
Both qubits are initialized with $\sqrt{X}$ gates, allowed to evolve freely for a time $\tau/2$ after which an $X$ gate is applied to both qubits. Then, the qubits evolve freely once again, before two final $\sqrt{X}$ gates are applied prior to measurement in the computational basis. The resulting $z$ component of the Bloch sphere after tracing out qubit $B$ is given by
\eq{
v^{XT}(\tau) = e^{-\l(\frac{\gamma_A+\gamma_B}{2}+\lambda_A\r) \tau} \left(\cos(J\tau)+\frac{\gamma_B}{2J}\sin(J\tau) \right)
}
to lowest order in the dissipative noise parameters.

\section{Quantum Noise Spectroscopy}
\label{app:sec:time-correlated}

Time-correlated noise is commonly observed in superconducting qubits~\cite{Gulacsi2023,vonLupke2020,Schlor2019,Zhang2022,Zhou2023, white_demonstration_2020}. Its presence is commonly detected via super-exponential decay in $T_2$ experiments and more precisely characterized via QNS~\cite{Paz-Silva2014,Paz-Silva2017}. In the case of the IBMQP devices, we find a substantial number of devices where qubits exhibit time-correlated dephasing and control noise. Below, we discuss the Filter Function Formalism, a mathematical framework used to investigate the effect of correlated noise, and QNS based on FTTPS. We briefly summarize FTTPS-based QNS in the presence of dephasing and multiplicative control noise.

\subsection{Filter Function Formalism}
\label{app:subsec:FFF}

The FFF takes a frequency domain perspective on the effect of spatio-correlated noise on a quantum system. Here, we provide an overview for the FFF focusing on a single qubit governed by the noise Hamiltonian $H_{N,1}(t)$ [Eq.~(\ref{eq:stoch-noise-Hamiltonian})] and control Hamiltonian $H_C(t)$ given by Eq.~(\ref{eq:control-H}), where $\Theta(t)=\int_0^t \Omega(s)\,ds$. In the toggling frame, the density matrix is transformed by $\tilde{\rho}(t)=U_C(t) \rho(t) U_C^\dagger(t)$, and the noise Hamiltonian becomes 
\eq{
\tilde{H}_N(t) &=U_C(t) H_{N}(t) U_C^\dagger(t)\nonumber\\
 &= \l[\cos\Theta(t) \sz + \sin\Theta(t)\sy\r]\beta(t) + \Omega(t)\epsilon(t)\frac{\sx}{2}.\nonumber
}
The time propagator of the noise dynamics in the toggling frame is given by 
\begin{eqnarray}
\tilde{U}_N(t) &=&\mathcal{T}_+\exp\l(\int_0^tds \tH_N(s)\r) \nonumber\\
&\approx& \exp\Big(-i \vec{a}(\tau)\cdot\vsigma\Big),
\end{eqnarray}
where the approximation is a result of employing the Magnus expansion, truncating the dynamics to first order, and introducing the so-called error vector $\vec{a}(t)=\int_0^t ds \l[\frac{1}{2}\Omega(s)\epsilon(s),\sin\Theta(s)\beta(s),\cos\Theta(s)\beta(s)\r]$~\cite{green_arbitrary_2013,norris2018slepqns}. The full dynamics are generated by $U(t)=U_C(t) \tilde{U}_N(t)$.

In anticipation of the dynamics generated during an FTTPS circuit, we will focus on the noise-averaged survival probability $p(\tau)=\braket{\braket{-y|U(\tau)|-y}}_{\epsilon, \beta}$, where $\ket{-y}=1/\sqrt{2}(\ket{0}-i\ket{1})$ and $\tau$ is the total time of one sequence within the FTTPS protocol. Accounting for the fact that the sequences generate identity evolution, the probability can be simplified further as 
\begin{eqnarray}
    p(\tau) &=& \braket{\cos^2 a(\tau) + (a^2_y(\tau)/a(\tau)) \sin^2 a(\tau)}_{\epsilon,\beta} \nonumber\\
    &=& \frac{1}{2} \left(1 + \left\langle\frac{a^2_y(\tau)+\left[a^2_x(\tau) + a^2_z(\tau)\right]\cos(2a^2(\tau))}{a^2(\tau)}\right\rangle_{\epsilon,\beta}\right) \nonumber\\
    &\stackrel{(1)}{=}& \frac{1}{2}\left(1 + e^{-\chi(\tau)} \cos\zeta(\tau)\right)
    \label{eq:py-error-vector}
\end{eqnarray}
where $a(\tau)=|\vec{a}(\tau)|$ and the $\braket{\cdot}_{\epsilon,\beta}$ represents the average over noise realizations. The expression in (1) results from assuming the pulses are instantaneous and thus, $a_y(\tau)=0$. The dynamics are characterized by a decay factor $\chi(\tau)\equiv \braket{a^2(\tau)}_{\epsilon,\beta}$ and rotation angle $\zeta(\tau)\equiv2\braket{a(\tau)}_{\epsilon,\beta}$.

Commonly, these expressions are transformed to the frequency domain, as this can provide greater intuition that draws on classical signal processing concepts. Expressing the error vector components in the frequency domain, we obtain
\seq{
\big\langle a_x^2(\tau)\big\rangle_{\epsilon,\beta} &= \frac{1}{\pi}\int_0^\infty F_\epsilon(\omega,\tau) S_\epsilon(\omega)d\omega,\\
\big\langle a_z^2(\tau) \big\rangle_{\epsilon,\beta} &= \frac{1}{\pi}\int_0^\infty F_\beta(\omega,\tau) S_\beta(\omega)d\omega,
}
where the control and dephasing filter functions are defined by
\seq{
F_\epsilon(\omega,\tau) &= \frac{1}{4}\l| \int_0^\tau \Omega(t) e^{i\omega t}\r|^2, \\
F_\beta(\omega,\tau) &= \l| \int_0^\tau \cos\Theta(t) e^{i\omega t}\r|^2 .
}
The PSDs are defined as given in Sec.~\ref{sec:noise-model-stoch} in the main text.

\subsection{Dephasing QNS with FTTPS}
\label{app:subsec:QNS}

In order to extract information about a device's dephasing noise PSD, we leverage the FTTPS. The advantage of using FTTPS for QNS lies in the large spectral concentration of their FFs~\cite{Murphy2022}. This feature results in a favorable condition number in the FF matrix and thus, reduces the chance of encountering an ill-posed inversion problem in the spectrum reconstruction procedure. Here, we will study FTTPS in the absence of control noise.

\begin{figure}[t]
    \centering
    \includegraphics[width=\columnwidth]{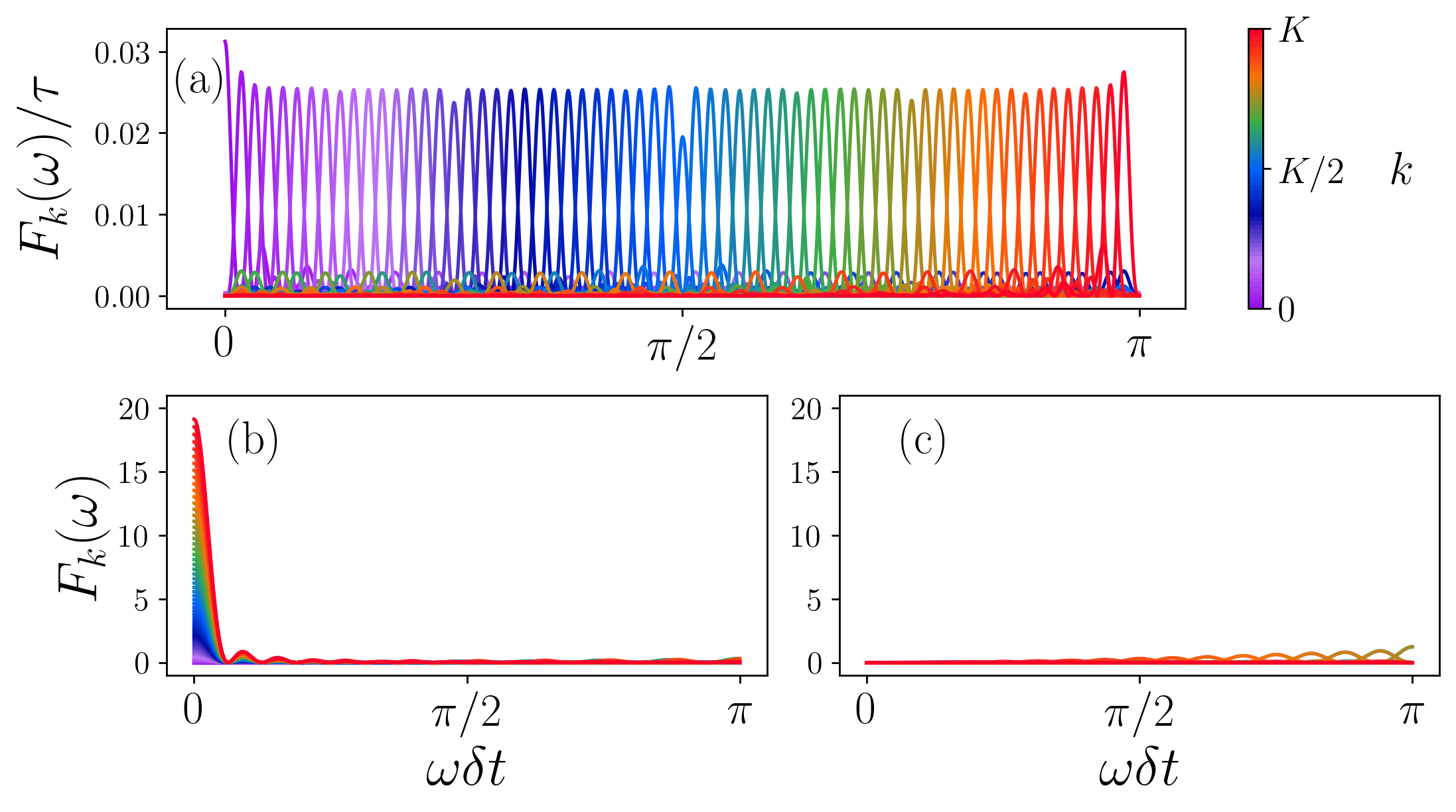}
    \caption{FTTPS filter functions: (a) dephasing and (b) control. (c) R-FTTPS control filter functions. $K=64, \delta t=0.035 \mu s$. Note that the FTTPS control FFs have large support in low-frequencies, whereas the R-FTTPS control FFs do not.}
    \label{fig:FFs}
\end{figure}

Starting from Eq.~(\ref{eq:py-error-vector}) and eliminating the contribution from control noise, $\zeta(\tau)=2\bar{\beta}\tau$ while 
\eq{
\label{eq:chi}
\chi(\tau) = \frac{1}{\pi} \int_0^\infty  S_\beta(\omega) F_\beta(\omega,\tau) d\omega.
}
In the large $\tau$ limit, the FF of the $k^\mrm{th}$ FTTPS $F_{\beta,k}(\omega,\tau)$ is well approximated by 
\begin{equation}
    F_{\beta,k}(\omega,\tau) \approx \tau^2 \delta(\omega - 2\pi k/\tau),
\end{equation}
where $\omega_k=2\pi k/\tau$ [see Fig.~\ref{fig:FFs}(a)]. The total FTTPS circuit duration is $\tau=2K\delta t$. Using Eq.~(\ref{eq:chi}), it is straightforward to estimate the PSD by computing $\chi(\tau)\approx \tau S(\omega_k)$, where to approximate the integral we used explicitly the discretized frequency step $\delta \omega=2\pi/\tau$. Note that estimates of $\chi$ are obtained from experimentally determined $p_{y,k}(\tau)$.

\subsection{FTTPS in the Presence of Stochastic Control Noise}
\label{app:subsec:corr_ctrl_noise}

Another type of correlated noise was observed via FTTPS: correlated control noise. Imperfect control can manifest via fluctuating fields in the control lines, and can be stochastic or coherent. In the case of stochastic noise, a common approach is to treat the noise fluctuations as white, uncorrelated noise. In the case of $X$ control, white noise can be well represented by a bit-flip channel (see Appendix~\ref{app:sec:stochastic_channel}). In the general case, stochastic fluctuations of the control fields can be correlated in time, requiring a more sophisticated modeling approach. In this section, we focus on evaluating the effect of stochastic control noise, specifically on the FTTPS. 

Consider the survival probability in Eq.~(\ref{eq:py-error-vector}) under the influence of strong control noise, such that the dephasing noise can be neglected. The probability oscillates with an angle $\zeta(\tau)=\bar{\epsilon}\Theta(\tau)$, while the decay is dictated by $\chi(\tau)=\big\langle a_x^2(\tau)\big\rangle_{\epsilon,\beta}$. Under the square pulse approximation and assuming the pulses are located at times $t=\tau_n \approx \frac{\tau}{2k}(n-1/2)$, mean of the accumulated angular error is $\zeta_k = \pi k \overline{\epsilon}$. The control filter functions (ctrl-FFs) are defined as 
\begin{equation}
    F_k(\omega)=\left| \frac{\pi}{2} \sum_{n=1}^{2k} e^{-i\omega\tau_k}\right|^2\approx \l( \frac{\pi}{2} \frac{\sin(\omega\tau/2)}{\sin(\omega\tau k/4)}\r)^2
\end{equation}
where the approximation comes from assuming the instantaneous pulse limit.   Roughly approximating the ctrl-FFs by $F_k(\omega)=4k^2 \delta(\omega)$ (see Fig.~\ref{fig:FFs}(b)), we obtain an expression for the second moment $\chi_k(\tau)\approx \sigma k^2$, with $\sigma = \pi^2 \delta \omega S_\epsilon(0)$. Thus, for predominantly low-frequency noise, $S_\epsilon(0)$ captures the relevant noise strength.

\subsubsection{R-FTTPS}
\label{app:subsec:RFTTPS}

To obtain further confirmation of the presence of correlated control noise, we define the Robust-FTTPS (R-FTTPS) sequences, by alternating the sign of every other $X$ pulse in the standard FTTPS, i.e. $X\rightarrow-X$. In practice, this is implemented by appending virtual $Z$ gates on each side of even pulses. In the absence of errors, this sign change is undetectable. However, in the presence of control errors, the sign change implies $e^{-\frac{i}{2}\Theta(t) \sigma_x}\rightarrow e^{+\frac{i}{2}\Theta(t) \sigma_x}$. Alternating signs will suppress the coherent and low-frequency contributions of the noise, as can be seen by computing the R-FTTPS accumulated phase: $\zeta_k^R = \pi\bar{\epsilon}\sum_{n=1}^{2k}(-1)^n =0$. The second moment is now characterized by the R-FTTPS ctrl-FFs 
\begin{equation}
    F_k^ R(\omega) = \l|  \frac{\pi}{2} \sum_{n=1}^{2k}  (-1)^{n} e^{-i\omega \tau_n} \r|^2\approx \l( \frac{\pi}{2} \frac{\sin(\omega\tau/2)}{\cos(\omega\tau k/4)}\r)^2
\end{equation}
[see Fig.~\ref{fig:FFs}(c)].

\section{Gaussian vs Square Pulses in FPW Experiments}
\label{app:subsec:gaussian_FPW}

Remarkably, we find in practice that most experimental phenomena observed can be well described within the constant pulse approximation. This model robustness to pulse-shape greatly simplifies the task of simulating driven dynamics. The LME solution can be used whenever $X,\sqrt{X}$ appear in a circuit of interest without the need of any Trotterization. One notable exception to this convenient simplification can be found in the FPW experiments. Since FPW experiments aim to amplify gate errors, it is perhaps not surprising that the results are largely influenced by pulse-shaping effects. In this section, we analyze the differences between Gaussian and constant pulses in FPW experiments.

The result of FPW experiments under Markovian noise can be parametrized with a decay rate $a$ and oscillation frequency $b$ as $f(t;a,b) = e^{-at}\cos(bt)$. In the constant pulse case, it can be shown analytically that $a_{\text{const}}=\frac{3}{4}\gamma+\frac{\lambda+\nu}{2}, b_{\text{const}}=\frac{2}{\pi}\beta$. On the other hand, when the qubit is driven with a Gaussian pulse, it is challenging to find fully analytical expressions for the survival probability of the FPW experiments in the presence of noise. 

\begin{figure}[t]
    \centering\includegraphics[width=\columnwidth]{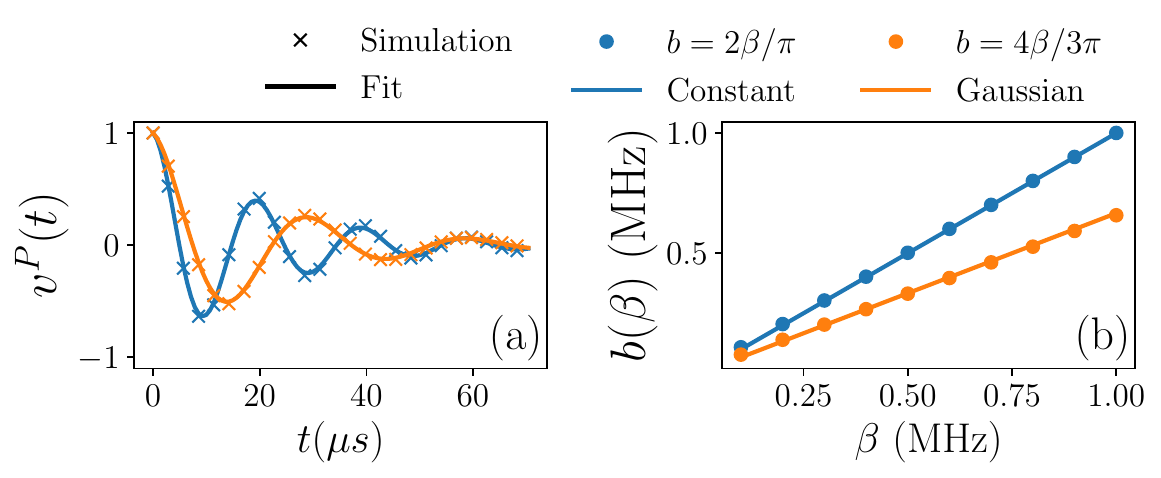}
    \caption{Noisy simulation of FPW circuits for constant (blue) and Gaussian pulses (orange). 
    (a) Simulation for $\beta=0.5~$MHz (crosses) and fit (solid lines) using $f(t;a=a_\text{const},b)= e^{-a t}\cos(bt)$. The decay rates are set to $\gamma=0.05~\text{MHz},\lambda=\nu=0.01~\text{MHz}$. (b) Oscillation frequency parameter $b$ as a function of detuning $\beta$, obtained from fitting $f(t;a=a_\text{const},b)$ to simulations. This shows that for a constant pulse, $v^P(t)$ oscillates with a frequency $2\beta/\pi$, whereas for a Gaussian pulse it oscillates with $4\beta/3\pi$.}
    \label{fig:gauss_const}
\end{figure}

A Gaussian pulse is characterized by $\Omega(t)=$ $A e^{-(t-\delta t/2)^2/2\sigma^2}+B$ for $t\in[0,\delta t)$, where the $B$ parameter is chosen such that the pulse starts and ends at zero, i.e., $\Omega(0)=\Omega(\delta t)=0$. A ubiquitous choice for the width of the Gaussian is $\sigma = \delta t/4$~\cite{wei2023characterizing,Motzoi2009}, while $A$ is chosen to produce the desired rotation angle $\int_0^{\delta t} \Omega(t) dt = \theta$. Under these conditions, we simulate the dynamics using a discretization of $N=160$ steps, consistent with IBM pulse characteristics. The simulation is performed by solving numerically the LME [Eq.~(\ref{eq:LE_v})], with $X$ rotation angles given by $\theta_n = \Omega(n\delta t) \delta t/N$, for $n=0,...,N-1$. In Fig.~\ref{fig:gauss_const}, we show numerically that fitting $f(t;a,b)$ to the simulation results of FPW experiments under Gaussian pulses yields the same decay rate as in the constant pulse case, i.e., $a_{\text{Gauss}}=a_{\text{const}}$, and an oscillation frequency $b_{\text{Gauss}}=\frac{4}{3\pi}\beta = \frac{2}{3} b_{\text{const}}$. This can be interpreted as Gaussian pulses providing suppression with a factor of $2/3$ against detuning processes. Since the experiments are implemented with Gaussian pulses, in order to account for this suppression, we modify the analytical fit expression with a factor of $2/3$, i.e., $v(t) = e^{-\tau \l(\frac{3}{4}\gamma+\frac{\lambda+\nu}{2}\r)}\cos\l(\frac{4}{3\pi}\beta \tau \r)$, as presented in Eq.~(\ref{eq:v_P}).

\section{Relating Stochastic Noise with Error Channels}
\label{app:sec:stochastic_channel}

In this section, we show how some common error channels, namely phase damping and depolarization, can be related to stochastic noise processes with simple statistical properties. 

\subsection{Dephasing Noise and Phase Damping}

Here, we establish a  Hamiltonian description of the  phase damping quantum channel. 
We start with the time-dependent dephasing Hamiltonian given in Eq.~(\ref{eq:stoch-noise-Hamiltonian}) with $\epsilon(t)\equiv0$. The noise-averaged state of the system after time $\tau$ is
\begin{eqnarray}
\mathcal{E}(\rho) &=& \langle U(\tau)\rho U^\dagger(T) \rangle_\beta \nonumber\\
&=& \left\langle \begin{pmatrix}
1 & 0 \\
0 & e^{-i\int_0^\tau \beta(t) dt}
\end{pmatrix}
\begin{pmatrix}
\rho_{00} & \rho_{01} \\
\rho_{10} & \rho_{11}
\end{pmatrix}
\begin{pmatrix}
1 & 0 \\
0 & e^{i\int_0^\tau \beta(t) dt}
\end{pmatrix}
\right\rangle_\beta \nonumber\\
&=& \left\langle
\begin{pmatrix}
\rho_{00} & \rho_{01}e^{i\int_0^\tau \beta(t) dt} \\
\rho_{10}e^{-i\int_0^\tau \beta(t) dt} & \rho_{11}
\end{pmatrix}
\right\rangle_\beta \nonumber\\
&\approx&
\begin{pmatrix}
\rho_{00} & \rho_{01} e^{i \bar{\beta} \tau - \chi(\tau)}  \\
 \rho_{01} e^{- i \bar{\beta} \tau - \chi(\tau)}  & \rho_{11}
\end{pmatrix},
\end{eqnarray}
where $\chi$ is given by Eq.~(\ref{eq:chi}). The decaying component of the off-diagonal elements become dependent on the control in general. However, if the noise is white, i.e., $S(\omega)=S_0$, it is straightforward to show (by following the steps in Sec.~\ref{app:subsec:FFF}) that $\chi(\tau)=S_0 \tau$. This indicates that the superoperator formalism is equivalent to the time-dependent Hamiltonian description when the noise is white. 

More concretely, the phase damping superoperator $\mathcal{E}_{PD}(\rho)$ can be written in terms of the Krauss operators $E_0 = \sqrt{1-p}I, E_1=\sqrt{p} Z$ with damping rate $p$. 
The effect on a general state $\rho$ is 
\seq{
\mathcal{E}_{PD}(\rho) &= E_0 \rho E_0 + E_1 \rho E_1 = \\
&= (1-\frac{p}{2}) \rho + \frac{p}{2} Z\rho Z = \\
&= (1-\frac{p}{2})\begin{pmatrix}
\rho_{00} & \rho_{01} \\
\rho_{10} & \rho_{11}
\end{pmatrix}+
\frac{p}{2}\begin{pmatrix}
\rho_{00} & -\rho_{01} \\
-\rho_{10} & \rho_{11}
\end{pmatrix} =\\
&= \begin{pmatrix}
\rho_{00} & \rho_{01}(1-p) \\
\rho_{10}(1-p) & \rho_{11}
\end{pmatrix}
}
and can be interpreted as acting on the state $\rho$ with a $Z$ operator with a given probability $p$. Adding a detuning noise $\bar{\beta} \sigma_z$ with propagator $U_\beta(t)=e^{-i\bar{\beta} t \sigma_z/2}$ leads to a density matrix
\seq{
U_\beta(t)&\mathcal{E}_{PD}(\rho)U_\beta^\dagger(t) =\\
&= \begin{pmatrix}
\rho_{00} & \rho_{01}e^{-i\bar{\beta} t}(1-p) \\
\rho_{10}e^{i\bar{\beta} t}(1-p) & \rho_{11}
\end{pmatrix},
}
from where the identification of the detuning parameter with the mean of $\beta(t)$ becomes evident. Lastly, by identifying $p = 1 - e^{-S_0 \tau}$ it becomes clear that the PD channel can be interpreted as the white noise limit of a time-dependent dephasing Hamiltonian, where the dephasing time can be defined as $T_2=1/S_0$.

\subsection{Isotropic Uncorrelated Noise and Depolarization}

The control noise case discussed in the main text is entirely analogous to the dephasing case. 
Rather, to showcase the versatility of the stochastic Hamiltonian description, we turn to depolarizing noise. In the superoperator formalism, the Krauss operators for depolarizing noise are $E=\{ \sqrt{1-3p/4}I, \sqrt{p}X/2, \sqrt{p}Y/2, \sqrt{p}Z/2 \}$, which can be interpreted as $\rho$ is left alone with probability of $1-p$, and the operators $X,Y,Z$ being applied with probability $p/3$. The effect of depolarization on a general state can be written as
\seq{
\mathcal{E}_D(\rho) &= p\frac{I}{2} + (1-p) \rho \\
&=
\begin{pmatrix}
\rho_{00} (1-p) + p/2 & \rho_{01} (1-p)  \\
\rho_{10}(1-p)  & \rho_{11} (1-p) + p/2
\end{pmatrix}
}

Next, we propose a general time-dependent Hamiltonian, and find the conditions that the parameters need to satisfy in order for the stochastic Hamiltonian to reproduce the effects of depolarizing noise. We write a Hamiltonian $H_{N}(t) = \vec{\eta}(t)\cdot \vsigma$, where we consider $\vec{\eta}(t)$ a Gaussian and stationary random variable with zero mean. The time evolution propagator is 
\seq{
U(\tau) &= e^{-i \int_0^\tau \vec{\eta}(t) dt \cdot \vec{\sigma}} = e^{-i \vec{\theta}_\eta \cdot \vsigma} \\
&= I \cos\theta_\eta - i (\hn_\eta \cdot \vsigma) \sin\theta_\eta
}
where we defined $\vtheta_\eta = \int_0^\tau \vec{\eta}(t) dt = \theta_\eta \hn_\eta$, with $\hn_\eta^i=\vtheta_\eta^i/\theta_\eta$ a unit vector. The state becomes,
\seq{
\label{eq:E_dep}
\mathcal{E}(\rho) &= \langle U(\tau)\rho U^\dagger(\tau) \rangle_\eta \\
&= \rho  \left\langle\cos^2\theta_\eta \right\rangle_\eta -i [\left\langle \cos\theta_\eta\sin\theta_\eta \hn_\eta\right\rangle_\eta\cdot\vsigma, \rho] + \\
&+ \left\langle ( \hn_\eta \cdot \vsigma ) \rho ( \hn_\eta \cdot \vsigma )  \sin^2\theta_\eta \right\rangle_\eta
}
In the weak noise approximation, the first two terms simplify greatly. Dropping the subscript $\eta$ from the noise-averaging, these terms can be expressed as
\eq{
\left\langle\cos^2\theta_\eta \right\rangle &\approx \left\langle\left(1-\frac{1}{2}\theta_\eta^2\right)^2 \right\rangle =   1-\left\langle\theta_\eta^2 \right\rangle,  \\
\left\langle \cos\theta_\eta\sin\theta_\eta \hn_\eta\right\rangle &\approx \left\langle \left(1-\frac{1}{2}\theta_\eta^2\right) \theta_\eta \frac{\vtheta_\eta}{\theta_\eta}\right\rangle = \left\langle \vtheta_\eta\right\rangle = 0.
}
Next, we make two additional assumptions regarding the statistical properties of the noise parameters: (1) we assume that the noise is uncorrelated noise along different axes and (2) noise isotropy, i.e.,
\eq{
\left\langle \vtheta_\eta^i \vtheta_\eta^j \right\rangle &= \left\langle \vtheta_\eta^i \vtheta_\eta^i \right\rangle \delta_{ij}, \\
\left\langle \vtheta_\eta^i \vtheta_\eta^i \right\rangle &= \frac{\left\langle \theta_\eta^2 \right\rangle}{3},
}
respectively. With these simplifying assumptions, the last term in Eq.~(\ref{eq:E_dep}) becomes
\begin{equation}
\langle ( \hn_\eta \cdot \vsigma ) \rho ( \hn_\eta \cdot \vsigma )  \sin^2\theta_\eta \rangle \approx
\sum_{i,j}  \left\langle \hn_\eta^i \hn_\eta^j \theta_\eta^2 \right\rangle \sigma_i \rho  \sigma_j. \nonumber
\end{equation}
This can be further simplified as
\begin{eqnarray}
\sum_{i,j}  \left\langle \hn_\eta^i \hn_\eta^j \theta_\eta^2 \right\rangle \sigma_i \rho  \sigma_j &=& \sum_{i,j}  \left\langle \frac{\vtheta_\eta^i \vtheta_\eta^j}{\theta_\eta^2} \theta_\eta^2 \right\rangle \sigma_i \rho  \sigma_j 
= \sum_{i,j}  \left\langle \vtheta_\eta^i \vtheta_\eta^j \right\rangle \sigma_i \rho  \sigma_j \nonumber\\
&=& \sum_{i}  \left\langle \vtheta_\eta^i \vtheta_\eta^i \right\rangle \sigma_i \rho  \sigma_i  
= \frac{\left\langle \theta_\eta^2 \right\rangle}{3} \sum_{i} \sigma_i \rho \sigma_i \nonumber\\
&=&  \frac{\left\langle \theta_\eta^2 \right\rangle}{3} \left( 2I-\rho \right).
\end{eqnarray}
Thus,
\begin{eqnarray}
\mathcal{E}(\rho) &=& \rho \left( 1-\left\langle\theta_\eta^2 \right\rangle \right) + \frac{\left\langle \theta_\eta^2 \right\rangle}{3} \left( 2I-\rho \right) \nonumber\\
&=& \frac{2}{3} \left\langle \theta_\eta^2 \right\rangle I + \left( 1- \frac{4}{3}\left\langle\theta_\eta^2 \right\rangle \right) \rho.
\end{eqnarray}
From here, the depolarizing error can be recovered by setting
\eq{
p &= \frac{4}{3}\left\langle\theta_\eta^2 \right\rangle = \frac{4}{3}\sum_i \int_0^\tau \int_0^\tau dt dt' \left\langle \eta_i(t) \eta_i(t')  \right\rangle
}
Analogously to the dephasing case, a constant $p$ implies that noise must be uncorrelated and isotropic in the Hamiltonian description. In this case, if $\l\langle \gamma_i(t) \gamma_i(t')  \r\rangle = C_{\eta} \delta(t-t')$, we get $p = 4C_\eta\tau/3$.

\section{Evidence of Correlated Dephasing Through DD}

\begin{figure}[t]
    \centering
    \includegraphics[width=\columnwidth]{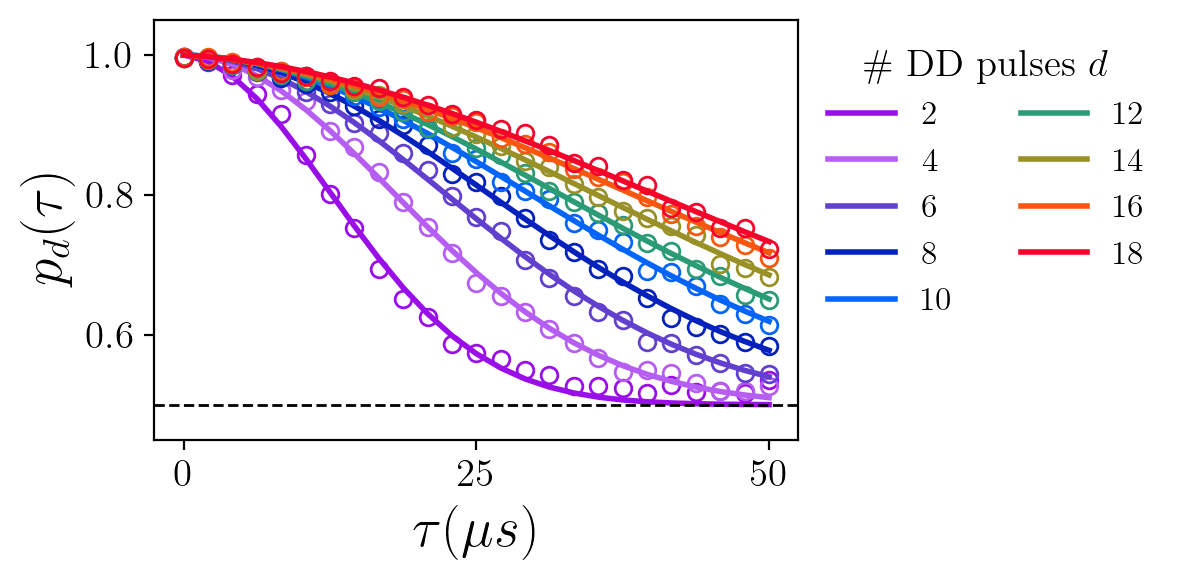}
    \caption{Experimental results (circles) obtained from running an increasing number of DD pulses on qubit 6 of \textit{ibmq\_guadalupe}. Improved performance with the number of DD pulses is observed. The experiment results are fit to a function $\chi_d(\tau) = A/d + B$, where the parameters $A,B$ are common to all DD experiments. The results of these fits are presented as solid lines, where we obtain $A=0.0063(2)$ and $B\approx 0$ (within error bars).}
    \label{fig:T2_DD}
\end{figure}

In Sec.~\ref{sec:experiment_results}, we explore the detection and modeling of correlated dephasing noise. By adding a number of DD pulses on T2 experiments, dephasing can be detected by comparing the relative performance. Figure~\ref{fig:T2_DD} shows experimental results obtained from running an increasing number of DD pulses on qubit 6 of \textit{ibmq\_guadalupe}, as function of total evolution time $\tau$. 

It is clear that the performance improves with the number of DD pulses. Based on the theory of DD, if this improvement is due to increased filtering of dephasing noise, the decay rate should decrease as $1/d$, where $d$ is the number of DD pulses. We fit the function $p_\mrm{D}(\tau) = (1+e^{-\chi_d(\tau)})/2$, where $\chi_d(\tau)\approx \int S(\omega) \mrm{sinc}((\omega-\omega_d)\tau)^2 d\omega$ is the overlap function. Note that the number of DD pulses shifts the center of the frequency response, given by $\omega_d=d\, 2\pi/\tau$. Then, we can fit this experiment results to a function $\chi_d(\tau) = A/d+B$, where the parameters $A,B$ are common to all DD experiments. The results of these fits are presented as solid lines with excellent agreement, obtaining further confirmation of the presence of correlated dephasing noise.

\section{Parametrizing the Markovian RB Decay Rate}
\label{app:sec:RB_dependence}

\begin{figure}[t]
    \centering
    \includegraphics[width=\columnwidth]{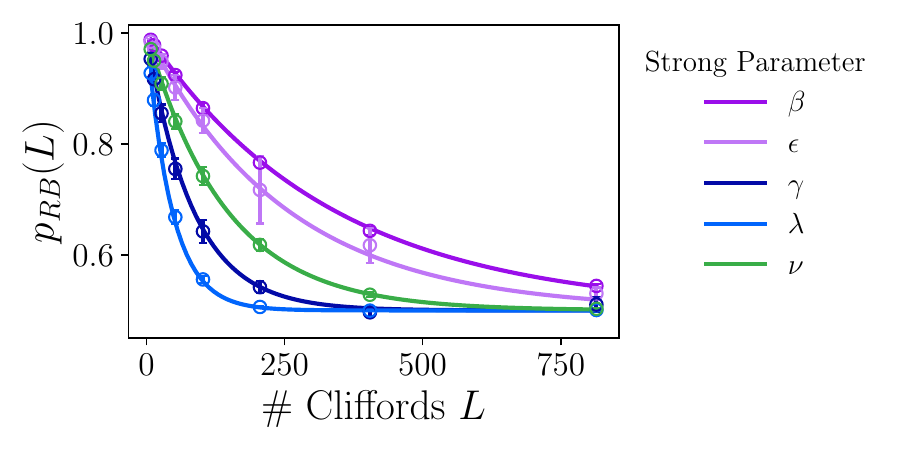}
    \caption{RB simulations (circles) with Markovian parameters chosen as shown in Table~\ref{table:noise_params}. Good agreement is found with the analytical predictions (solid lines) obtained from Eq.~(\ref{eq:RB_decay_rate_analytical}).}
    \label{fig:RB_params}
\end{figure}

We perform numerical studies aimed to examine the impact of Markovian errors on the RB decay rate $r_{RB}$. In the absence of SPAM errors, the RB experiment result can be modeled as $p_{RB}(L) = (1+e^{-r_{RB} L})/2$, where $L$ is the number of Cliffords in a given RB sequences. We perform simulation of multiple combinations of the Markovian parameters, and find that, to first order, the decay rate can be approximated as
\eq{
\label{eq:RB_decay_rate_analytical}
r_{RB} \approx \delta t(\gamma+\lambda+\nu) + \l(\frac{\pi \epsilon}{2}\r)^2.
}
Thus, the impact of dissipative Markovian errors on the decay rate increases linear as a function of the time elapsed in the sequence. Coherent errors, on the other hand, enter quadratically. Note that these results are consistent with the literature, see for example Ref.~\cite{Hashim2021}. Figure~\ref{fig:RB_params} presents examples of simulations performed with Markovian parameters near values obtained from real devices. The noise parameters were chosen to be ``weak'' by default, while varying one parameter set as ``strong'' (see left and right columns of Table~\ref{table:noise_params}, respectively). The exponential decays (solid lines) are analytically computed by combining Eq.~(\ref{eq:RB_decay_rate_analytical}) and $p_{RB}(L)$.

\section{Resonances in FTTPS}
\label{sec:Sw_resonance}

\begin{figure}[t]
    \centering
    \includegraphics[width=\columnwidth]{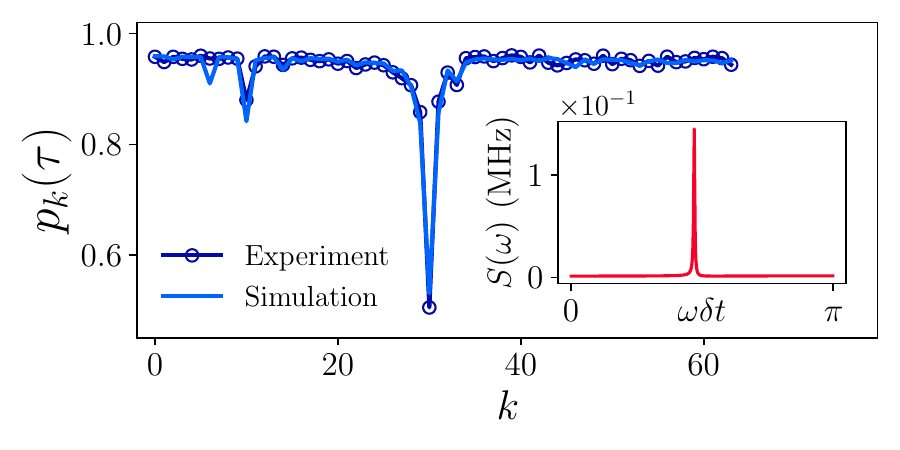}
    \caption{FTTPS experiment  results (circles) and simulation (solid lines) on qubit 13 of \textit{ibm\_algiers}. A resonance peak is observed at $k=30$, consistent with a frequency of approximately $41\pm1\,$MHz. Inset: PSD obtained from the FTTPS experiment, using the QNS protocol outlined in Sec.~\ref{app:subsec:QNS}. This PSD is then used to generate the noise trajectories required for simulation. Excellent agreement is found between experiment and simulation, as quantified by the MSE defined in Sec.~\ref{subsec:overview}.}
    \label{fig:FTTPS_resonance}
\end{figure}

As discussed in Sec.~\ref{subsubsec:corr_deph_exp}, low-frequency dephasing noise is prominent in the superconducting qubit devices examined in this study. Despite the common spectral features of low-frequency noise accompanied by a white noise floor, exceptions do exist. In particular, we find a subset of qubits that exhibit resonance peaks at high frequency. An example of this phenomenon can be seen in Fig.~\ref{fig:FTTPS_resonance}, detected in an FTTPS experiment. 

\begin{figure}[t]
    \centering    
    \includegraphics[width=\columnwidth]{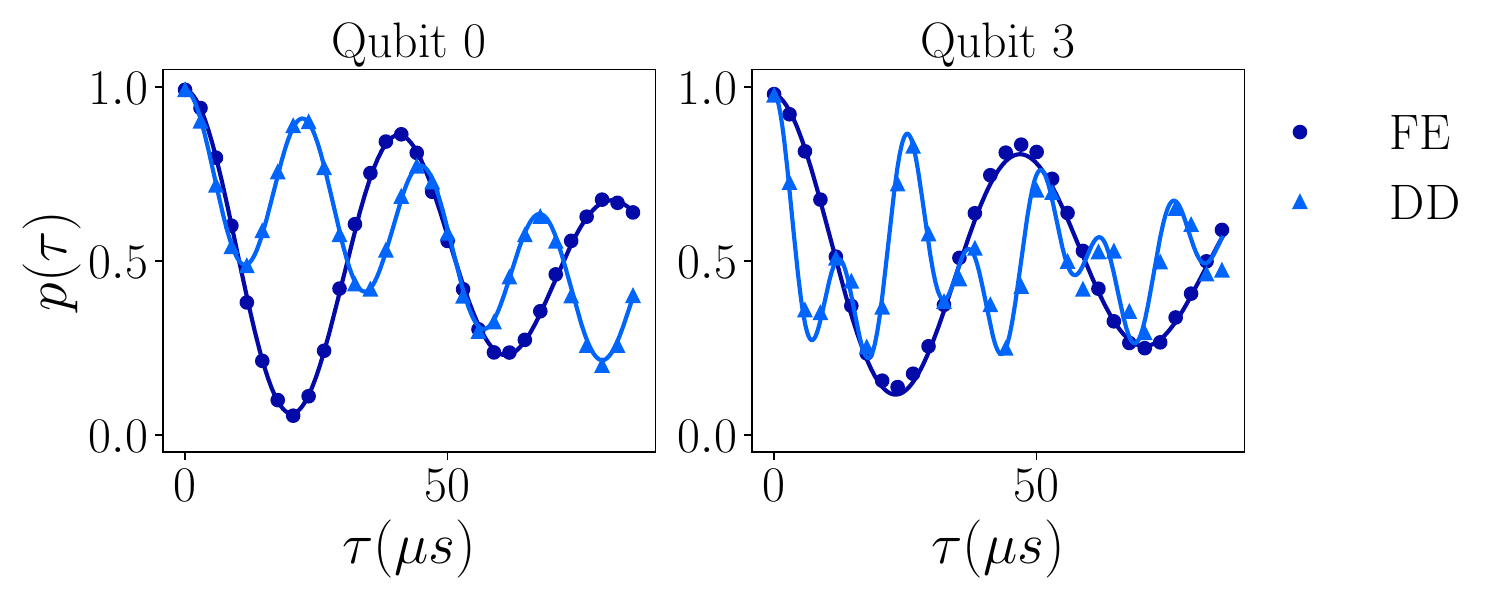}
    \caption{Ramsey experiments on qubits 0 and 3 of \textit{ibm\_cairo}, where spectator qubits are left to evolve freely (dark-blue circles; FE) and decoupled via XY4 (light-blue triangles; DD). Note that XY4 is applied exclusively on spectator qubits. Solid lines correspond to multi-qubit LME simulations including TLS and crosstalk couplings, where excellent agreement with the experiments is found. }
    \label{fig:TLS_XT}
\end{figure}

The large resonance at $k=30$ can be mapped to a frequency value by drawing upon the FFF (see Sec.~\ref{app:subsec:FFF}). The $k^\mrm{th}$ FF $F_k(\omega)$ has a narrow frequency sensitivity peak centered at $\omega=\frac{2\pi k}{\tau}$ (see Fig.~\ref{fig:FFs}(a)), where $\tau=2K\delta t$ is the FTTPS duration. Consequently, the resonance observed in Fig.~\ref{fig:FTTPS_resonance} corresponds to a frequency of $\approx 41\pm1\,$MHz. This frequency value is consistent with that observed in other qubits, typically in the range of 25 to 70$\,$MHz. A possible origin of this phenomenon is frequency collisions with neighboring qubits, since IBMQP Eagle generation devices have typical resonance frequency differences between nearest-neighbor qubits in the order of 10-100 MHz. A thorough study of the physical origin of this resonant phenomenon is left for future work.

Note that, as is the case with low-frequency noise, the PSD of this noise can be fit following the same procedure outlined in Sec.~\ref{app:subsec:QNS}. The resulting PSD is shown in the inset. The PSD can be used alongside \textit{mezze} in a stochastic noise simulation, showing excellent agreement with experiment.

\section{TLS and Crosstalk}
\label{app:sec:TLS_XT}

An interesting phenomenon was identified within the TLS model in the presence of crosstalk and local detuning. As introduced in Sec.~\ref{subsubsec:TLS_exp}, many qubits exhibit the TLS behavior in the Ramsey experiments; this is exemplified in four different qubits in Fig.~\ref{fig:TLS}. The multi-qubit model introduced in this paper has three relevant frequencies associated to the Ramsey experiment: the TLS-to-qubit and qubit-qubit crosstalk coupling strength $\xi$ and $J$, and the local single-qubit detuning frequency $\beta$. When all these are taken into account, the LME model yields
\seq{
\label{eq:v_TLS_XT}
v(\tau) = e^{- \alpha \tau} \cos[(\beta+J)\tau]\cos(\xi \tau),
}
where $\alpha=\gamma/2+\lambda$. Note that this Bloch vector corresponds to the single-qubit Ramsey experiments, where the spectator qubit remains in the $\ket{0}$ state and evolving freely.

\begin{figure}[t]
    \centering    
    \includegraphics[width=\columnwidth]{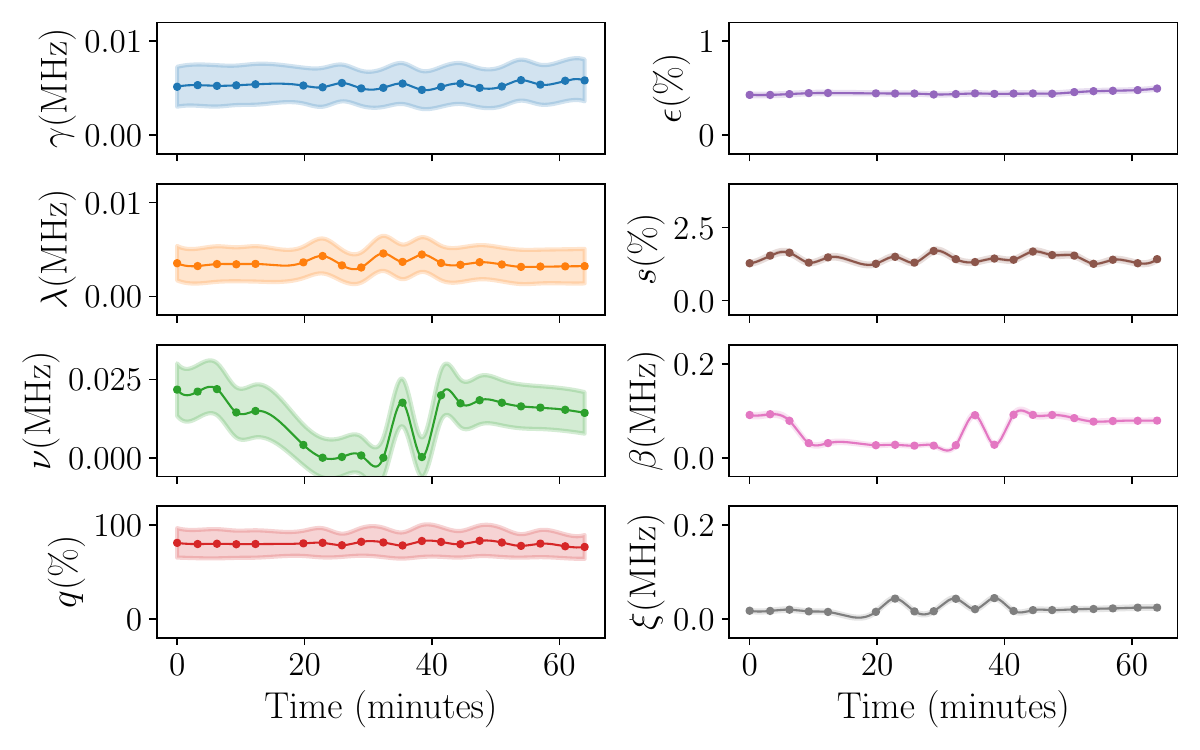}
    \caption{Results (dots) obtained from characterization experiments on qubit 117 of \textit{ibm\_osaka}, which fits well a Markovian model. The same batch of experiments was run 20 times in the span of one hour, on February 10th, 2024. Time on the horizontal axis is measured from the first experiment run, at 22:23hs. Solid lines are obtained from cubic spline interpolations. Shaded regions represent uncertainties obtained from the fit.  }
    \label{fig:model_evol}
\end{figure}

Examples of these Ramsey experiments can be seen for two different qubits in Fig.~\ref{fig:TLS_XT} (dark-blue circles; denoted as FE). Both of these qubits were selected due to their seemingly close-to-uniform oscillations, where the oscillations can be well described by a single relevant frequency.
Looking at Eq.~(\ref{eq:v_TLS_XT}), we notice that this poses two alternatives: either $\xi\approx0$, or $\beta=-J$. In order to resolve this uncertainty, we performed another experiment. Immediately after the single-qubit Ramsey experiment is run, the experiment is repeated with the XY4 decoupling sequence applied to all spectator qubits in order to suppress quantum crosstalk between the primary qubit and its spectators. The results of this experiment are shown as light blue triangles in Fig.~\ref{fig:TLS_XT} (denoted as DD). The remaining oscillation suggests that the TLS is present. As such, the relevant equation that describes these experiments is Eq.~(\ref{eq:v_TLS}). From the both experiments, $\xi$ and $\beta$ can be fit, from where we find that one of the values is consistent with the single frequency oscillation obtained in the original Ramsey experiments. Consequently, we can positively state that, within our model, $\beta\approx-J$, and that the frequency observed in the pure Ramsey experiment corresponds to the TLS coupling $\xi$. 

The interpretation of these results is that the qubits's driving frequencies are calibrated to precisely cancel the combined effect of neighboring crosstalk. Thus, this result adds another layer of complexity to the modeling of dephasing noise, since the TLS frequencies are seen to oscillate on a timescale much shorter than detuning frequencies. A detailed study of the consequences of the presence of TLS on stochastic dephasing noise is left for future work. Finally, we believe that, in principle, it is a reasonable assumption to consider all qubits to have non-zero $\beta,J,\xi$.

\begin{figure}[t]
    \centering
    \includegraphics[width=1\linewidth, trim={0cm 4cm 4cm 0cm},clip]{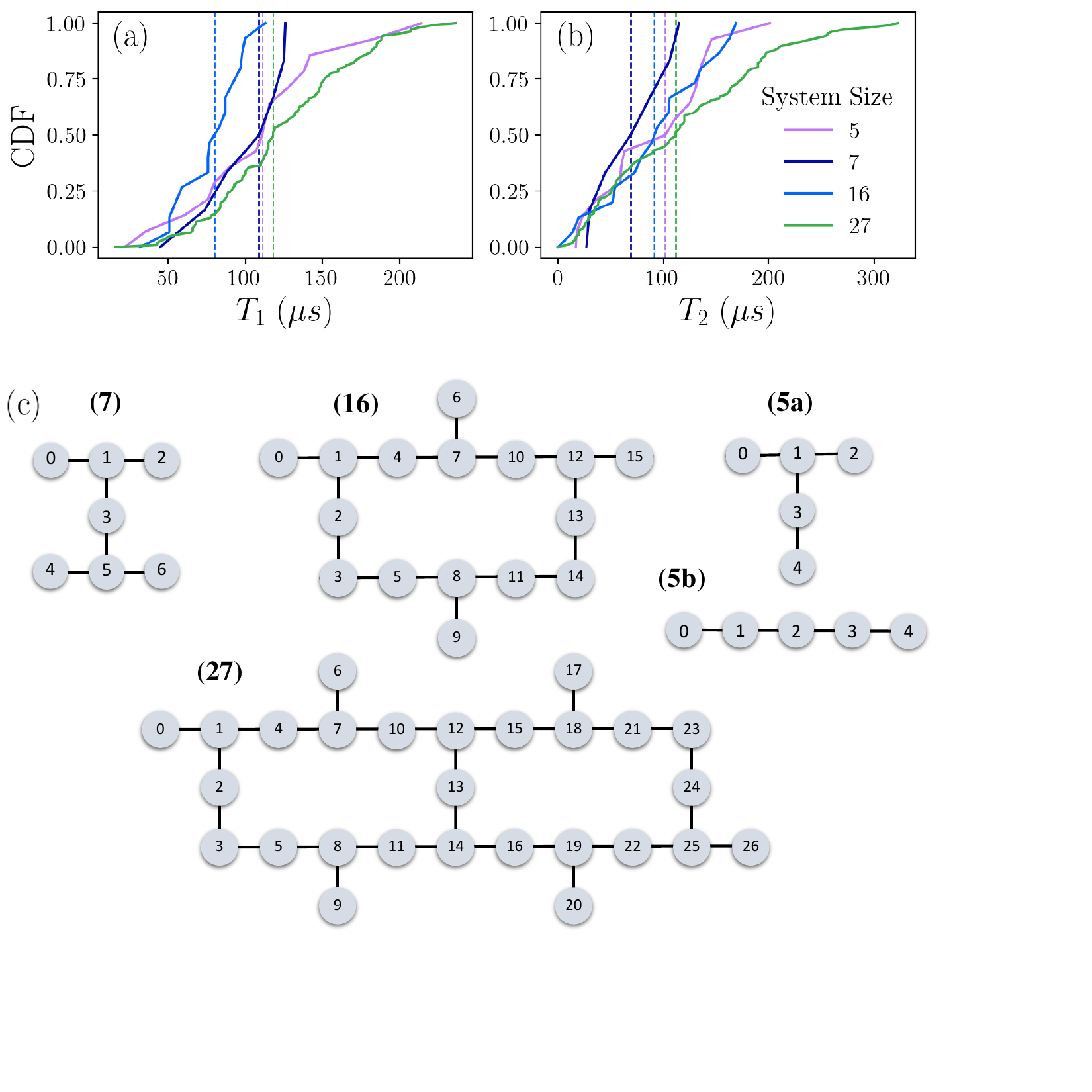}
    \caption{Device connectivity and coherence times properties. We show cumulative distribution functions (CDF) of (a) $T_1$ and (b) $T_2$ times, collected by system size in number of qubits: (5a) \textit{ibmq\_lima}, \textit{ibmq\_belem}, (5b) \textit{ibmq\_manila}, (7) \textit{ibmq\_lagos}, (16) \textit{ibm\_guadalupe}, (27) \textit{ibm\_auckland}, \textit{ibm\_cairo}, \textit{ibm\_hanoi}, \textit{ibm\_algiers}. 
    (c) Device layouts for each system size.}
    \label{fig:device_cdf}
\end{figure}

\section{Model Stability}
\label{app:sec:model_stability}

We investigate the stability of the model parameters over time. Figure~\ref{fig:model_evol} presents experimental data obtained from fitting the results of the characterization experiments over a window of one hour. As shown, most parameters have a good degree of stability within this window. A thorough statistical analysis of the stability over many qubits and longer time windows is required to more adequately determine if these trends continue or if significant drift occurs. These time sensitive experiments likely require dedicated access to hardware that differs from the standard IBMQP access. For that reason, we leave this analysis for future work, but provide preliminary insight here.

\section{Device Properties}
\label{sec:device_cdf}

In this section, we present device topologies and cumulative distribution functions (CDFs) of IBMQP devices used throughout this work. See Fig.~\ref{fig:device_cdf} for device layouts and information regarding $T_1$ and $T_2$ times of qubits in devices studied, obtained from IBM's calibration data.


%

\end{document}